\documentclass[12pt]{article}
\pdfoutput=1
\usepackage[top=30truemm,bottom=30truemm,left=25truemm,right=25truemm]{geometry}
\usepackage{authblk}
\usepackage{amsthm}
\usepackage{amsmath,amssymb,bm,mathtools}

\DeclareMathOperator{\arccosh}{arccosh}
\DeclareMathOperator{\arcsinh}{arcsinh}
\DeclareMathOperator{\arctanh}{arctanh}

\usepackage{cite}
\usepackage[]{hyperref}
\hypersetup{colorlinks,linkcolor={blue},citecolor={blue},urlcolor={blue}}  
\usepackage[titletoc]{appendix}
\usepackage{url}
\usepackage{graphicx,color}
\usepackage{braket}
\usepackage{qcircuit}
\usepackage{here}
\usepackage{mathrsfs}
\usepackage{anyfontsize}
\usepackage{subcaption}
\usepackage{tikz}
\usepackage{pgfplots}
\usepackage{titlesec}
\pgfplotsset{compat=1.18}
\usepgfplotslibrary{fillbetween}

\newtheorem*{definition*}{Definition}
\newtheorem*{assumption*}{Assumption}

\def\coth{{\mathrm{coth}}}
\def\sinh{{\mathrm{sinh}}}
\def\cosh{{\mathrm{cosh}}}
\def\tanh{{\mathrm{tanh}}}

\def\CB{{\cal B}}

\def\CD{{\cal D}}

\def\CS{{\cal S}}

\def\CM{{\cal M}}

\def\be{\begin{equation}}
\def\ee{\end{equation}}
\def\ba{\begin{eqnarray}}
\def\ea{\end{eqnarray}}

\numberwithin{equation}{section}

\def\Z2{\mathbb{Z}_2}

\AtBeginDocument{%
  {\footnotesize\global\footnotesep=0.8\baselineskip}%
}

\definecolor{cardinal}{rgb}{0.6,0,0}
\definecolor{darkgreen}{rgb}{0,0.4,0}
\definecolor{golden}{rgb}{0.92, 0.7, 0}
\definecolor{midnight}{rgb}{0, 0, 0.5}
\definecolor{darkblue}{rgb}{0, 0, 0.7}
\definecolor{purple}{rgb}{0.5, 0, 0.5}

\newcommand{\Blue}{\color{blue}}

\usepackage{tikz}
\usetikzlibrary{calc,decorations.pathreplacing,intersections,arrows.meta,angles,quotes}

\begin{document}

\begin{titlepage}
\thispagestyle{empty}

\vspace*{-2.5cm}

\begingroup
    \fontsize{11}{11}\selectfont
    \begin{flushright}
MIT-CTP/5899
\\
YITP-26-37
\\
IPMU26-0015 
\end{flushright}
\endgroup

\medskip

\begin{center}
\noindent{\bf \Large A Tale of Two Hartle--Hawking Wave Functions:
}\\
\vspace{0.1cm}
\noindent{\bf \Large Fully Gravitational vs Partially Frozen}\\
\vspace{1.0cm}

{\bf Galit Anikeeva$^{\,a}$, Rapha\"el Dulac$^{\,b}$, Zixia Wei$^{\,c}$, and Mengyang Zhang$^{\,d}$}
\vspace{0.8cm}\\

{\it 
$^a$MIT Center for Theoretical Physics—a Leinweber Institute, \\ Massachusetts Institute of Technology,
Cambridge, MA 02139, USA
}\\[1.5mm]

{\it
$^b$Institut de Physique Th\'eorique,
Universit\'e Paris Saclay, 
CEA, CNRS, \\
Orme des Merisiers, Gif sur Yvette, 91191 CEDEX, France
}\\[1.5mm]

{\it 
$^c$Yukawa Institute for Theoretical Physics, Kyoto University, Kyoto 606-8502, Japan}\\[1.5mm]

{\it 
$^c$Society of Fellows, 
Harvard University, Cambridge, MA 02138, USA
}\\[1.5mm]

{\it 
$^d$Kavli Institute for the Physics and Mathematics of the Universe (WPI), \\
The University of Tokyo, Kashiwa, Chiba 277-8583, Japan}\\[1.5mm]



\vspace{0.3cm}


\medskip

\end{center}

\begin{abstract}

We revisit the Hartle--Hawking wave function in AdS spacetime, where natural spatial slices are open and require an additional spacetime boundary. This leads to two constructions: a fully gravitational wave function, in which the boundary configuration is integrated over, and a partially frozen one, in which it is fixed, as in AdS/CFT. 
To illustrate the fully gravitational construction, we explicitly analyze it in AdS$_3$ Einstein gravity and AdS$_2$ Jackiw-Teitelboim gravity. 
We then evaluate the one-loop correction to the hyperbolic-ball partition function in $D$-dimensional AdS Einstein gravity, expected to give the leading contribution to the wave-function norm.
We demonstrate that the fully gravitational hyperbolic ball partition function—where the boundary fluctuates—develops a nontrivial one-loop phase of $(\mp i)^{D+1}$, analogous to that of the sphere partition function in dS gravity. 
By contrast, the partially frozen partition function, where the boundary is fixed, remains real and positive.
Motivated by this AdS comparison, we conversely investigate a partially frozen dS sphere partition function where the metric on an equator is fixed, finding that its one-loop phase cancels nontrivially.
Our results suggest that the phase problem is controlled by whether the gravitational path integral is fully dynamical or partially frozen.
\end{abstract}

\end{titlepage}

\newpage
\setcounter{page}{1}
\begingroup
    \fontsize{11}{11}\selectfont
    \linespread{1.0}
    \tableofcontents
\endgroup

\newpage

\section{Introduction}\label{sec:introduction}

\begin{figure}[ht]
    \centering
    \includegraphics[width=15cm]{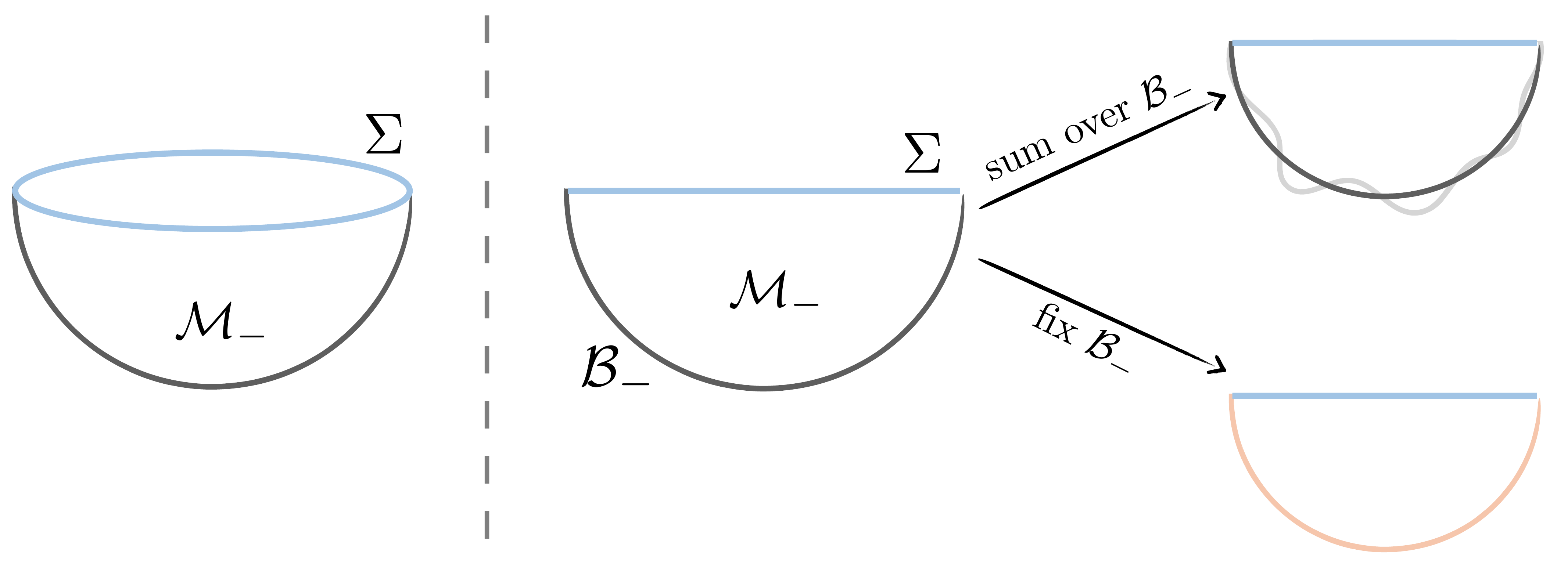}
    \caption{(Left) In the original HH proposal, data are fixed on a closed spatial slice $\Sigma$ (blue circle), and one sums over $\CM_-$ compatible with those data, represented here by a half-sphere.
    (Right) If $\Sigma$ is open, $\partial \Sigma \neq \varnothing$, a Euclidean history ending on $\Sigma$ (blue interval) necessarily contains an additional spacetime boundary component $\CB_-$ (gray curve). 
    This leads to two possible prescription: either summing over configurations of $\CB_-$ or keeping $\CB_-$ fixed. 
    In the upper (lower) figure, the wavy curve (orange curve) indicates a dynamical (frozen) $\CB_-$. 
    }
    \label{fig:HH_intro}
\end{figure}

The Hartle–Hawking (HH) proposal \cite{HH83} provides an elegant definition for the wave function of the entire universe via the gravitational path integral (GPI). In its original form, one considers a closed $(D-1)$-dimensional spatial slice $(\Sigma,\gamma_{ab})$ with fixed metric in a $D$-dimensional gravitational theory with Euclidean action $I_{\rm E}[g_{\mu\nu}]$. The HH wave function is formally defined by summing over compact Euclidean geometries $(\CM_-,g_{\mu\nu})$ whose only boundary is $\Sigma$: 
\begin{align}\label{eq:original_HH}
    \Psi_{\rm HH}[\gamma_{ab}] \equiv \int_{\partial \CM_- = \Sigma} \mathcal{D}g_{\mu\nu} ~e^{-I_{\rm E}[g_{\mu\nu}] }\,,
\end{align}
Each $\CM_-$ can be regarded as a Euclidean history interpolating between ``nothing" and $\Sigma$. Therefore, the HH proposal is often called the ``no-boundary'' proposal. See the left panel of figure \ref{fig:HH_intro} for a sketch. 

Most studies of \eqref{eq:original_HH} have focused on de Sitter (dS) gravity. 
However, despite more than four decades after its proposal, many fundamental aspects of the HH wave function remain unresolved. For reviews from a modern perspective, see for example \cite{Lehners23,Maldacena24-1}. 
Many of these issues are rooted in the fact that dS gravity remains poorly understood as a quantum theory.
By contrast, our understanding of quantum gravity in asymptotically anti-de Sitter (AdS) spacetime has advanced substantially following the discovery of the AdS/CFT correspondence \cite{Maldacena97,GKP98,Witten98}. 
This motivates a careful reconsideration of the HH wave function in asymptotically AdS spacetime: do the problems associated with the dS HH wave function also appear in AdS, and if so, how should they be understood in light of AdS/CFT?

\paragraph{HH wave function on an open spatial slice}~\par

A new feature appears immediately in AdS. Natural spatial slices are open rather than closed, and have a nontrivial boundary $\partial \Sigma \neq \varnothing$. Consequently, a Euclidean history $\CM_-$ ending on such a slice cannot have only $\Sigma$ as its boundary. Instead, it must also contain a spacetime boundary component $(\CB_-,h_{ij})$ satisfying 
\begin{align}
    \partial \CB_- = \partial \Sigma, ~~~~\partial \CM_- = \Sigma \cup \CB_-. 
\end{align}
The spatial slice $\Sigma$ and the spacetime boundary $\CB_-$ meet along the codimension-two corner $\partial \Sigma$, as shown in figure \ref{fig:HH_intro}.
This simple feature leads to two inequivalent generalizations of the HH wave function. The distinction between them will be central in this paper. 

The first possibility is to follow the spirit of the original HH proposal as closely as possible and sum over all Euclidean geometries compatible with the spatial data $\Sigma$. In particular, the metric on the spacetime boundary $\CB_-$ is not fixed, but is instead integrated over. This gives
\begin{align}\label{eq:open_HH}
    \boxed{\Psi_{\rm HH}[\gamma_{ab}] \equiv  
    \int_{\CB_-} \CD h_{ij}
    \int_{\partial \CM_- = \Sigma \cup \CB_-} \mathcal{D}g_{\mu\nu} ~e^{-I_{\rm E}[g_{\mu\nu}] }= \int_{\partial \CM_- \supset \Sigma} \mathcal{D}g_{\mu\nu} ~e^{-I_{\rm E}[g_{\mu\nu}] }}~, 
\end{align}
In the final expression, the notation $\partial \CM_- \supset \Sigma$ emphasizes that the metric on $\Sigma$ is fixed in the GPI, while any additional spacetime boundary components and their metrics are summed over. We will refer to \eqref{eq:open_HH} as the {\it fully gravitational HH wave function}\footnote{Unfortunately, we can no longer call this a no-boundary proposal. It might be, however, called an ``all-boundary'' proposal or a ``no-boundary-condition" proposal. }, because the spacetime boundary $\CB_-$ is itself gravitating.

\begin{figure}
    \centering
    \includegraphics[width=0.9\linewidth]{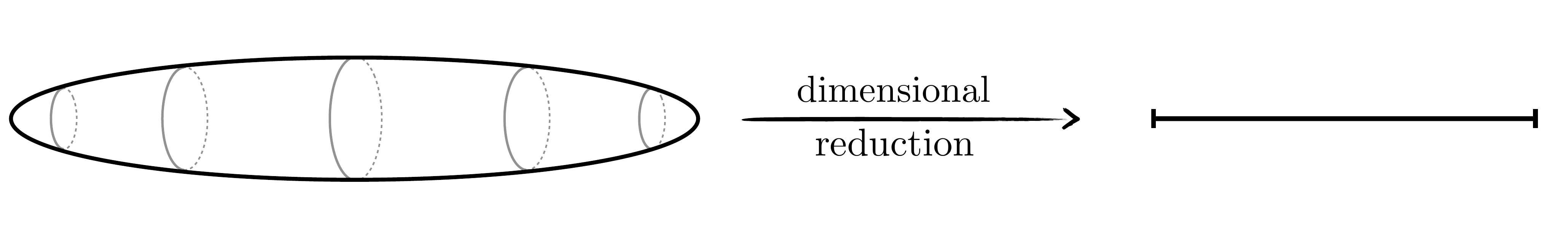}
    \caption{The HH computation \eqref{eq:open_HH} for open spatial slices $\Sigma$ naturally arises from the original HH proposal for closed spatial slices after dimensional reduction. 
    Here this figure shows how a boundary can appear from a dimensional reduction of a closed manifold in higher dimensions. 
    From the lower-dimensional point of view, $\CB_-$ is dynamical and should be summed over in the GPI.}
    \label{fig:dim_reduction}
\end{figure}

This is a natural extension of the original no-boundary proposal. Even if one starts from the strict no-boundary path integral \eqref{eq:original_HH} on a closed spatial slice, spacetime boundary components like $\CB_-$ can appear after dimensional reduction, as illustrated in figure \ref{fig:dim_reduction}. 
{For example, a compact higher-dimensional geometry with a shrinking internal cycle may reduce to a lower-dimensional geometry with an apparent end-of-the-world or asymptotic boundary component. From the lower-dimensional point of view, the data on this component are not fixed externally and should therefore be included in the gravitational path integral.}

The second possibility is to keep the spacetime boundary $\CB_-$ fixed and impose the Dirichlet boundary condition (DBC) on it. This is the version most naturally associated with AdS/CFT and is the one studied in much of the previous literature on AdS HH wave functions \cite{HJ18,BCT20,BCGT21,CJ23}. One fixes an open spatial slice $(\Sigma,\gamma_{ab})$ together with the spacetime boundary $(\CB_-,h_{ij})$, and defines\footnote{In this sense, \eqref{eq:partially_frozen_HH} might be called a ``Dirichlet-boundary-condition proposal".}
\begin{align}\label{eq:partially_frozen_HH}
    \boxed{\Psi_{\rm HH}^{\CB_-}[\gamma_{ab}] \equiv \int_{\partial \CM_- = \Sigma \cup \CB_-} \mathcal{D}g_{\mu\nu} ~e^{-I_{\rm E}[g_{\mu\nu}] }}~, 
\end{align}
Here the superscript indicates that $\CB_-$ is frozen in this GPI. We will refer to \eqref{eq:partially_frozen_HH} as the {\it partially frozen HH wave function}. Schematically, the two HH wave functions are related by
\begin{align}\label{full_HH}
    \Psi_{\rm HH}[\gamma_{ab}] = \int_{\CB_-} \mathcal{D}h_{ij}~  \Psi_{\rm HH}^{\CB_-}[\gamma_{ab}].  
\end{align}
Note that $\Psi_{\rm HH}^{\CB_-}[\gamma_{ij}]$ is morally different from the original HH proposal \eqref{eq:original_HH}: in addition to the spatial geometry on $\Sigma$,  the GPI fixes extra data on $\CB_-$, away from $\Sigma$. 

While it would be great to perform the full GPI, saddle-point approximations are often useful.
The saddle-point problems associated with \eqref{eq:open_HH} and \eqref{eq:partially_frozen_HH} are correspondingly different. In the partially frozen construction, the metric on $\CB_-$ is fixed, as in the usual AdS/CFT GPI \cite{Witten98}. In the fully gravitational construction, by contrast, the boundary metric is varied, as in \cite{CM08,Takayanagi11}. Therefore, the boundary condition on $\CB_-$ is not imposed by hand but rather arises as an equation of motion (EOM). For example, in AdS$_3$ Einstein gravity, extremizing the renormalized Euclidean action with respect to the metric on $\CB_-$ gives the Neumann-type boundary condition 
\begin{align}
    K_{ij}|_{\mathcal{B}_-} - K_{\mathcal{B}_-} h_{ij} + h_{ij} = 0\,.
\end{align}
We will analyze this fuly gravitational saddle-point problem explicitly in AdS$_3$ Einstein gravity and in AdS$_2$ Jakiw-Teitelboim gravity in section \ref{sec:AdS3} and section \ref{sec:JT-noboundary}, respectively.

\paragraph{The phase issue in the norm computation}~\par

We aim to use the comparison between $\Psi_{\rm HH}$ and $\Psi_{\rm HH}^{\CB_-}$ to address some fundamental perspectives of the HH proposal. In particular, we will study the {\it phase problem} associated to the norm of the HH wave function in AdS, and its implication in dS gravity. The distinction between the two HH wave functions turn out to be especially important for the norm computation.

In dS gravity, the sphere partition function evaluated around an $S^D$ saddle is often regarded as the leading contribution to the norm of the HH wave function \eqref{eq:original_HH}.\footnote{See \cite{CLM11,MTY19,CJM19,Maldacena24,TW25,ST25} for examples where this interpretation is adopted. See also \cite{CJ25,Cotler26,Godet2026} for arguments that the leading contribution to the norm of the HH wave function \eqref{eq:original_HH} may not be the sphere partition function.} 
However, the one-loop corrected sphere partition function is known to have a nontrivial phase \cite{Polchinski88,ADLS20}, 
\begin{align}
     Z^{\text{1-loop}}_{\rm sphere} = (\mp i)^{D+2}\times \left(\text{positive number}\right)\,.
\end{align}
This phase is in tension with interpreting the sphere partition function as a norm, and also with the state counting interpretation associated with the cosmological horizon \cite{GH77}. Related phases in gravitational partition functions on the sphere and other backgrounds, and how they may be removed, have been extensively discussed recently \cite{Maldacena24,ST25,IMS25,CJ25,CSTY25,GS26,IT26,STW26}. See \cite{AABILS26,Harlow26,Zhao26} for some other recent interesting discussions on different perspectives of the HH wave function. 

Another main goal of this paper is to understand whether an analogous phase problem appears in AdS. The answer depends on which AdS HH wave function one studies. Consider first the partially frozen wave function $\Psi_{\rm HH}^{\CB_-}$, with $\CB_-$ fixed to be a $(D-1)$-dimensional hemisphere $S^{D-1}/Z_2$. The leading contribution to its norm is expected to be the partition function on the hyperbolic ball $H^D$, with DBC imposed on the spacetime boundary. From the AdS/CFT point of view, this object is expected to be positive, since it is dual to the CFT partition function on $S^{D-1}$. Thus 
\begin{align}\label{eq:HB_DBC_intro}
     Z^{\text{1-loop}}_{\rm hyperbolic\,ball\,DBC} = \left(\text{positive number}\right)\,.
\end{align}
This can also be verified directly from the bulk gravitational analysis, as discussed in \cite{ST25}. 

How about the fully gravitational HH wave function \eqref{eq:open_HH}? 
The situation turns out to be different.
Its norm is naturally associated with a hyperbolic-ball path integral in which the boundary geometry is also integrated over. We study the one-loop correction to this free-boundary hyperbolic-ball partition function in section \ref{sec:One_loop_norm_GPI}, and find it also possesses a nontrivial phase 
\begin{align}\label{eq:HB_FBC_intro}
     Z^{\text{1-loop}}_{\rm hyperbolic\,ball,\,free\,bdy} = (\mp i)^{D+1}\times \left(\text{positive number}\right)\,,
\end{align}
though the dimension dependence differs from the sphere partition function in dS gravity. 

The comparison between \eqref{eq:HB_DBC_intro} and \eqref{eq:HB_FBC_intro} suggests that the phase problem is not intrinsic to positive cosmological constant gravity alone. Rather, our results indicate that the phase is tied to whether the norm GPI is fully gravitational or partially frozen. In AdS, the fixed-boundary GPI has the positivity expected from AdS/CFT, whereas the fully gravitational one develops a phase.\footnote{Here is one intuition behind this distinction. The positivity of the norm computation in the path integral language is usually associated with the reflection positivity with respect to a notion of Euclidean time. The frozen spacetime boundary $\CB$ provides a place where such a reflection structure can be imposed, whereas this structure is absent in the fully gravitational setup \cite{Wei25}.} 

Conversely, in dS gravity one may ask whether the phase can be removed by freezing suitable data in a spacetime subregion. Motivated by this question, we study in section \ref{sec:paper2-fixed-equator} a GPI on a sphere in which the metric on an equator is fixed. We find that the one-loop phase cancels in a nontrivial way: 
\begin{equation}
    Z^{1-\text{loop}}_{\text{sphere},\text{fixed-equator}} =  (\text{positive number})\,.
\end{equation}
This is possibly interpreted as an inner product in a non-gravitational theory living on the equator. 
This perspective is reminiscent of the dS/dS correspondence \cite{AKST04,DHST10}, where a holographic screen is placed at the equator of the dS$_{D}$ spacetime, although we will not assume or advocate a specific realization of that proposal here.

The rest of the paper is organized as follows. 
In section \ref{sec:AdS3}, we study the fully gravitational HH wave function in AdS$_3$ Einstein gravity using a minisuperspace saddle-point approximation. In section \ref{sec:JT-noboundary}, we analyze the analogous construction in AdS$_2$ JT gravity. In section \ref{sec:other_interpretations}, we discuss alternative interpretations of the gravitational path integral in \eqref{eq:open_HH}, including its interpretation as a no-boundary wave function for a spatial subregion and as a partition function associated with subregion Cauchy-slice holography. In section \ref{sec:One_loop_norm_GPI}, we analyze the one-loop correction to the hyperbolic-ball partition function with gravitating boundary geometry and exhibit the resulting phase. In section \ref{sec:paper2-fixed-equator}, we study the fixed-equator sphere partition function and show that its one-loop phase cancels nontrivially. We will close with conclusions and some remarks in section \ref{sec:conclusion}. 

Each section is self-contained and technically independent of the others, allowing readers to start with whichever section is most relevant to them.

\section{Hartle--Hawking wave function in \texorpdfstring{AdS$_3$}{AdS3}}
\label{sec:AdS3}

In this section, we study the fully gravitational HH wave function \eqref{eq:open_HH}
on a spatial slice $\Sigma$ with the disk topology in AdS$_3$ Einstein gravity. For simplicity and without risk of ambiguity, we will just call it the HH wave function in this section. 

\begin{figure}[ht]
    \centering
    \includegraphics[width=0.6\linewidth]{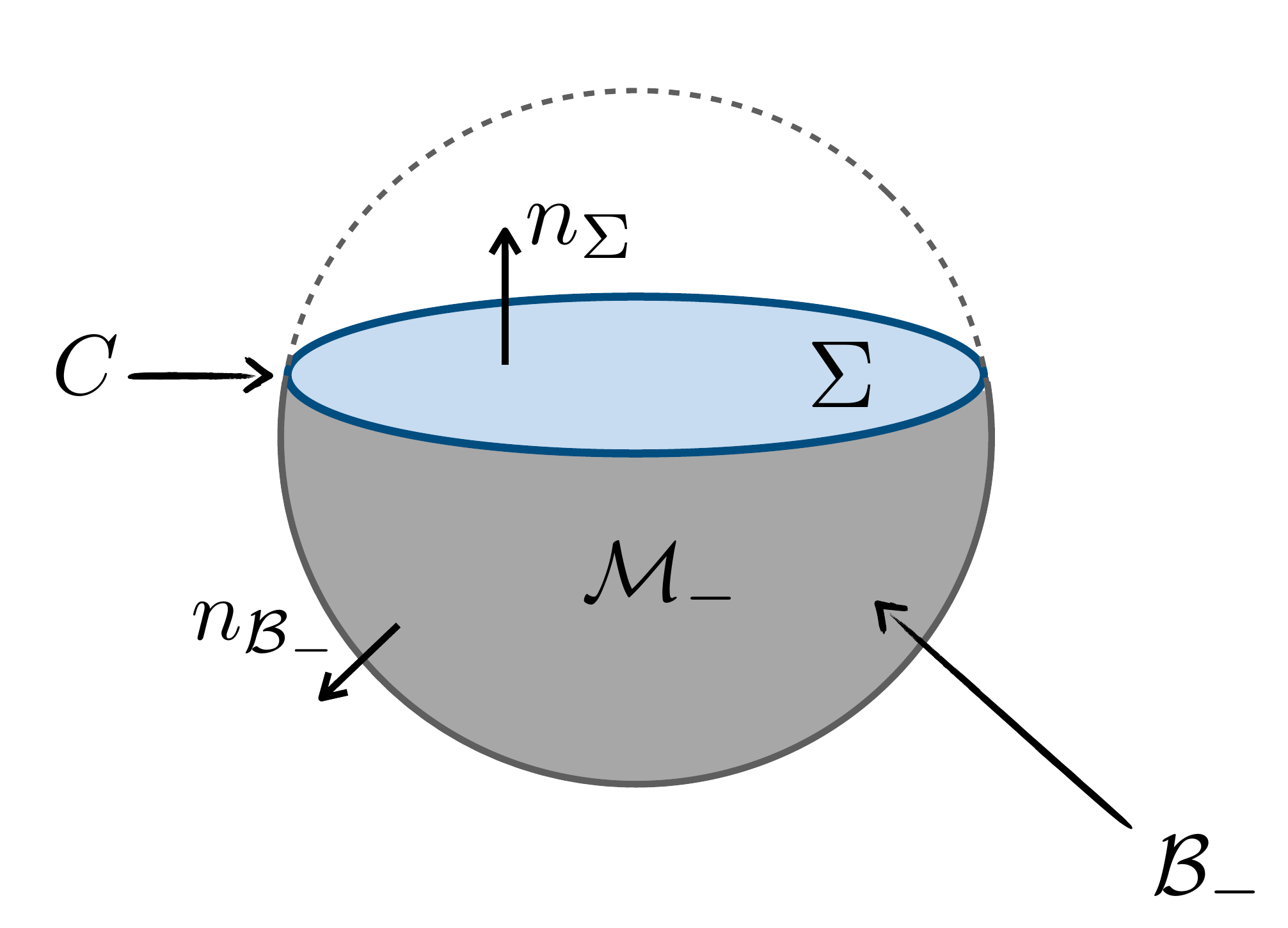}
    \caption{Setup for computing the HH wavefunction in AdS$_3$. The metric is fixed on $\Sigma$ (light blue), while the dynamical spacetime boundary is $\mathcal{B}_{-}$ (gray).
    $\CM_-$ denotes the region bounded by $\Sigma \cup \CB_-$, and $C$ denotes the interface (dark blue) between $\Sigma$ and $\CB_-$, which is a codimension-2 corner.  
    The unit vectors normal to $\Sigma$ and $\CB_-$ are respectively called $n_{\Sigma}$ and $n_{\mathcal{B}_{-}}$. Their scalar product is the dihedral angle which enters in the definition of the corner term needed for a well defined variational principle with a non smooth boundary. On-shell, The spacetime boundary $\CB_-$ lives on an $S^2$ (shown in the dashed line) embedded in $H^3$ at a position fixed by the tension of the spacetime boundary.}
    \label{fig:def_HH_wavefunc_AdS3}
\end{figure}

Let $C=\partial\Sigma$. We fix a Riemannian metric $\gamma_{ab}$ on $\Sigma$ and sum over compact Euclidean 3-manifolds $\CM_-$ with
\begin{equation}
    \partial\mathcal{M}_-=\Sigma\cup \mathcal{B}_-,\qquad \Sigma\cap \mathcal{B}_-=C\,,
\end{equation}
where $\mathcal{B}_-$ is the gravitating spacetime boundary and $\partial \mathcal{B}_-=C$. The metric on $\CB_-$ is denoted as $h_{ij}$. See figure \ref{fig:def_HH_wavefunc_AdS3} for a sketch of the setup.  
The path integral is\footnote{Note that, since both $\gamma_{ab}$ and $h_{ij}$ can be induced from $g_{\mu\nu}$, it is sufficient to denote the Euclidean action as $I_{\rm E}[g_{\mu\nu}]$ rather than $I_{\rm E}[g_{\mu\nu},h_{ij},\gamma_{ab}]$. Here, we adopt the later notation in this section since it will turn out to be more transparent in some discussions.}
\begin{align}\label{eq:HHdef1}
    \Psi_{\rm HH}[\gamma_{ab}] \equiv  
    \int_{\CB_-} \CD h_{ij}
    \int_{\partial \CM_- = \Sigma \cup \CB_-} \mathcal{D}g_{\mu\nu} ~\exp
    \left({-I_{\rm E}[g_{\mu\nu},h_{ij},\gamma_{ab}] }\right)\,, 
\end{align}
We consider pure Einstein gravity on a manifold $\mathcal M_-$ with a dynamical spacetime boundary $\mathcal B_-$ of tension $T$ and a spatial slice $\Sigma$.
The Euclidean action entering includes the usual Gibbons--Hawking--York terms for each smooth boundary component, together with a Hayward term on the corner $C=\Sigma\cap \mathcal{B}_-$. 
\begin{equation}
\begin{aligned}
    I_{\rm E}[g_{\mu\nu},h_{ij},\gamma_{ab}]=&-\frac{1}{16 \pi G_N}\int_{\mathcal M_-}\!\sqrt g\,(R+2)
    -\frac{1}{8\pi G_N}\int_{\Sigma}\!\sqrt\gamma\,K_{\Sigma}
    -\frac{1}{8\pi G_N}\int_{\mathcal{B}_-}\!\sqrt h\,(K_{\mathcal{B}_-}-T)\\
    &{+\frac{1}{8\pi G_N}\int_{C}\!\sqrt\sigma\,\Big(\Theta-\frac{\pi}{2}\Big)\,.}
\end{aligned}
\label{eq:EuclActionFull}
\end{equation}
We have set the cosmological constant to $\Lambda = -1$, such that the AdS radius is $1$. 
We focus on $T>1$, for which the spacetime boundary $\CB_-$ is compact on-shell as we will see later. 
By taking the $T\rightarrow1$ limit, one can recover the usual asymptotic boundary. 
The induced metric on $\Sigma$ is fixed, while the configurations on $\CB_-$ are summed over off-shell and on-shell geometry of $\mathcal{B}_-$ is determined by the Neumann condition coming from the EOM on the spacetime boundary. 

Besides, $\sigma$ is the induced metric on the codimension-two curve $C$, and $\Theta$ is the dihedral angle between $\Sigma$ and $\mathcal{B}_-$,
\begin{equation}
    \Theta\equiv \arccos(-n_{\Sigma}\cdot n_{\mathcal{B}_-}),\qquad 0<\Theta<\pi.
\end{equation}
This dihedral angle is the quantity that enters the Hayward term, and the angle between the outward-pointing normal vectors is its supplement $\pi-\Theta$.
The joint corner term makes the variational problem well posed when the induced metrics on $\Sigma$ and $\mathcal{B}_-$ are fixed: without it, varying \eqref{eq:EuclActionFull} produces an uncancelled contribution localized on $C$.\footnote{The choice of $\frac{\pi}{2}$-counterterm in the Hayward term is fixed by requiring the corner contribution to vanish when the bra and ket are glued together smoothly through $C$.} 

For later use, we record the bulk and spacetime-boundary EOM. Varying with respect to the bulk metric gives
\begin{equation}
    R_{\mu\nu}+2 g_{\mu\nu}=0\,.
\end{equation}
The solutions are locally hyperbolic three-manifolds. Varying with respect to the induced metric on $\mathcal{B}_-$ gives
\begin{equation}
    \delta I_{\rm E}\big|_{\mathcal{B}_-}=-\frac{1}{16 \pi G_N}\int_{\mathcal{B}_-}\!\sqrt h\,\big(K_{ij}|_{\mathcal{B}_-}-K_{\mathcal{B}_-} h_{ij}+T h_{ij}\big)\,\delta h^{ij}\,,
\end{equation}
so stationarity with respect to the boundary variation implies the Neumann-type boundary condition
\begin{equation}
    K_{ij}|_{\mathcal{B}_-}-K_{\mathcal{B}_-} h_{ij}=-T h_{ij} \qquad\Longleftrightarrow\qquad K_{ij}|
  _{\mathcal{B}_-}=T h_{ij} \quad (D=3).
    \label{eq:NeumannBC}
\end{equation}
For $T>1$, an on-shell $\mathcal{B}_-$ can be described by an $S^2$ embedded in $H^3$. Using the following coordinates on $H^3$:
\begin{equation}
    ds^2=du^2+\sinh^2 u\,d\Omega_2^2\,,\qquad d\Omega_2^2=d\theta^2+\sin^2\theta\,d\phi^2\,,
    \label{eq:H3polar}
\end{equation}
the constant $u=u_0$ surface satisfies \eqref{eq:NeumannBC} for $T>1$, where 
\begin{equation}
    T=\coth u_0\,,\qquad R_{\mathcal{B}_-}^{(2)}=2(T^2-1)\,.
    \label{eq:u0fromT}
\end{equation}
 See figure \ref{fig:def_HH_wavefunc_AdS3} for a sketch.

At leading semiclassical order, \eqref{eq:HHdef1} is approximated by the dominant saddle $(g^*_{\mu\nu},h^*_{ij})$,
\begin{equation}
    \Psi_{\rm HH}[\gamma_{ab}]\;\approx\;\exp\big(-I_{\rm E}[g^*_{\mu\nu},h^*_{ij},\gamma_{ab}]\big).
    \label{eq:HHsaddle}
\end{equation}
Below, we focus on looking for the dominant saddle for a given spatial slice $\Sigma$.

\subsection{Minisuperspace approximation and  hyperbolic slicing}

While the spatial slice $\Sigma$ can have any metric in principle, we would like to restrict it to a special class for convenience. This is called the minisuperspace approximation. In dS gravity, the most common minisuperspace approximation is to restrict the spatial slice to be a sphere, and the only free parameter is the radius \cite{HH83}. 

In the AdS$_3$ case we are considering, a natural minisuperspace family is to restrict $\Sigma$ to be a hyperbolic disk with finite cutoff. Such a $\Sigma$ is parameterized by two independent parameters: the area $A$ of $\Sigma$ and the proper length $L$ of its boundary circle $\partial\Sigma = C$. The metric of such a hyperbolic disk can be written as 
\begin{equation}
    ds^2_{\Sigma}=\ell^2\Big(d\rho^2+\sinh^2\rho\,d\varphi^2\Big)\,,\qquad 0\le \rho\le \rho_c\,,\qquad \varphi\sim\varphi+2\pi\,,
    \label{eq:diskmetric}
\end{equation}
where $\ell$ is the intrinsic hyperbolic radius of $\Sigma$ (hence $R^{(2)}_{\Sigma}=-2/\ell^2$) and $\rho_c$ is a dimensionless cutoff. $A$ and $L$ turn out to be 
\begin{equation}
    L\equiv\int_C\!ds=2\pi\ell\,\sinh\rho_c,\qquad
    A\equiv\int_{\Sigma}\!\sqrt\gamma=2\pi\ell^2\,(\cosh\rho_c-1)\,,
    \label{eq:ALdisk}
\end{equation}
accordingly. 
One can also choose another two parameters for the minisuperspace. 
It will later turn out to be useful to write $A$ in terms of $L$ and $\ell$, 
\begin{equation}
    A(L,\ell)=\ell\sqrt{L^2+4\pi^2\ell^2}-2\pi\ell^2\,.
    \label{eq:A-of-L-ell}
\end{equation}

\paragraph{Hyperbolic slicing of $H^3$}~\par
To build a solution with fixed data \eqref{eq:diskmetric} on $\Sigma$, it is convenient to write $H^3$ in coordinates in which constant ``Euclidean-time'' slices are hyperbolic disks. Consider the embedding of $H^3$ as the hyperboloid:
\begin{equation}
    -X_0^2+X_1^2+X_2^2+X_3^2=-1,\qquad X_0>0\,,
    \label{eq:embeding_coordinates}
\end{equation}
in $\mathbb{R}^{1,3}$ with metric $ds^2=-dX_0^2+dX_1^2+dX_2^2+dX_3^2$.  The parametrization:
\begin{equation}
\begin{aligned}
    X_0&=\cosh\eta\,\cosh\rho\,, &&X_3=\sinh\eta\,,\\
    X_1&=\cosh\eta\,\sinh\rho\,\cos\varphi\,,&&X_2=\cosh\eta\,\sinh\rho\,\sin\varphi\,,
\end{aligned}
\label{eq:H2slicing}
\end{equation}
gives the ``$H^2$ slicing'' metric:
\begin{equation}
    ds^2=d\eta^2+\cosh^2\eta\,\Big(d\rho^2+\sinh^2\rho\,d\varphi^2\Big)\,.
    \label{eq:H3H2sliceMetric}
\end{equation}
The surface $\eta=\eta_0$ is a hyperbolic plane of hyperbolic radius $\ell=\cosh\eta_0$.  Therefore a natural minisuperspace parameterization choice is to take: 
\begin{equation}
    \Sigma:\ \eta=\eta_0\,,\qquad 0\le \rho\le \rho_c\,,
    \label{eq:SigmaChoice}
\end{equation}
so that \eqref{eq:diskmetric} holds with $\ell=\cosh\eta_0$.  In this parametrization the two parameters of the disk are $(\eta_0,\rho_c)$, and \eqref{eq:ALdisk} becomes:
\begin{equation}
    L=2\pi\cosh\eta_0\,\sinh\rho_c\,,\qquad
    A=2\pi\cosh^2\eta_0\,(\cosh\rho_c-1)\,.
    \label{eq:ALetac}
\end{equation}

Note that the above expression for $L$ and $A$ are invariant under the flip of the sign of $\eta_0 \leftrightarrow -\,\eta_0$. It suggests that the two hyperbolic disks, placed at $\eta = \pm \eta_0$ with the same dimensionless cutoff $\rho_c$, have the identical $(L,\,A)$ parameters.  These two hyperbolic disks can be distinguished by their extrinsic curvature $K_\Sigma = \pm 2\tanh{\eta_0}$ with respect to the unit normal vector $n_\Sigma=+\partial_\eta$.

\paragraph{Solutions with boundary $\partial\mathcal M_-=\Sigma\cup \mathcal{B}_-$}~\par

We now construct a smooth compact solution $\mathcal M_-$ bounded by the spatial slice $\Sigma$ and by the spacetime boundary $\mathcal{B}_-$ of tension $T>1$ \eqref{eq:u0fromT}. Since the bulk is locally $H^3$, the only nontrivial part is the relative positioning of $\mathcal{B}_-$ and $\Sigma$ when embedded in $H^3$.

Let $P$ denote the ``center'' of the hyperbolic ball bounded by the sphere $\mathcal{B}$ in the embedding description (i.e. the point $u=0$ in the coordinate \eqref{eq:H3polar}). Without loss of generality we may place $P$ on the axis $\rho=0$ in the \eqref{eq:H3H2sliceMetric} coordinate ($X_1=X_2=0$ in embedding coordinates). Thus the most general position of $P$ in \eqref{eq:embeding_coordinates} is parametrized by the value of $\eta=s$, so that
\begin{equation}
    P=(\cosh s,0,0,\sinh s),\qquad P\cdot P=-1\,.
\end{equation}
By definition of $P$, using coordinates \eqref{eq:H3polar}, the geodesic distance $u$ between a point $X$ and $P$ satisfies $\cosh u=-X\cdot P$, and the on-shell 2-sphere $\CB$ centered at $P$ is therefore
\begin{equation}
    \mathcal{B}:\qquad -X\cdot P=\left(\cosh u_0\right).
\end{equation}
Using \eqref{eq:H2slicing}, this condition becomes an explicit equation for the boundary profile in $(\eta,\rho)$,
\begin{equation}
    \cosh\eta\,\cosh\rho\,\cosh s-\sinh\eta\,\sinh s=\cosh u_0\,,
    \label{eq:braneImplicit}
\end{equation}
with $u_0$ determined by $T$ through \eqref{eq:u0fromT}. Given the Hartle--Hawking data $(L,A)$, the boundary curve $C=\partial\Sigma=\Sigma\cap \mathcal{B}_-$ sits at $(\eta,\rho)=(\eta_0,\rho_c)$. Requiring that $C$ lies on $\mathcal{B}$ fixes the displacement parameter $s$ implicitly by
\begin{equation}
    \cosh\eta_0\,\cosh\rho_c\,\cosh s-\sinh\eta_0\,\sinh s=\cosh u_0\,.
    \label{eq:sCondition}
\end{equation}
There is a solution for $s$ in terms of $(T,\eta_0,\rho_c)$ provided the circumference of the disk is not too large compared to the size of the great circle of $\CB$:
\begin{equation}
    \cosh(s-\alpha)=\frac{\cosh u_0}{\sqrt{(\cosh\eta_0\cosh\rho_c)^2-(\sinh\eta_0)^2}},\qquad \tanh\alpha=\frac{\tanh\eta_0 }{\cosh\rho_c}\,.
    \label{eq:ssolution}
\end{equation}

For $L<L_{\text{crit}}(T)$, i.e. the circumference of the great circle of $\CB$, \eqref{eq:ssolution} has two real roots,
\begin{equation}
    s_{\rm small}=\alpha+\delta,\qquad s_{\rm large}=\alpha-\delta,\qquad
    \cosh\delta=\frac{\cosh u_0}{\sqrt{1+(L/2\pi)^2}}\,.
    \label{eq:s-two-real-branches}
\end{equation}
 These correspond to the two Euclidean fillings compatible with the same data on $\Sigma$. The Euclidean saddle associated with the root $s_{\mathrm{small}}$ has the spacetime boundary $\CB_-$ that covers less than half of the 2-sphere $\CB$, while the other one associated with $s_{\mathrm{large}}$  has $\CB_-$ that covers more than half of $\CB$. The two roots become degenerate when $L$ takes its critical value $L_{\text{crit}}(T) = \frac{2\pi}{\sqrt{T^2-1}}$. This corresponds to the case when the spacetime boundary is exactly half of the boundary sphere $\CB$.
 
As commented before, we can also place the spatial slice $\Sigma$ at $\eta=-\eta_0$, which will lead to two distinctive roots $s_{\mathrm{small}}'$ and $s_{\mathrm{large}}'$. By flipping the sign of $\eta_0$ in Equation \eqref{eq:ssolution}, we immediately read out two new roots
 \begin{equation}
     s_{\mathrm{small}}' = -\alpha+\delta,\quad s_{\mathrm{large}}' = -\alpha-\delta,
 \end{equation}which, similarly, become degenerate when $L = L_{\mathrm{crit}}(T)$. The subscripts ``small'' and ``large'' are chosen consistently to indicate whether the spacetime boundary $\CB_-'$ covers less or more than half of the boundary sphere $\CB'$. It seems that we obtain two more Euclidean saddles compatible with the spatial slice $\Sigma$, while these two new saddles can be related to the former ones in a simple way. Provided that $s_{\mathrm{small}}= -s_{\mathrm{large}}
 '$ and $s_{\mathrm{large}}=-s_{\mathrm{small}}'$ respectively, we can map $\CB'$ with the center labeled by $s'$ to $\CB$ by reversing the $\eta$-axis. The spatial slice $\Sigma$ cuts the hyperbolic ball $\CM$ and its boundary sphere $\CB$ into two parts, which we label by $\CM_\pm$ and $\CB_{\pm}$ respectively. The reversal of the axis swaps these two parts. Therefore, the primed saddles $\CM_-'$ can be equivalently interpreted as the complement of $\CM_-$ (let us denote it as $\CM_+$ here) with respect to the hyperbolic ball $\CM$. More precisely, we have $\CM_{-,\mathrm{small}/\mathrm{large}}' = \CM_{+,\mathrm{large}/\mathrm{small}}$. Note that $\CM_+$ has the outward normal vector on $\Sigma$ point in the opposite direction compared to $\CM_-$, which changes the sign of the extrinsic curvature $K_{\Sigma}$. This agrees with the expectation that $K_\Sigma' = -K_{\Sigma}=-2\tanh{\eta_0}$. Since four saddles form two pairs, $\CM_{\pm,\text{small}}$ and $\CM_{\pm,\text{large}}$ ,and each pair can be glued together smoothly to be the hyperbolic ball $\CM$, the Euclidean actions evaluated on the saddles in the same pair admit the following relation 
  \begin{equation}
     I_{\text{E},\CM_-}^{\text{on-shell}}+I_{\text{E},\CM_+}^{\text{on-shell}} = I_{\text{E},\CM}^{\text{on-shell}}.
 \end{equation} 
 \begin{figure}[ht]
     \centering
     \includegraphics[width=16.5cm]{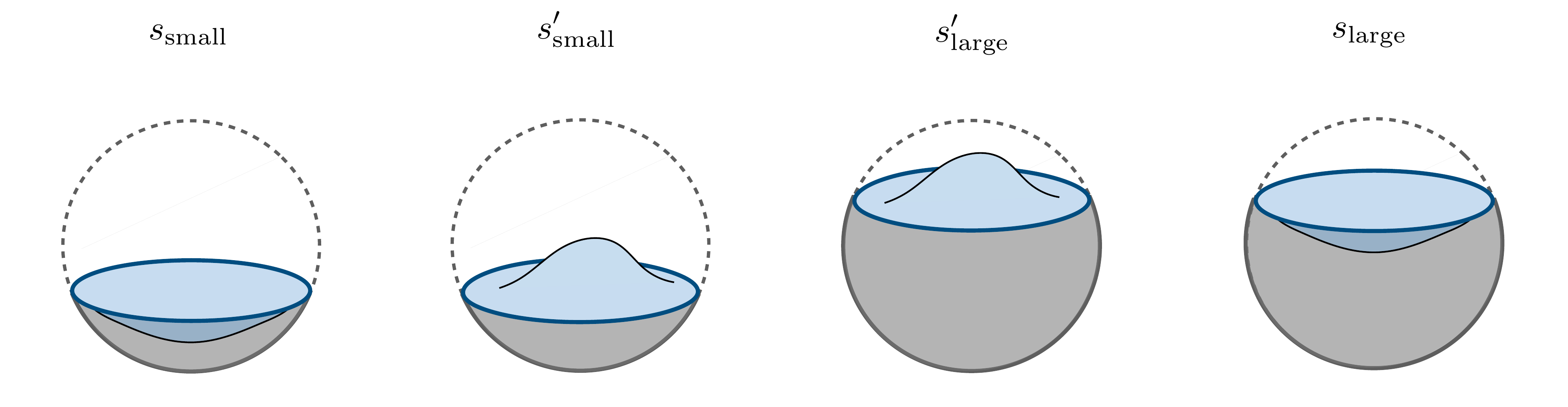}
     \caption{For a given choice of $\Sigma$ (light blue), four different saddles $\CM_-$ enclosed by $\Sigma$ and the spacetime boundary $\CB_-$ (gray) are possible, we represent each of them in this figure. By construction $s_{\text{small}}$ is the complementary of $s_{\text{large}}'$ and similarly for $s_{\text{small}}'$ and $s_{\text{large}}$. We argue in the text which are the saddles of interest. The extrinsic curvature on the primed quantity $s_{\text{small}}'$ and $s_{\text{large}}'$ are the same and of opposite sign compare to the un-primed ones.}
     \label{fig:four saddles AdS3}
 \end{figure}

In this section, we will evaluate the HH wave function semiclassically by choosing the appropriate saddle from the four saddles discussed above, see figure \ref{fig:four saddles AdS3} for a sketch of the different saddles. The choice of the saddle does not only depend on a direct comparison on the on-shell actions, but also relies on the prescription of the steepest-descent contour. We will present a comprehensive analysis on the contour prescription and the saddle choice in the latter subsection when the on-shell action is evaluated.

When the disk is ``too large,'' the two  roots, $s_{\mathrm{small}}$ and $s_{\mathrm{large}}$, move off the real axis and the on-shell $\CM_-$ bounded by $\Sigma \cup \CB_-$ becomes complex.
In the next subsection we briefly study the consequences of such a complex saddle.

\subsection{Large spatial slices and complex saddles}
The construction above assumed that \eqref{eq:ssolution} admits a \emph{real} solution for $s$, so that both $P$, the center of the hyperbolic ball $\CM$, and the spacetime boundary profile \eqref{eq:braneImplicit} lie in real Euclidean $H^3$. This is not always the case: for spatial slices $\Sigma$ with sufficiently large $(A,L)$, the two pairs of real Euclidean saddles, labeled by $(s_{\text{small}},s_{\text{large}}')$ and $(s_{\text{large}},s_{\text{small}}')$, merge at $L=L_{\text{crit}}(T)$ and continue to complex saddles.

\begin{figure}[t]
\centering
\begin{tikzpicture}
\begin{axis}[
  width=12cm, height=7.5cm,
  xlabel={$T$}, ylabel={$L$},
  xmin=1.05, xmax=4,
  ymin=0, ymax=10,
  domain=1.05:4, samples=250,
  legend style={at={(0.98,0.98)},anchor=north east},
]
\addplot[name path=Lmaxcurve, thick] {2*pi/sqrt(x^2-1)};
\addlegendentry{$L_{\text{crit}}(T)=\frac{2\pi}{\sqrt{T^2-1}}$}

\path[name path=bottom] (axis cs:1.05,0) -- (axis cs:4,0);
\addplot[fill opacity=0.12] fill between[of=Lmaxcurve and bottom, soft clip={domain=1.05:4}];

\node[anchor=west] at (axis cs:1.18,1.2) {real Euclidean saddle};
\node[anchor=west] at (axis cs:1.3,8.7) {complex saddles};
\end{axis}
\end{tikzpicture}
\caption{Existence domain of the real Euclidean saddle. For fixed supercritical tension $T>1$, a real Euclidean saddle exists only for $L\le L_{\text{crit}}(T)$. Beyond this, the steepest-descent evaluation requires complex saddles.}
\label{fig:phase_TL}
\end{figure}
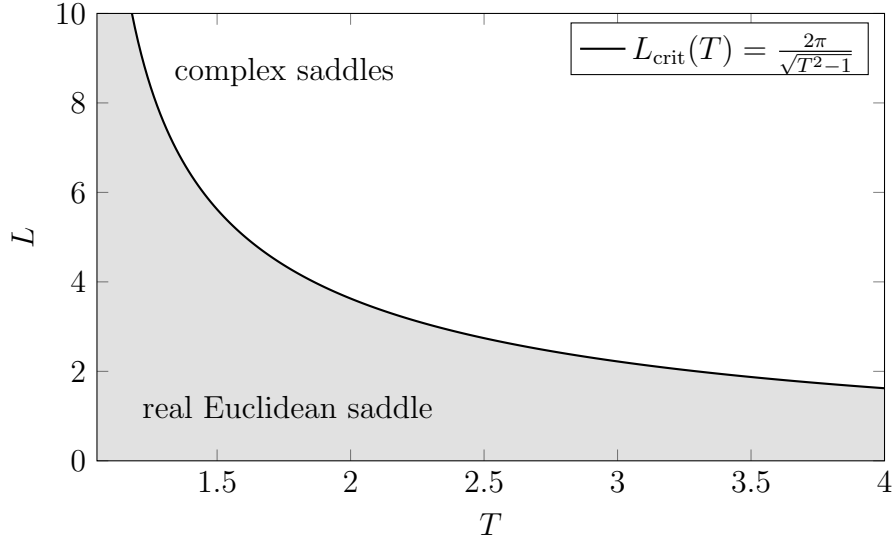

Real solutions to \eqref{eq:ssolution} exist if and only if the right-hand side is greater than or equal to $1$. Equivalently, the real Euclidean saddles discussed above are present precisely when:
\begin{equation}
    \frac{\cosh u_0}{\sqrt{1+(L/2\pi)^2}}\ge 1
    \qquad\Longleftrightarrow\qquad
    L\le L_{\text{crit}}(T)\equiv 2\pi\sinh u_0=\frac{2\pi}{\sqrt{T^2-1}}\,,
    \label{eq:LmaxCondition}
\end{equation}
as shown in figure \ref{fig:phase_TL}.
Geometrically, \eqref{eq:LmaxCondition} indicates that the intersection curve $C=\Sigma\cap \mathcal{B}_-$, as a circle on the spacetime-boundary 2-sphere $\CB$ of radius $\sinh u_0$, has the circumference bounded by $2\pi\sinh u_0$, the length of the equator of $\CB$.

It is also useful to note that within our hyperbolic-disk minisuperspace, $A$ and $L$ can not be chosen arbitrarily.  Eliminating $\rho_c$ from \eqref{eq:ALetac} gives
\begin{equation}
    L^2-4\pi A = 4\pi^2\cosh^2\eta_0\,(\cosh\rho_c-1)^2>0,
\end{equation}
which imposes an upper bound on the area $A$ of the hyperbolic disk $\Sigma$ in terms of $L$.  Thus the requirement \eqref{eq:LmaxCondition} on $L$ also constrains the value of $A$ implicitly.

When $L>L_{\mathrm{crit}}(T)$, \eqref{eq:ssolution} has no real solution. Although the GPI defining $\Psi_{\rm HH}$ is an integral over real geometries,  its saddle point approximation can be dominated by a complex solution. 
In this complexified problem, \eqref{eq:ssolution} is solved by taking $s$ complex. Defining:
\begin{equation}
    \kappa\equiv\frac{\cosh u_0}{\sqrt{1+(L/2\pi)^2}},
\end{equation}
we have $\kappa\in(0,1)$ in the regime $L>L_{\mathrm{crit}}(T)$, so we may write $\kappa=\cos\beta$ with $\beta\in(0,\pi/2)$.  The solutions of \eqref{eq:ssolution} are then:
\begin{equation}
    s=\alpha\pm i\beta\,,\qquad \tanh\alpha=\frac{\tanh\eta_0}{\cosh\rho_c}\,.
    \label{eq:sComplex}
\end{equation}
Substituting \eqref{eq:sComplex} into \eqref{eq:braneImplicit} describes the embedding of the spacetime boundary $\CB_-$ with the same boundary condition \eqref{eq:NeumannBC} into a complexified geometry that solves the bulk Einstein equations.
The induced metric on $\Sigma$ remains the same as the real hyperbolic disk \eqref{eq:diskmetric}, while the bulk metric and the embedding map of $\mathcal{B}_-$ become complex. 
This complex geometry can be constructed by gluing half of the Euclidean hyperbolic ball to the Lorentzian AdS$_3$ spacetime, where the gluing surface is a hyperbolic disk with $L=L_{\text{crit}}(T)$ and hyperbolic radius $\ell=1$. The gluing operation can be realized via analytical continuation of the spacetime coordinates. Instead of using the ``$H^2$ slicing'' metric introduced in the previous discussion, it's tempting to demonstrate the analytical continuation in the following global AdS$_3$ coordinates
\begin{equation}
    \label{eqn:GlobalAdS3}
    ds^2= \cosh^2{\rho_{\rm G}} dt_{\rm E}^2+d^2\rho_{\rm G}+\sinh^2{\rho_{\rm G}} d^2\varphi\,,
\end{equation}
 where we simply analytically continue the Euclidean time $t_{\rm E}$ to the Lorentzian time $t_{\rm L}$ via Wick rotation $t_{\rm E}= -i t_{\rm L}$. The coordinate transformation from $(\eta,\,\rho)$ coordinates to the global coordinates is given as follows
 \begin{equation}
     \cosh{\eta}\,\cosh{\rho} = \cosh{t_{\rm E}}\,\cosh{\rho_{\rm G}},\quad \sinh{\eta} = \sinh{t_{\rm E}}\,\cosh{\rho_{\rm G}}\,.
 \end{equation} 
 The angular variable $\varphi$ is the same in both coordinates. By rewriting the equation \eqref{eq:braneImplicit} of $\CB$ in the global coordinates, we find the embedding map of $\CB$
 \begin{equation}
     \label{eqn:embeddingBglobal}
     \cosh{\rho_{\rm G}}\,\cosh{(t_{\rm E}-s)}=\cosh{u_0}\,.
 \end{equation}
 Every constant-$t_{\rm E}$ slice of $\CB$ is a circle with radius $\sinh{\rho_{\rm G}}$ solved in terms of $t_{\rm E}$ from the above equation. Therefore, the spacetime boundary $\CB_-$ can be depicted as a 1D trajectory within the complex-$t_{\rm E}$ plane. The two endpoints of this trajectory correspond to $\rho_{\rm G} = 0$ and $\rho_{\rm G} = \arcsinh{(L/2\pi)}$ respectively. For $\rho_{\rm G} =0$, there are two solutions for $t_{\rm E}$ at $t_{\rm E}= s\pm u_0$. The circular slice with $\rho_{\rm G} = \arcsinh{(L/2\pi)}$ corresponds to the intersection circle $C= \Sigma\,\cap\,\CB_- $, and by placing $C$ on the real $t_{\rm E}$ axis, we obtain its Euclidean time coordinate $t_{\rm E} =\alpha$. If we further requires $\rho_{\rm G}(t_{\rm E})$ to be real along the trajectory and to pass through the ``center'' of $\CB$ at $t_{\rm E}=s$, the trajectory in the complex-$t_{\rm E}$ plane will be determined completely.
\begin{figure}[t]
    \centering
    \includegraphics[width=14cm]{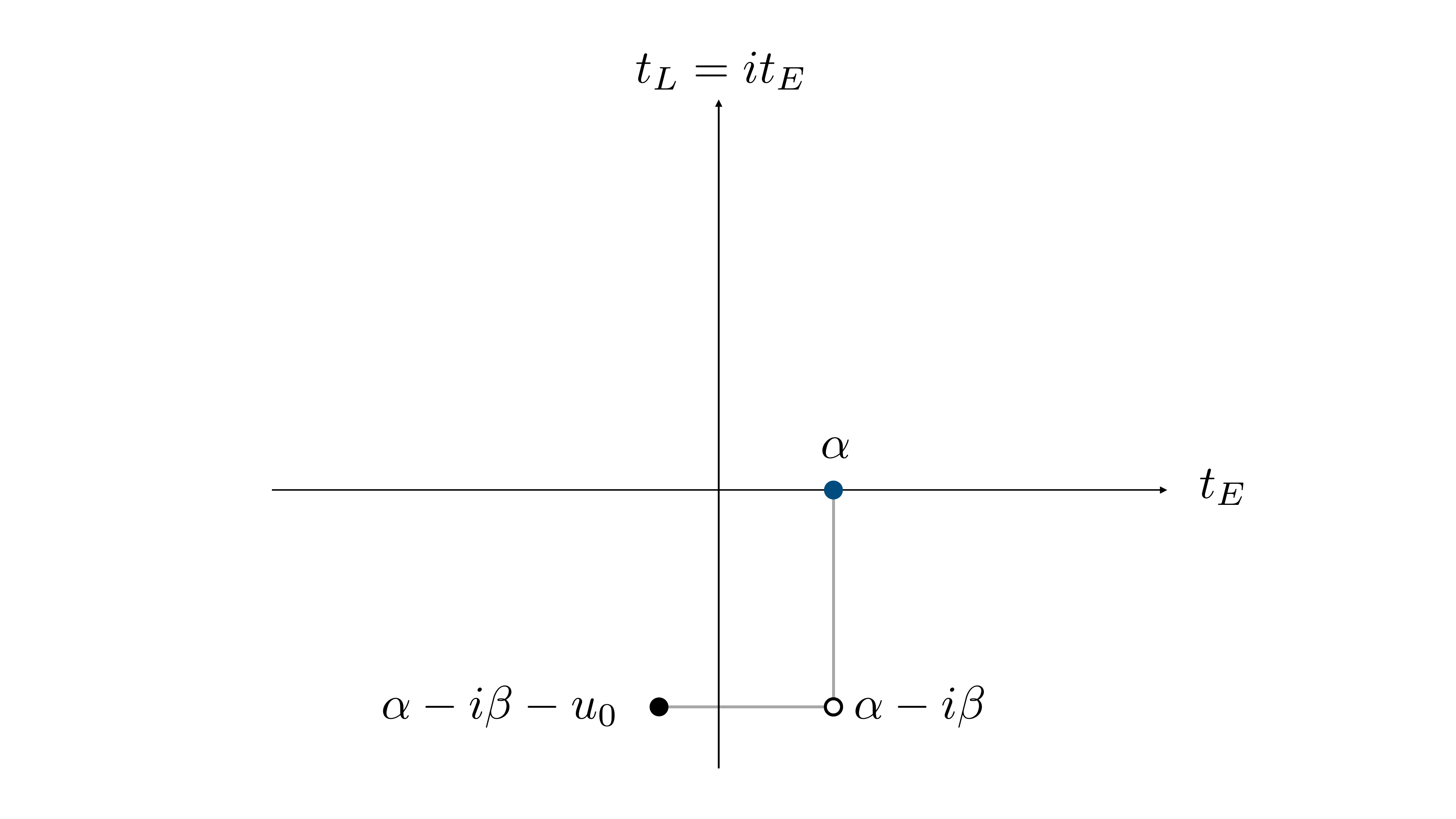}
    \caption{A trajectory (gray) of spacetime boundary $\CB_-$ in the complex-$t_{\rm E}$ plane. Here we choose the saddle with $s=\alpha-i\beta$ and set the $\rho_{\rm G}=0$ endpoint at $t_{\rm E}=s-u_0$. Other saddles can be drawn in the similar way. The dark blue dot represents the intersection circle $C$ at $t_{\rm E}=\alpha$. The unfilled dot at $t_{\rm E}=s$ corresponds to the boundary of the gluing surface between Euclidean and Lorentzian spacetime, which is a circle on $\CB_-$.}
    \label{fig:Btractory}
\end{figure}
In figure \ref{fig:Btractory}, we provide a demonstration of the trajectory of $\CB_-$ by choosing one specific saddle geometry.
\begin{figure}[ht]
    \centering
    \includegraphics[width=0.7\linewidth]{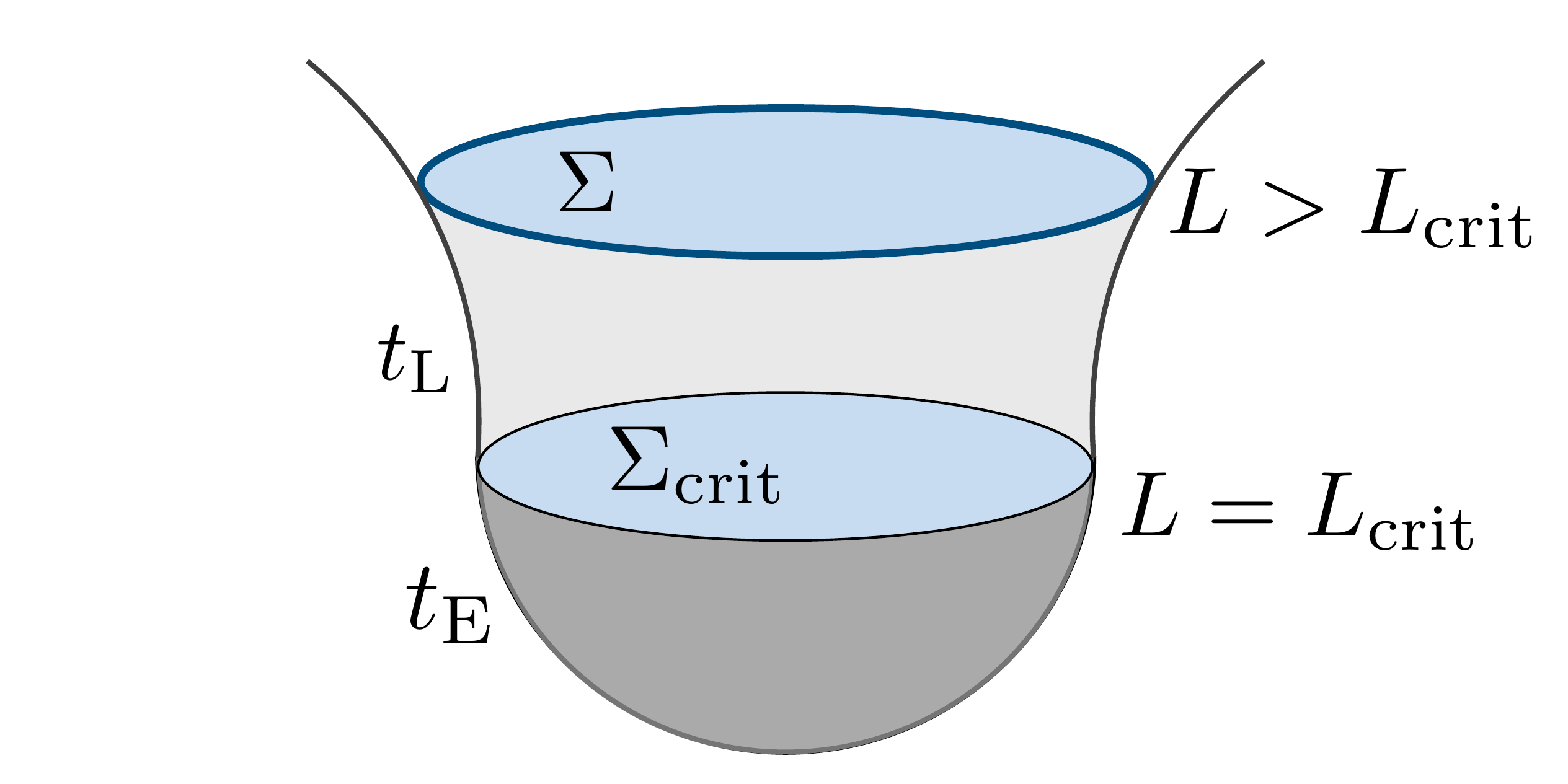}
    \caption{When the boundary length $L$ is bigger than the critical value $L_{\text{crit}}(T)=\frac{2\pi}{\sqrt{T^2-1}}$, the saddle point is necessarily complex. An example which can be realized by gluing a Euclidean part and a Lorentzian part together is shown.  
    The Lorentzian evolution $t_{\rm L}$ depicted in a lighter color, in addition to the half-sphere contribution in Euclidean signature $t_{\rm E}$ (the darker part).}
    \label{fig:Lorentzian evolution}
\end{figure}
We show the gluing of Euclidean and Lorentzian geometries explicitly in figure \ref{fig:Lorentzian evolution}. Note that in figure \ref{fig:Lorentzian evolution}, for simplicity, we demonstrate the spatial slice $\Sigma$ as a constant-$t_{\rm L}$ slice, while this is not generally true. The spatial slice $\Sigma$, as a constant-$\eta$ slice in the $(\eta,\,\rho)$ coordinates, does not always map to a constant-$t_{\rm L}$ slice in the global coordinates unless $\eta_0=0$, i.e. $K_\Sigma =0$. The spatial slice $\Sigma$ with nonzero $\eta_0$ covers the following codimension-1 surface in the global coordinates
\begin{equation}
    \Sigma = \{\alpha\leq t_{\rm E}\leq \eta_0,\, \cosh{\rho_G(t_{\rm E})}=\frac{\sinh{\eta_0}}{\sinh{t_{\rm E}}},\,0\leq \varphi <2\pi\}\,.
\end{equation}
The region enclosed between $t_{\rm E}=\alpha$ plane and $\Sigma$ is in Euclidean signature, since the value of $t_{\rm E}$ coordinate lies on the real axis. This can be viewed as adding a Euclidean ``cap'' on top of the Lorentzian geometry in figure \ref{fig:Lorentzian evolution}.

The similar phenomenon also appears in the HH wave function computation in dS space \cite{HH83}. When the spatial slice, which is a sphere with its radius as the minisuperspace parameter, is too large to be filled by the real Euclidean saddle (half-sphere), the steepest-descent contour will pass through complex saddles that include a Lorentzian evolution via the analytic continuation of the Euclidean time, see also \cite{Witten21-2} for a criterion determining which complex saddles should be included in the gravitational path integral.

For real-$s$ saddles, \eqref{eq:braneImplicit} fully determines the embedding of the boundary sphere $\CB$ in $H^3$. The boundary sphere $\CB$ reaches the axis $\rho=0$ at $\eta = s\pm u_0$. In the following calculation of the HH wave function, we define our spacetime boundary $\CB_-$ to be the region on the boundary sphere $\CB$ that is bounded by the curve $C$ and contains the lower endpoint $\eta_{\text{min}} = s-u_0$. Therefore, the saddle $\CM_-$ in between $\CB_-$ and $\Sigma$ is the region \begin{equation}
    \mathcal M_-=\Big\{\eta_{\rm min}\le \eta\le \eta_0\,,\ \ 0\le \rho\le \rho_Q(\eta)\,,\ \ 0\le \varphi<2\pi\Big\}\,,
    \label{eq:regionM}
\end{equation}
where $\rho_Q(\eta)$ can be solved from \eqref{eq:braneImplicit} as follows \begin{equation}
    \cosh\rho_Q(\eta)=\frac{\cosh u_0+\sinh\eta\,\sinh s}{\cosh\eta\,\cosh s}\,.
    \label{eq:rhoQofeta}
\end{equation} $\rho_Q(\eta)$ stands for the radial coordinate of the point on the spacetime boundary $\CB_-$.

For complex-$s$ saddles, $\CM_-$ can be conveniently decomposed into three pieces $\CM_- = \CM_{\rm E}\,\cup \CM_{\rm L}\,\cup\CM_{\text{cap}}$, which correspond to the Euclidean part of $\CB_-$ trajectory, the Lorentzian part of $\CB_-$ trajectory and the cap between $t_{\rm E}=\alpha$ and $\Sigma$. In summary, we write down the expressions of each subregion explicitly in the global AdS$_3$ coordinates
\begin{equation}
    \begin{aligned}
        \CM_{\rm E} &=\{t_{\rm E}=s+\lambda,\,1\leq \cosh{\rho_{\rm G}}\leq \frac{\cosh{u_0}}{\cosh{\lambda}},\,0\leq\varphi<2\pi;\,\lambda \in [-u_0,0]\}\\
        \CM_{\rm L}&=\{t_{\rm E}=s+i \lambda,\,1\leq \cosh{\rho_{\rm G}}\leq \frac{\cosh{u_0}}{\cos{\lambda}},\,0\leq\varphi<2\pi;\,\lambda \in [0,\beta]\}\\
        \CM_{\text{cap}}&=\{\alpha\leq t_{\rm E}\leq \eta_0,\,1\leq\cosh{\rho_{\rm G}}\leq \frac{\sinh{\eta_0}}{\sinh{t_{\rm E}}},\,0\leq\varphi<2\pi\}\,.
    \end{aligned}
    \label{eqn:complexsaddleprofile}
\end{equation}
Here, we choose the saddle $s=\alpha-i\beta$ and assume $\eta_0>\alpha>0$.

\subsection{Semiclassical Hartle--Hawking wave function}
\begin{figure}[ht]
    \centering
    \includegraphics[width=0.9\linewidth]{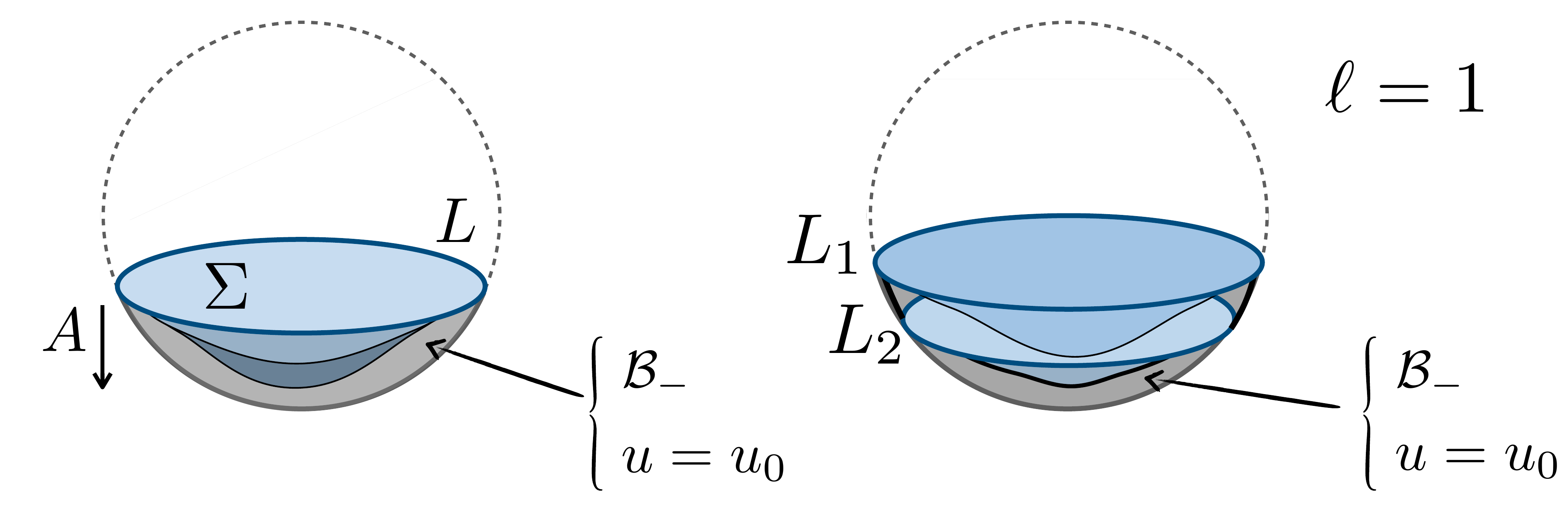}
    \caption{The computation of the HH wave function in AdS$_3$. The HH wave function is computed thanks to different slicing of the hyperbolic ball controlled by two parameters $L,A$. On the left is depicted the HH wave function for a same $L$ and different $A$ increasing along the arrow, note that on each slice the induced metric is that of the hyperbolic disk with different hyperbolic radius $\ell$. On the right is represented the slicing for $L_2>L_1$ with the same unit hyperbolic radius $\ell=1$. In each case $\mathcal{M}_{-}$ is the region bounded by $u=u_0$ and the surface which represents the $H^2$ slicing. In particular on the right figure, the black region should be included to $\mathcal{B}_{-,\,1}$ but not in $\mathcal{B}_{-,\,2}$.}
    \label{fig:HH_wavefunction_in_AdS3}
\end{figure}

We now evaluate the Euclidean action \eqref{eq:EuclActionFull} on the real minisuperspace saddle \eqref{eq:regionM}, term by term. At this stage, we will perform the most general calculation without specifying which saddle is selected. We schematically represent the computation and the minisuperspace approximation in figure \ref{fig:HH_wavefunction_in_AdS3}.

\paragraph{Bulk term.}
On shell $R=-6$, so the bulk contribution reduces to a term proportional to the bulk volume:
\begin{equation}
    -\frac{1}{16 \pi G_N}\int_{\mathcal M_-}\!\sqrt g\,(R+2)=\frac{1}{4\pi G_N}\,\mathrm{Vol}(\mathcal M_-).
\end{equation}
Using \eqref{eq:H3H2sliceMetric} and \eqref{eq:rhoQofeta}, the volume of the saddle geometry $\CM_-$ is
\begin{equation}
\begin{aligned}
    \mathrm{Vol}(\mathcal M_-)
    &=2\pi\int_{\eta_{\rm min}}^{\eta_0}\!d\eta\,\cosh^2\eta\int_0^{\rho_Q(\eta)}\!d\rho\,\sinh\rho
    =2\pi\int_{\eta_{\rm min}}^{\eta_0}\!d\eta\,\cosh^2\eta\,\big(\cosh\rho_Q(\eta)-1\big)\\
    &= \frac{2\pi}{\cosh s}\Bigg[\cosh u_0\,\sinh\eta+\frac{\sinh s}{2}\,\sinh^2\eta-\cosh s\Big(\frac14\sinh 2\eta+\frac{\eta}{2}\Big)\Bigg]_{\eta=\eta_{\rm min}}^{\eta=\eta_0}.
\end{aligned}
\label{eq:VolResult}
\end{equation}

\paragraph{The $\Sigma$ term.}
For the constant-$\eta$ slice \eqref{eq:SigmaChoice}, the outward unit normal is $n_{\Sigma}=+\partial_{\eta}$, it contributes to the on-shell action through:
\begin{equation}
    K_{\Sigma}=2\tanh\eta_0\,,\quad -\frac{1}{8\pi G_N}\int_{\Sigma}\!\sqrt\gamma\,K_{\Sigma}=-\frac{A}{4\pi G_N}\tanh\eta_0\,,
    \label{eq:SigmaTerm}
\end{equation}
where $A$ is the area of the hyperbolic disk $\Sigma$ in \eqref{eq:ALetac}.

\paragraph{The $\CB_-$ term.}
On shell $K_{\mathcal{B}_-}=2T$, so the spacetime-boundary action becomes a pure area term:
\begin{equation}
    -\frac{1}{8\pi G_N}\int_{\mathcal{B}_-}\!\sqrt h\,(K_{\mathcal{B}_-}-T)=-\frac{T}{8\pi G_N}\,\mathrm{Area}(\mathcal{B}_{-})\,.
    \label{eq:braneTermOnShell}
\end{equation}
The $\mathcal{B}_{-}$ is the portion of the boundary sphere $\CB$ with $u=u_0$ lying in the region $\eta_{\text{min}}\leq\eta\le\eta_0$ (equivalently $X_3\le \sinh\eta_0$). One may relate  $u_0$ to the tension $T$ by $\sinh{u_0}=1/\sqrt{T^2-1}$ and $\cosh u_0=T/\sqrt{T^2-1}$. In geodesic polar coordinates around the center $P$, the induced metric on $\mathcal{B}_-$ is $ds^2_{\mathcal{B}_-}=\sinh^2u_0\,d\Omega_2^2$, hence
\begin{equation}
    \mathrm{Area}(\mathcal{B}_{-})=2\pi\sinh^2u_0\,\big(1+\cos\theta_0\big)\,,
    \label{eq:capArea}
\end{equation}
where $\theta_0$ is fixed by matching the length of the latitude circle at $\theta_0$ with the circumference $L$ of the intersection circle $C=\Sigma \,\cap\,\CB_-$.
Using the embedding map of $C$ in equation \eqref{eq:sCondition} one finds
\begin{equation}
    \cos\theta_0=\frac{\sinh\eta_0-\cosh u_0\,\sinh s}{\sinh u_0\,\cosh s}\,.
    \label{eq:costheta0}
\end{equation}
In principle, there are two latitude circles on the sphere $\CB$ at $\theta_0$ and $\pi-\theta_0$ that are identical in length, and their cosine values differ by sign. To justify the overall sign of \eqref{eq:costheta0}, we consider $\eta_0=\eta_{\text{min}}=s-u_0$, and the formula \eqref{eq:costheta0} gives $\cos{\theta_0} =-1$ as expected, since when $\eta_0=\eta_{\text{min}}$, the area of $\CB_-$ vanishes.

Here we briefly comment on the relation among $\theta_0$ angles of four aforementioned real saddles. The expression $\eqref{eq:costheta0}$ flips the sign when $\eta_0\rightarrow -\eta_0$ and $s\rightarrow-s$. For the pairs of complementary saddles, we have $s_{\text{small}} = -s_{\text{large}}'$ and $s_{\text{large}}=-s_{\text{small}}'$ with $\eta_0'=-\eta_0$, which suggests that the $\theta_0$ angles of these complementary saddles sum to $\pi$ and their spacetime boundary $\CB_-$ together cover the whole sphere $\CB$. Moreover, for fixed $\eta_0$, $\cos\theta_0(\eta_0,\,s_{\text{small}})$ and $\cos{\theta_0(\eta_0,s_{\text{large}})}$ can be shown to differ only by sign from \eqref{eq:costheta0}. Although the $s_{\text{small}}$ saddle and $s_{\text{large}}$ saddle are not complement to each other in the sense that the extrinsic curvatures on their spatial slice $K_{\Sigma}$ are the same, the union of their spacetime boundaries $\CB_-$ still have an area $\text{Area}(\CB)$ identical to that of two complementary saddles. Therefore, there are only two distinctive spacetime boundaries $\CB_-$ with the areas given by $\text{Area}(\CB_-) = 2\pi \sinh^2{u_0}(1\pm\cos{\theta_0})$.

Since the corner circle $C$ has the circumference
\begin{equation}
    L=2\pi \sinh u_0\,\sin\theta_0=L_{\text{crit}}(T)\sin\theta_0\,,
    \label{eq:L-theta0}
\end{equation}
the two distinctive spacetime boundaries $\CB_-$ for fixed minisuperspace parameters $(A,L)$ can be equivalently distinguished by
\begin{equation}
    \cos\theta_0^{\rm small}=-\sqrt{1-\left(\frac{L}{L_{\text{crit}}}\right)^2},\qquad
    \cos\theta_0^{\rm large}=+\sqrt{1-\left(\frac{L}{L_{\text{crit}}}\right)^2}.
    \label{eq:costheta0-branches}
\end{equation}
The notions of ``small'' and ``large'' in the superscript share the same meaning as those in $s_{\text{small}}$ and $s_{\text{large}}$, which indicates if the spacetime boundary $\CB_-$ covers less than or more than half of the sphere $\CB$. When $L$ takes its critical value $L_{\text{crit}}$, both spacetime boundaries $\CB_-$ become the exact half of the sphere.

\paragraph{The corner term.}
Finally, the joint term in \eqref{eq:EuclActionFull} depends on the dihedral angle $\Theta$ between $\Sigma$ and $\mathcal{B}_-$ along $C$. Since $n_{\Sigma}=\partial_{\eta}$, we have
\begin{equation}
    \cos\Theta=-n_{\Sigma}\cdot n_{\mathcal{B}_-},
\end{equation}
with $n_{\mathcal{B}_-}$ the outward unit normal to $\mathcal{B}_-$.
From \eqref{eq:braneImplicit}, a convenient unnormalized normal is $\nabla F$ with
\begin{equation}
    F(\eta,\rho)\equiv \cosh\eta\,\cosh\rho\,\cosh s-\sinh\eta\,\sinh s-\cosh u_0.
\end{equation}
Using the metric \eqref{eq:H3H2sliceMetric} one finds at the corner $(\eta,\rho)=(\eta_0,\rho_c)$,
\begin{equation}
    \cos\Theta=-\frac{\sinh\eta_0\,\cosh\rho_c\,\cosh s-\cosh\eta_0\,\sinh s}{\sqrt{\big(\sinh\eta_0\,\cosh\rho_c\,\cosh s-\cosh\eta_0\,\sinh s\big)^2+\cosh^2 s\,\sinh^2\rho_c}}.
    \label{eq:cosvartheta}
\end{equation}
Note that for complementary saddles, we can easily verify that $\Theta_{\text{small}} = \pi - \Theta_{\text{large}}'$ and $\Theta_{\text{large}} = \pi - \Theta_{\text{small}}'$. However, $\Theta_{\text{small}}$ and $\Theta_{\text{large}}$ are in general irrelevant. Unless $\eta_0=0$, we have $\Theta_{\text{small}} = \pi- \Theta_{\text{large}}$. 
Given the value of the dihedral angle $\Theta$, the corner contribution to the on-shell action is then
\begin{equation}
    {\frac{1}{8\pi G_N}\int_C\!\sqrt\sigma\,\Big(\Theta-\frac{\pi}{2}\Big)=\frac{L}{8\pi G_N}\Big(\Theta-\frac{\pi}{2}\Big),}
    \label{eq:jointTerm}
\end{equation}
where $L$ is the boundary length in \eqref{eq:ALetac}. 

\paragraph{Semi-classical Hartle--Hawking wave function.}
Combining \eqref{eq:VolResult}, \eqref{eq:SigmaTerm}, \eqref{eq:braneTermOnShell} and \eqref{eq:jointTerm}, the minisuperspace saddle gives
\begin{equation}
    I_{\rm E}^{\rm on-shell}(\eta_0,\rho_c;T)
    =\frac{1}{4\pi G_N}\mathrm{Vol}(\mathcal M_-)
    -\frac{A}{4\pi G_N}\tanh\eta_0
    -\frac{T}{8\pi G_N}\mathrm{Area}(\mathcal{B}_{-})
    {+\frac{L}{8\pi G_N}\Big(\Theta-\frac{\pi}{2}\Big),}
    \label{eq:IEfinal}
\end{equation}
with $\mathrm{Vol}(\mathcal M_-)$ and $\mathrm{Area}(\mathcal{B}_{-})$ given above and with $s$ fixed by \eqref{eq:sCondition}. Equation \eqref{eq:IEfinal} applies to all four real Euclidean saddles upon substituting the corresponding root for $s$ and choosing the sign for $\eta_0$. 

As in the computation of the Hartle–Hawking wave function in dS spacetime \cite{HH83}, the choice of saddle is sensitive to the choice of steepest-descent contour. In dS, the spatial slice $\Sigma$ is a $(D-1)$-dimensional sphere, with its radius as the minisuperspace parameter, embedded in $D$-dimensional sphere and the spatial slice $\Sigma$ cuts the $D$-dimensional sphere into two pieces which lead to two distinctive saddles. As these two saddles can be glued together through the interface $\Sigma$, they form a pair of complementary saddles. These two saddles can be simply distinguished by the sign of the extrinsic curvature $K_\Sigma$ on the spatial slice, which can be viewed as the canonical conjugate to the minisuperspace parameter. In the dS discussion, it was shown that the naive comparison between the on-shell actions of two saddles suggests choosing the one with $K_{\Sigma}<0$, which is a geometry that fills more than half of the $D$-dimensional sphere. However, a careful analysis on the steepest-descent contour of the GPI indicates that the contour will only pass through the saddle with positive $K_\Sigma$, so the semiclassical HH wave function should be approximated by the saddle which looks subdominant when comparing the actions. 

In the AdS$_3$ calculation, we find more saddles, while they still form pairs of complementary saddles. We are taking a similar minisuperspace approximation labeled by $(L,\,A)$. Therefore, it is tempting to adopt the same contour prescription introduced in dS gravity and impose positivity on the quantities that are canonical conjugates to the AdS minisuperspace parameters. The area parameter $A$ is determined by the induced metric on the spatial slice $\Sigma$, and the suitable conjugate variable is then also the extrinsic curvature $K_\Sigma$. Given the dependence of $K_\Sigma$ on $\eta_0$, the positivity constraint on $K_\Sigma$ requires $\eta_0$ also to be positive. This constraint immediately filters out two saddles with $\eta_0'= -\eta_0$. The other minisuperspace parameter, $L$, depends on how the intersection circle $C$ is embedded in the spacetime boundary $\CB_-$. Since we treat our spacetime boundary as a fully gravitating one, we can interpret the path integral on the spacetime boundary as a GPI on a dS$_2$ brane world\footnote{Note, however, this is different from end-of-the-world branes studied in the AdS$_3$/BCFT$_2$ context \cite{KR00,Takayanagi11,FTT11}, whose world volume is AdS$_2$.
By contrast, similar dS$_2$ branes have been considered in the holographic dual of boundary CFT with spacelike boundaries \cite{AKTW20,AKRTW21,AKRTW22,FKKT25} and the holographic dual of crosscap CFT \cite{Wei24,WY24,LXH25}.} \cite{RS99-1,RS99-2,SMS99,GT99,Gubser99} with $C$, the corresponding spatial slice. From this perspective, $L$ is exactly the same minisuperspace parameter as that discussed in \cite{HH83}, and the natural conjugate variable is the extrinsic curvature $K_C$ of $C$ with respect to the spacetime boundary $\CB_-$. In the previous calculation, we show that the intersection circle $C$ is a latitude circle at $\theta_0$ inside the boundary sphere $\CB$. The unit outward normal vector of $C$ with respect to $\CB_-$ is $n=-\frac{1}{\sinh{u_0}}\partial_{\theta}$, which gives the extrinsic curvature $K_C= -\frac{\cot{\theta_0}}{\sinh{u_0}}$ in terms of $\theta_0$. The positivity of $K_C$ implies that $\theta_0$ must be greater than or equal to $\frac{\pi}{2}$, i.e. $\cos{\theta_0}<0$. Following the contour prescription learned from dS gravity calculation, we single out one real saddle out of four, which is labeled by $s_{\text{small}}$ with $\eta_0\ge 0$, and the spacetime boundary $\CB_-$ in this case covers less than half of the sphere $\CB$. 

Finally, the HH wave function in the saddle point approximation is
\begin{equation}
    \Psi_{\rm HH}(A,L)\ \approx\ \exp\Big[-I_{\rm E,\CM_{-,\text{small}}}^{\rm on-shell}(A,L;T)\Big],
\end{equation}
where $(A,L)$ are related to $(\eta_0,\rho_c)$ through \eqref{eq:ALetac} and the on-shell Euclidean action is evaluated at the $s_{\text{small}}$ saddle using the general formula \eqref{eq:IEfinal}. When $L>L_{\text{crit}}(T)$, the saddles become complex, and the steepest-descent contour will not only pass through one single saddle. The above semiclassical wave function will become a sum over complex saddles.

\subsection{Further restricted minisuperspace by fixing \texorpdfstring{$\ell$}{l} }
\label{sec:restricted-minisuperspace}

In the two-parameter minisuperspace used above, the boundary disk $\Sigma$ is a constant-curvature
hyperbolic disk with intrinsic curvature radius
\begin{equation}
    \ell \;\equiv\; \cosh\eta_0\ \ge 1,
\end{equation}
together with a cutoff $\rho_c$.  The induced boundary data $(A,L)$ are related to $(\ell,\rho_c)$ by
\eqref{eq:ALetac}.  Fixing $\ell$ (equivalently fixing $\eta_0$) reduces minisuperspace to a single parameter,
which we may take to be $L$ (or $\rho_c$). We will focus on this further restricted minisuperspace by choosing $\ell=1$ in this subsection.

The simplest restricted minisuperspace is the one with $\ell=1$, i.e.\ $\eta_0=0$, so that $\Sigma$ is totally geodesic and
$K_\Sigma=0$. In this case, $A$ becomes an implicit function of $L$, using \eqref{eq:ALetac}:
\begin{equation}
    L=2\pi\sinh\rho_c,\qquad A=2\pi(\cosh\rho_c-1),\quad  A(L)=2\pi\left(\sqrt{1+\Big(\frac{L}{2\pi}\Big)^2}-1\right).
    \label{eq:AofL-ell1}
\end{equation}
Since $\eta_0 = 0$, the four real saddles degenerate to two labeled by $s_{\rm small}$ and $s_{\rm large}$. It is straightforward to obtain the two real saddles in terms of the minisuperspace parameter $L$
\begin{equation*}
    \begin{aligned}
        s_{\rm small}(L)&=+\operatorname{arccosh}\!\left(\frac{\cosh u_0}{\sqrt{1+(L/2\pi)^2}}\right),\\
        s_{\rm large}(L)&=-\operatorname{arccosh}\!\left(\frac{\cosh u_0}{\sqrt{1+(L/2\pi)^2}}\right),\qquad (L<L_{\text{crit}}(T))\,.
    \end{aligned}
\end{equation*}
In this special case, we have $s_{\rm small} = -s_{\rm large}$, so the remaining two real saddles are exactly complements to each other.

The expression of the corner angle $\Theta$ in between $\Sigma$ and $\CB_-$ simplifies drastically in the $\ell=1$ ensemble. By setting $\eta_0=0$ and using the relation $T=\coth u_0$ to rewrite the equation \eqref{eq:cosvartheta}, we obtain the following solution for the corner angle $\Theta$
\begin{equation}
    \Theta(L)=\arcsin\!\left(T\,\frac{(L/2\pi)}{\sqrt{1+(L/2\pi)^2}}\right),
    \qquad (L\le L_{\text{crit}}(T)),
    \label{eq:vartheta-of-L-arcsin}
\end{equation}
where the $s_{\rm small}$ saddle  has $\Theta\in(0,\pi/2)$, and the complementary $s_{\rm large}$ saddle has $\Theta\in(\pi/2,\pi)$. 

\begin{figure}[ht]
    \centering
    \includegraphics[width=0.65\linewidth]{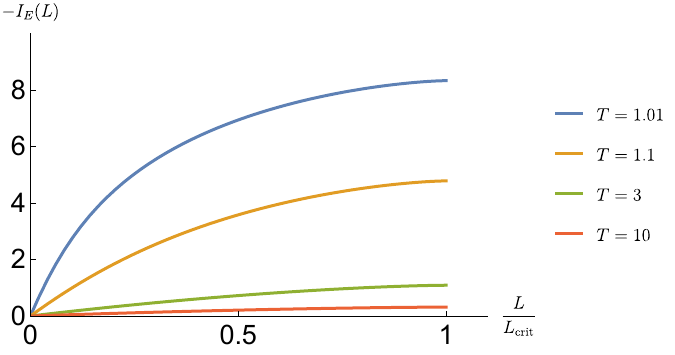}
    \caption{On-shell action $I_{\rm E}^{(\ell=1)}(L)$ as a function of $L$ for different tensions of the spacetime-boundary. We took $4\pi G_N=1$ for this plot. The maximum of the Hartle-Hawking wave function is exactly half of the on shell action evaluated on $\mathcal{M}$, the manifold bounded by the spacetime boundary $\mathcal{B}$. }
    \label{fig:AdS3action}
\end{figure}

Given the contour prescription proposed in the previous subsection, we choose the real saddle with positive $K_\Sigma$ and $K_C$ to evaluate the semiclassical wave function. As $\ell = 1$, the extrinsic curvature $K_\Sigma$ vanishes, while the positivity condition on $K_C$ still helps us select out the $s_{\rm small}$ saddle.
We can now evaluate the on-shell action \eqref{eq:IEfinal} explicitly at $\eta_0=0$ for the $s_{\rm small}$ saddle.  In this case the
$\Sigma$-term vanishes by \eqref{eq:SigmaTerm}, and we obtain a closed form for
$L\le L_{\rm crit}(T)$:
\begin{equation}
    \begin{aligned}
    I_{\rm E}^{(\ell=1)}(L)
    &=
    \frac{1}{4 G_N}\left[
        \operatorname{arccosh}\!\left(\frac{\cosh u_0}{\sqrt{1+(L/2\pi)^2}}\right)-u_0
    \right]
    {-\frac{L}{8\pi G_N}\left[
        \frac{\pi}{2}-\arcsin\!\left(T\,\frac{(L/2\pi)}{\sqrt{1+(L/2\pi)^2}}\right)
    \right]\,.}
    \end{aligned}
\label{eq:IE-ell1-closed}
\end{equation}
We also demonstrate a plot of the on-shell action $-I_{\rm E}^{(\ell=1)}(L)$ as a function of the ratio between $L$ and $L_{\rm crit}$ in figure \ref{fig:AdS3action} for various tension $T$ choices.

The maximum of $-I_{\rm E}^{(\ell=1)}(L)$, that indicates the peak of the semiclassical HH wave function, is attained at $L=L_{\rm crit}(T)$, where it evaluates to $I_{\rm E}(L_{\rm crit}(T))=-\frac{u_0}{4G_N}$. This is exactly one half of the on-shell action evaluated on the full hyperbolic ball $\CM$ bounded by the sphere $\mathcal{B}$ at $u=u_0$. This is consistent with the expectation that the norm of the HH state in AdS space, which can be well approximated by the partition function on the hyperbolic ball $\CM$, semiclassically should match the square of the HH wave function at its peak. 

For $L>L_{\rm crit} (T)$, one needs to analytically continue the on-shell action \eqref{eq:IE-ell1-closed} (or equivalently to evaluate the on-shell action using the complex saddle described in \eqref{eqn:complexsaddleprofile}) so that $I_{\rm E}^{(\ell=1)}$ becomes complex. We shall now explore this in detail.

The explicit analytic continuation of the on-shell action \eqref{eq:IE-ell1-closed} is carried out in Appendix \ref{app:analytic_continuation_l=1}. The on-shell action can be written as:
\begin{equation}
    I_{\rm E}^{(\ell=1)}(L)\;=\; I_R^{(\ell=1)}(L)\ \pm\ i\,I_I^{(\ell=1)}(L)\,,
    \label{eq:IE-ell1-complex-decomp}
\end{equation}
with real and imaginary parts
\begin{equation}
\boxed{
\begin{aligned}
    I_R^{(\ell=1)}(L)
    &{= -\frac{1}{4 G_N}\,u_0,}
    \\
    I_I^{(\ell=1)}(L)
    &= \frac{1}{4 G_N}\,\arccos\!\Bigg(\frac{\cosh u_0}{\sqrt{1+(L/2\pi)^2}}\Bigg)\;-\;\frac{L}{8\pi G_N}\,\arccosh\!\Bigg(T\,\frac{(L/2\pi)}{\sqrt{1+(L/2\pi)^2}}\Bigg)\,.
\end{aligned}}
\label{eq:IR-II-ell1}
\end{equation}
The $\pm$-sign ahead of the imaginary part comes from the two distinctive branches of the analytical continuation, which correspond to the two complex saddles $s = \pm i\beta$ defined in \eqref{eq:sComplex} respectively ($\alpha$ in \eqref{eq:sComplex} is zero for $\ell =1$ case). At the transition point $L=L_{\rm crit}(T)$, the imaginary part of the action vanishes, in agreement with the fact that the selected $s_{\rm small}$ saddle remains real for $L\le L_{\rm crit}(T)$.
Different from the real saddle case, the steepest-descent contour will pass through both complex saddles, which leads to a real oscillatory wave function proportional to:
\begin{equation}
    \Psi^{(\ell=1)}_{\rm HH}(L)\ \propto\ 2\,\exp\!\big[-I_R^{(\ell=1)}(L)\big]\,
    \cos\!\big(I_I^{(\ell=1)}(L)\big)\,.
\end{equation}

At large $L\to\infty$, the oscillatory phase behaves as follows
\begin{equation}
    I_I^{(\ell=1)}(L)\;=\; \frac{\arccosh T}{8\pi G_N}\,L\;-\; \frac{\pi}{8 G_N}\,,
\end{equation}
which grows linearly with $L$ in magnitude. This is the limit when the boundary $C$ of the spatial slice $\Sigma$ approaches the asymptotic boundary of global AdS$_3$ spacetime. The first term comes from the Hayward term. The $\arccosh{T}$ factor shows how the spacetime boundary $\CB_-$ bends away from the asymptotic boundary with the limit $\lim_{T\rightarrow 1^+}\arccosh{T} = 0$, when the spacetime boundary $\CB_-$ coincides with the asymptotic boundary.

\subsection{Probabilistic interpretation of Hartle-Hawking wave function}
Once we evaluate the semiclassical HH wave function, we might want to interpret it in the same way as we understand the ordinary quantum-mechanical wave function, which provides a probabilistic distribution on the phase space of the corresponding quantum system. In order to proceed in this way, one immediate question we have to address is what the gravity phase space on the spatial slice $\Sigma$ looks like. 
 
The naive answer is that the phase space consists of the induced metric $\gamma_{ab}$ and its conjugate momentum $\pi^{ab}$ on the spatial slice $\Sigma$. However, this is not the final answer, since the gravity theory has a large gauge symmetry given by the diffeomorphisms. To get the true quantum gravity phase space, we have to identify the configurations $(\gamma_{ab},\,\pi^{ab})$ and $(\gamma_{ab}',\,\pi'^{ab})$, which are connected via the diffeomorphism transformation. Equivalently, we need to impose the Hamiltonian (time-reparametrization) and Momentum (spatial-diffeomorphism) constraints when performing the canonical quantization and deriving the quantum gravity Hilbert space. In general, it is difficult to explicitly write down the gravity phase space after gauging the full diffeomorphism symmetry. Solving the Hamiltonian constraints completely is also technically intractable. This obstructs us from extracting a meaningful probabilistic distribution over the quantum gravity phase space.

In order to obtain an approachable phase space where we can perform canonical quantization, we learn a lesson from AdS/CFT that we should break some gauge symmetries. In AdS/CFT, there is a canonical choice of the set of diffeomorphism symmetry to break, which are those so-called ``large-diffeomorphisms''. These are diffeomorphisms that do not vanish at the asymptotic boundary and preserve the Dirichlet boundary condition fixed there. These broken gauge symmetries form the conformal symmetry algebra and become the physical symmetry of the conformal field theory living on the asymptotic boundary. In this setup, we can possibly map the partially-frozen AdS HH wave function to the CFT Hilbert space and interpret it as giving a probabilistic distribution on the CFT phase space.

Regarding the HH wave function in dS gravity and fully-gravitational AdS HH wave function defined in this paper, we miss a canonical choice of the broken gauge symmetries, since either we do not have boundary or the boundary becomes fully gravitational. Then the phase spaces obtained from breaking gauge symmetries will not be unique. For instance, we can treat the minisuperspace parameter $(L,\,A)$ used in this section and their canonical conjugates as one suitable set of phase space variables. Even if we determine the phase space variables, we need to know the measure $\mu(A,\,L)$ over the phase space (usually given by the symplectic structure on the phase space) in order to define the probabilitistic distribution 
 \begin{equation}
     P_{\rm HH}(A,\,L) = \mu(A,\,L)\,|\Psi_{\rm HH}(A,\,L)|^2\,.
 \end{equation}
Because we do not have a canonical choice of the phase space from breaking partial gauge symmetries, we decide not to pursue a probabilistic interpretation of the HH wave function evaluated in this paper.

\section{Hartle--Hawking wave function in \texorpdfstring{AdS$_2$}{AdS2}}
\label{sec:JT-noboundary}

\begin{figure}[ht]
    \centering   \includegraphics[width=0.6\linewidth]{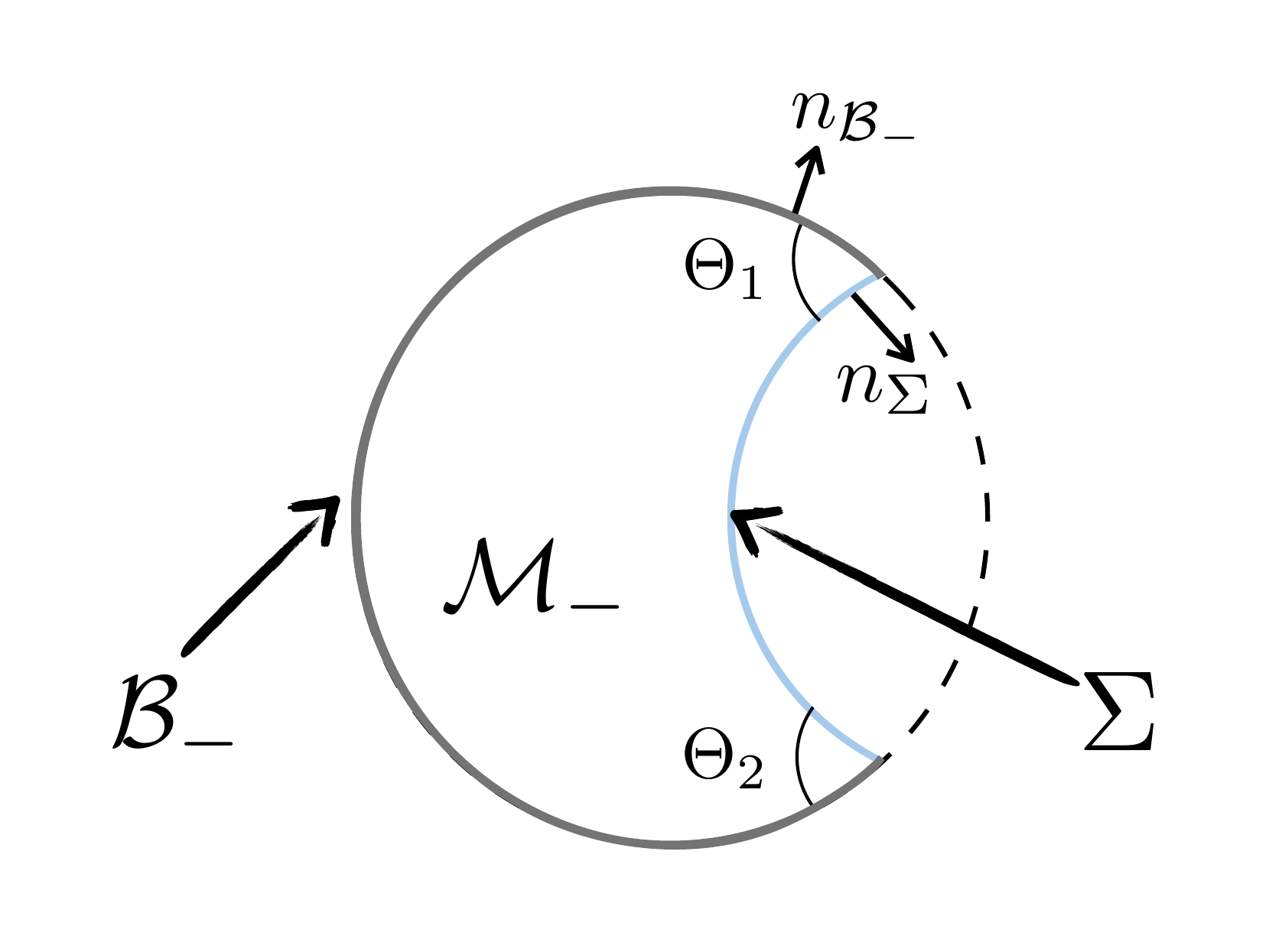}
    \caption{A 2D manifold $\mathcal{M}_-$ is represented with a spacetime boundary $\mathcal{B}_{-}$ in gray. The spatial slice $\Sigma$ is represented in light blue. The angles $\Theta_1$ and $\Theta_2$ appearing in the Hayward term are also shown in the figure. The unit vectors normal to $\Sigma$ and $\CB_-$ are labeled by $n_\Sigma$ and $n_{\CB_-}$ respectively.}
    \label{fig:JT introduction parameters}
\end{figure}

As a second example of a fully gravitational HH wave function \eqref{eq:open_HH} in AdS, we consider the following action for AdS$_2$ JT gravity
\begin{equation}
\label{eqn:JTaction}
\begin{aligned}
    I_{\text{E}} = &-\frac{S_0}{2\pi}(\frac{1}{2}\int d^2x\,\sqrt{g}\, R+\int_{\mathcal{B_-}} du\,\sqrt{h}\,K+(\frac{\pi}{2}-\Theta_1)+(\frac{\pi}{2}-\Theta_2))\\&-\frac{1}{2}\int d^2x\, \sqrt{g}\, \Phi\,(R+2) -\int_{\mathcal{B_-}} du\, \sqrt{h}\,(\Phi\,(K-T)-\mu)-\int_{\Sigma}dv\, \sqrt{\gamma}\,\Phi\,K\\
    &-\Phi(x_1)(\frac{\pi}{2}-\Theta_1)-\Phi(x_2)(\frac{\pi}{2}-\Theta_2)\,,
\end{aligned}
\end{equation}
and compute the Euclidean gravitational path integral on 2D manifolds $\CM_-$ with boundary $\Sigma$, the spatial slice (i.e. the HH slice), and $\mathcal{B}_{-}$, a cutoff surface, referred to as the “spacetime boundary.” The parameter $T$ associated to $\CB_-$ is called the tension because the boundary action on $\mathcal{B}_{-}$ mimics that of the spacetime boundaries in higher dimensions, and $T$ corresponds to the tension of the spacetime boundary in higher dimensions. In particular, $T=1$ reproduces the usual holographic renormalization counterterms, and the spacetime boundary $\mathcal{B}_{-}$ reaches the standard asymptotic AdS boundary when on shell. We also introduce an additional mass parameter $\mu$ for $\CB_-$ compared to the usual 2D JT action.\footnote{The counterpart of this parameter in higher dimensions is brane-localized perfect fluid.} See figure \ref{fig:JT introduction parameters} for a sketch of the setup.

In the action \eqref{eqn:JTaction}, we include additional Hayward terms, which couple the dilaton $\Phi(x)$ to the angles $\Theta_1$ and $\Theta_2$. These angles represent the intersection angles between the spacetime boundary $\CB_-$ and the spatial slice $\Sigma$,  as depicted in figure \ref{fig:JT introduction parameters}. The Hayward terms are required to fix the boundary values $\Phi(x_1)$ and $\Phi(x_2)$ of the dilaton field.\footnote{Alternatively, one may instead impose Neumann boundary conditions $\Theta_1=\Theta_2=\frac{\pi}{2}$.}

The GPI, by definition, evaluates a wave functional depending on the specified boundary conditions on both the spatial slice $\Sigma$ and the spacetime boundary $\CB_-$. By varying the given actions, we find that the valid Dirichlet boundary conditions to be fixed on $\Sigma$ are
\begin{equation}
\label{eqn:JTCauchybc}
    \begin{aligned}
        \Phi|_{\Sigma}&=\phi(\lambda)\\
        \gamma_{vv}|_{\Sigma} &= 1\,. 
    \end{aligned}
\end{equation}
The spatial slice $\Sigma$ is a curve inside the Euclidean AdS$_2$ geometry with a fixed dilaton profile as a function of the affine parameter $\lambda$ on it, and the variable $v$ measures the proper length of this curve. The affine parameter $\lambda$ gives a coordinate-independent parameterization for the dilaton profile. By fixing the range of $v$, we fix the proper length of the spatial slice to be $L$. $(L, \phi(\lambda))$ together specify the data needed for the evaluation of the wave function.\footnote{In principle, we can consider an arbitrary dilaton profile on the spatial slice, which is similar to considering an arbitrary induced metric $\gamma_{ab}$ on the spatial slice in higher dimensions. An analogy to the minisuperspace ansatz can be applied to JT gravity, where we only consider dilaton profiles $\phi(\lambda)$ that can be fixed on constant-extrinsic-curvature $K$ curves. An explicit expression for the dilaton profile and its dependence on $K$ is demonstrated in Section \ref{sec:Cauchycalculation}.}

Provided that higher topologies are suppressed by the Gauss–Bonnet term proportional to $S_0$, the leading contribution comes from the disk topology. The bulk equations of motion of JT gravity are given by:
\begin{equation}
\label{eqn:JTEOM}
    \begin{aligned}
        R+2=& 0\\
        \nabla_\mu\nabla_\nu\Phi -g_{\mu\nu}\nabla^2\Phi+g_{\mu\nu}\Phi=&0\,.
    \end{aligned}
\end{equation}
The admissible on-shell geometry with disk topology must be a submanifold of the Euclidean AdS$_2$ space, which has a metric with constant negative curvature. Let us introduce some useful coordinates for Euclidean AdS$_2$ space.

\paragraph{Poincar\'e disk coordinate}~\par
The metric of the Euclidean AdS$_2$ space can be written in the Poincar\'{e} disk coordinate $(r,\theta)$ 
 \begin{equation}
     \label{eqn:Poincaredisk metric}
     ds^2= \frac{4(dr^2+r^2 d\theta^2)}{(1-r^2)^2}\,,
 \end{equation}
 where $r$ ranges between $0$ and $1$, and $\theta$ has a period of $2\pi$. The Poincar\'{e} disk coordinate covers the whole Euclidean AdS$_2$ space.

\paragraph{Embedding in $\mathbb{R}^{2,1}$}~\par
The Euclidean AdS$_2$ space can be embedded into the $(2+1)$-dimensional Minkowski space. It is useful to specify the embedding map from the Poincar\'e disk into the Minkowski space:
 \begin{equation}
     ds^2 = -d(X^0)^2+d(X^1)^2+d(X^2)^2\,,
 \end{equation}
 where the Euclidean AdS$_2$ space covers the surface
 \begin{equation}
     -(X^0)^2+(X^1)^2+(X^2)^2 =-1\,,
 \end{equation}
 where we chose the bulk AdS-radius to be $1$. The embedding map is given as follows
 \begin{equation}
 \label{eqn:embeddingmap}
     X^0 = \frac{1+r^2}{1-r^2},\quad X^1 = \frac{2r}{1-r^2}\sin{\theta},\quad X^2 = \frac{2r}{1-r^2}\cos{\theta}\,.
 \end{equation}
 Note that $r$ on the Poincar\'{e} disk is between 0 and 1, so the Poincar\'{e} disk only covers the $X^0>0$ branch. The isometry group $\text{PSL}(2,\mathbb{R})$ of the Poincar\'{e} disk can be realized linearly as the Lorentz group $\text{SO}^+(2,1)$ acting on the embedding surface.

\paragraph{Euclidean Schwarzchild (Rindler) coordinate}~\par
The Euclidean AdS$_2$ geometry can also be written in the Euclidean Schwarzschild (Rindler) coordinate
\begin{equation}
\label{eqn:EuclideanSchwarzschild}
    ds^2 = (r_{\text{Sch}}^2-r_s^2)\,dt_{\rm E}^2+\frac{dr_{\text{Sch}}^2}{(r_{\text{Sch}}^2-r_s^2)}\,,
\end{equation}
which contains two parameters $(\beta,\, r_s)$. $\beta$ is the periodicity around Euclidean time $t
_{\rm E}$ direction and $r_s$ can be viewed as the horizon radius. The geometry covers the region $r_{\text{Sch}}\ge r_s$. Notice that the geometry is smooth only if  $r_s=\frac{2\pi}{\beta}$. For reference, we also write down the map between the Euclidean Schwarzschild and the Poincar\'{e} disk coordinates:
 \begin{equation}
 \label{eqn:poincaretorindler}
     r_{\text{Sch}} = r_s \frac{1+r^2}{1-r^2},\quad t_{E} = \frac{\theta}{r_s}\,.
 \end{equation}

~\par
The EOM for the dilaton field can be conveniently solved in the embedding Minkowski space, where the slices of constant dilaton value $\phi$ form hyperplanes 
 \begin{equation}
     \Phi(X^\mu) = C_\mu X^\mu\,,
 \end{equation}
 with the normal vector $C^\mu$. Therefore, the general solution of dilaton field on the Poincar\'e disk is given as 
 \begin{equation}
 \label{eqn:dilatonsolution}
     \Phi(r,\theta) = C_0 \frac{1+r^2}{1-r^2}+\frac{2 r}{1-r^2}(C_1\sin{\theta}+C_2\cos{\theta})\,,
 \end{equation}
 which are parameterized by three real coefficients $(C_0,C_1,C_2)$. The coefficient $C_0$ labels the dilaton value at the origin of the disk. Once we fix the dilaton profile on the spatial slice $\Sigma$, these three coefficients will be fully determined.

Remember although we are doing a computation where configurations of the spacetime boundary $\CB_-$ are summed over off shell, the saddle point approximation leads to EOM for the spacetime boundary $\CB_-$ which looks like Neumann boundary conditions\footnote{See \cite{GIKY20} for a comprehensive classification of JT boundary conditions and their associated actions.}:
\begin{equation}
    \label{eqn:JTspacetimebc}
    \begin{aligned}
        K|_{\mathcal{B}_-}\,=&\,T,\\
        \partial_n\, \Phi|_{\mathcal{B_-}}-T\,\Phi|_{\mathcal{B_-}}\,=&\,\mu\,.
    \end{aligned}
\end{equation}
At the leading order of the saddle point approximation,
\begin{equation}
    \Psi_{\text{HH}}^{\text{JT}}[L,\phi(\lambda)]\approx \exp{\{-I_{\text{E}}[g^*,\Phi^*]\}}|_{\gamma|_{\Sigma},~\Phi|_{\Sigma},~K|_{\mathcal{B_-}},~\partial_n\Phi|_{\mathcal{B_-}}}\,,
\end{equation}
realized by introducing a gravitating spacetime boundary $\CB_-$ with tension $T$ and mass $\mu$ coupled to the bulk JT gravity. 
Recall that $L$ is the proper length of the spatial slice $\Sigma$, and $\phi(\lambda)$ is the dilaton profile.
In the following, we consider in this regime. 

Note that the partially frozen counterpart \eqref{eq:partially_frozen_HH} in JT gravity, with an asymptotic spacetime boundary $\mathcal{B}_-$ and fixed Dirichlet boundary conditions on it, was studied in \cite{HJ18}.

\subsection{JT minisuperspace ansatz and maximally-symmetric bubble}
\label{sec:JTminsuperspaceansatzandmaximallybubble}
In this subsection, we pave the way for the evaluation of the semiclassical HH wave function by introducing a minisuperspace approximation for the data imposed on the spatial slice $\Sigma$, which we parameterize by its on-shell extrinsic curvature $K$. We also aim to determine the position of the spacetime boundary $\mathcal{B}_{-}$ by solving the on-shell EOMs \eqref{eqn:JTspacetimebc} on it. In particular, the first equation in \eqref{eqn:JTspacetimebc} indicates that the spacetime boundary $\CB_-$ can also be derived from a curve with constant extrinsic curvature $K_{\CB_-} = T$. As the minisuperspace-approximated spatial slice $\Sigma$ and the spacetime boundary $\CB_-$ are both realized by constant-$K$ curves on-shell, we first need to obtain the general solutions for curves with constant extrinsic curvature $K$ within the Poincaré disk. 

In the Poincar\'e disk coordinate \eqref{eqn:Poincaredisk metric}, the coordinate $r$ is defined in the segment $[0,1)$. Let us extend its definition to the semi-infinite line $[0,\infty)$ and regard the Poincar\'e disk as a ``unit" disk embedded in a $\mathbb{R}^2$ up to a Weyl rescaling. In this treatment, the $r\geq 1$ region is unphysical, but it turns out to be useful when classifying curves (with constant extrinsic curvature) on the Poincar\'e disk. 

\begin{figure}[ht]
    \centering
    \includegraphics[width=0.5\linewidth]{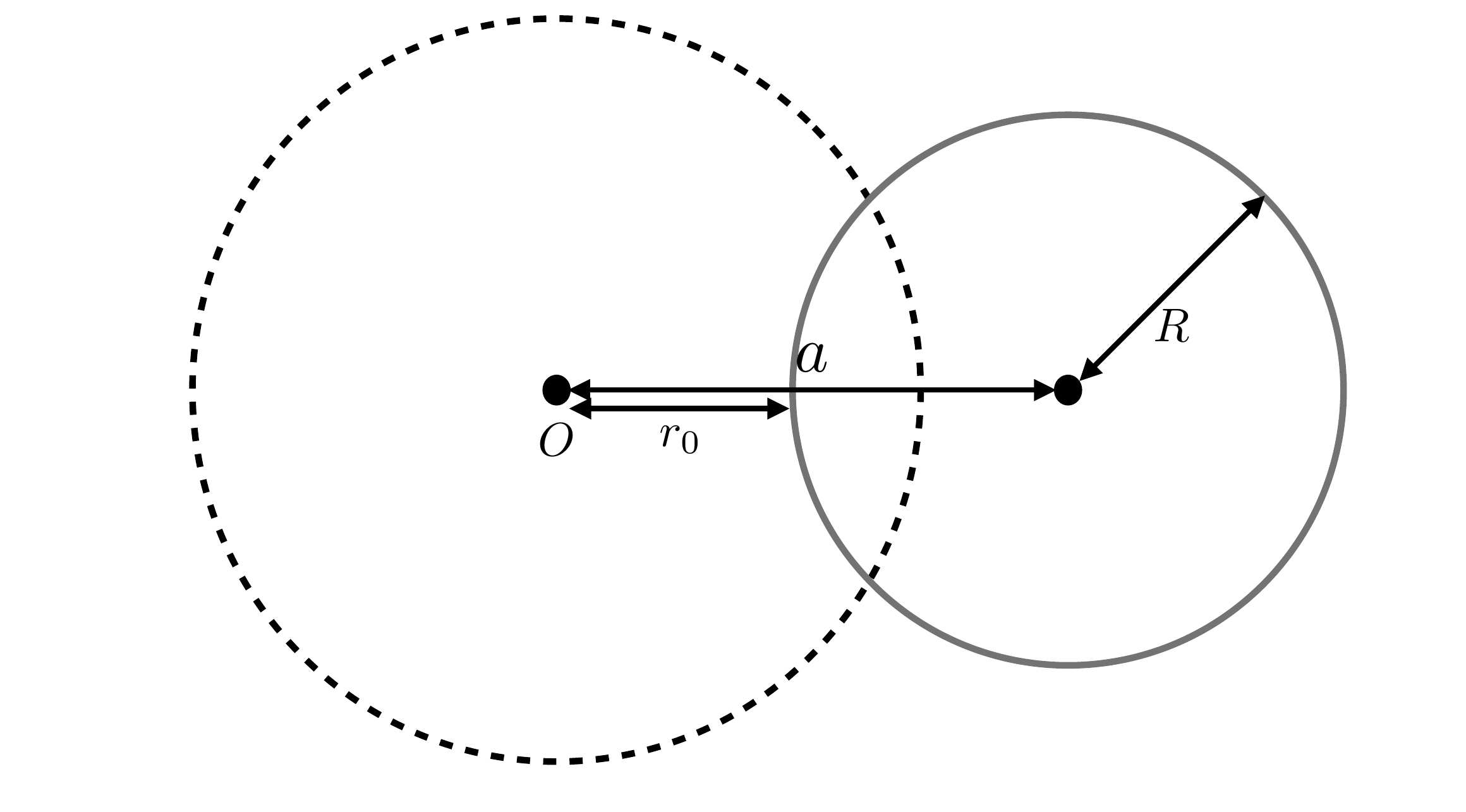}
    \caption{A typical constant-$K$ curve (solid) is depicted in the two-dimensional plane $\mathbb{R}^2$. The boundary of Poincar\'e disk is drawn in dashed line, which is a unit circle with the center at the origin $O$ of the plane $\mathbb{R}^2$. The Poincar\'e disk coordinate covers the region inside the dashed circle. The constant-$K$ curve, as a circle within the plane, is classified by two parameters, the radial coordinate $r=a$ of its center (at $\theta =0$), and its radius $R$. Note that here the radius is measured by the flat metric on $\mathbb{R}^2$. 
    The radial coordinate of the point on the constant-$K$ curve closest to the origin $O$ is marked as $r_0$, where $r_0$ is related to the parameters $a$ and $R$ by $|a-R|$.}
    \label{fig:PoincarediskConstantK}
\end{figure}

More precisely, these curves are classified as circles in the $\mathbb{R}^2$ parameterized by $r\in [0, \infty)$. Provided that the rotation around the angular coordinate $\theta$ in the Poincar\'e disk is an isometry, we assume that the centers of these circles all lie on the $\theta=0$ axis. They intersect or are contained within the Poincar\'e disk parameterized by $r\in [0, 1)$, which is a unit disk at the origin $r=0$ of the plane. 
They are classified by two parameters $a$ (the radial coordinate of the center of the circle) and $R$ (the radius of the circle), see figure \ref{fig:PoincarediskConstantK} for a depiction of these curves. The extrinsic curvature of the curve in terms of these two parameters can be written as
\begin{equation}
\label{eqn:KaRformula}
    K = -\frac{1-a^2+R^2}{2R}\,,
\end{equation}
which is computed with respect to the unit normal vector pointing towards the center of the circle. In Appendix \ref{sec:generalKcurve}, we provide a comprehensive analysis of this constant-curvature family of curves, specify their equations, and evaluate their proper length.

Once the solutions for these curves are obtained, the next step is to carve out the spatial slice $\Sigma$ and the spacetime boundary $\mathcal{B}_-$ from these curves. In this section, we focus on the case where the tension of spacetime boundary $T$ is greater than $1$. By requiring the absolute value of $K$ in the expression \eqref{eqn:KaRformula} to be greater than $1$, we find that the parameters $a$ and $R$ must satisfy the following conditions: $a+R<1$ or $a-R>1$ or $a-R<-1$. When $a-R>1$ or $a-R<-1$, the curves will be completely outside the Poincar\'e disk, i.e. within the unphysical region, which lead to invalid solutions. On the other hand, when $a+R<1$, the constant-$K$ curve becomes a circle fully contained in the Poincar\'e disk\footnote{The expression \eqref{eqn:KaRformula} leads to an extrinsic curvature $K<-1$ for $a+R<1$. In order to obtain $K_{\CB_-}=T>1$, we need to choose the outward normal vector to the spacetime boundary $\CB_-$ to point away from the center of the circle, and the 2D bulk of $\CM_-$ to be the interior of this constant-$K$ circle.}, so that the spacetime boundary is compact. We regard this circle as the boundary circle $\mathcal{B}$, where the spacetime boundary $\mathcal{B}_-$ lives. When $T\rightarrow 1^+$, the boundary circle $\CB$ approaches the usual asymptotic boundary.

The next step is to find another curve with extrinsic curvature $|K_\Sigma|<1$ to intersect the boundary circle $\mathcal{B}$ at two points\footnote{The reason to choose the absolute value of $K_\Sigma$ to be less than 1 is because only in this case the curve can intersect with the asymptotic boundary at two distinctive points, and we are allowed to take the $T\rightarrow 1^+$ limit of our calculation to compare with the known fixed DBC result in \cite{HJ18}.}. The part of the $|K_\Sigma|$-curve in between these two intersection points becomes our spatial slice $\Sigma$, while the arc of the boundary circle $\mathcal{B}$ between the two intersection points is the spacetime boundary. Note that there are two arcs on the boundary circle $\mathcal{B}$ with the same endpoints, which are complements to each other. This already signals the existence of two saddle-point geometries, each of which is enclosed by the spatial slice $\Sigma$ and one of these two arcs. We will provide arguments on the choice between these two saddles when we evaluate the semiclassical HH wave function explicitly in the latter subsection.

In order to locate the relative positions between the spatial slice $\Sigma$ and the spacetime boundary $\CB_-$, we take the following procedures. We first write down a one-parameter family of solutions for the spatial slice $\Sigma$, with definite extrinsic curvature $K$ and a proper length $L$, labeled by $r_{\text{Sch},0}$ in the Schwarzchild coordinate (or equivalently $r_0$ in the Poincar\'e disk coordinate as explained in figure \ref{fig:PoincarediskConstantK}), which parameterizes the radial coordinate of the point on the constant-$K$ curve, where the spatial slice $\Sigma$ is derived from, closest to the center of the Poincar\'e disk, i.e. the origin $O$ in figure \ref{fig:PoincarediskConstantK}. We parameterize the proper length of the spatial slice by the affine parameter $\lambda \in (-\frac{L}{2},\frac{L}{2})$ and fix the dilaton values of the following three points on the spatial slice 
\begin{equation}
    \phi\left(\frac{L}{2}\right) = \phi_l\,,\quad\phi\left(-\frac{L}{2}\right) = \phi_r\,,\quad \phi(0) = \phi_0\,.
    \label{eq:boundary_condition_dilaton}
\end{equation}
We then write down the explicit expression for a restricted family of the dilaton profiles on the spatial slice in terms of $K$, $L$, $r_0$ and the three fixed dilaton values $\phi_l,\,\phi_r,\,\phi_0$. At the same time, we can determine the three coefficients $C_0$, $C_1$ and $C_2$ in the general solution of the dilaton field on the Poincar\'e disk in terms of the same set of parameters by the three equations imposed by \eqref{eq:boundary_condition_dilaton}. Finally, we solve the EOMs \eqref{eqn:JTspacetimebc} of the spacetime boundary $\CB_-$ to find the location of the boundary circle $\CB$ classified by the parameters $(a_\CB,\, R_\CB)$ in terms of $C_\mu$ coefficients, which implicitly depend on the parameter $r_0$. By requiring the endpoints of the spatial slice to be on the boundary circle $\CB$, we can fix the value $r_0$, which tells us the position of the spatial slice and the spacetime boundary (through the relation between $(a_\CB,\,R_\CB)$ and $r_0$). We demonstrate this procedure step by step.

Note that the parameter $r_0$ is related to the two parameters $a$ and $R$, introduced before to classify the constant-$K$ curves, by $r_0 = |a-R|$. See again figure \ref{fig:PoincarediskConstantK}.
Given $|K|<1$ and the expression of $K$ in \eqref{eqn:KaRformula}, we have $|a-R|<1$, so the radial coordinate $r_0$ is guaranteed to lie within the Poincar\'e disk. When $a>R$, we set the angular coordinate of the point closest to the origin in both coordinate systems to be $\theta = 0$ and $t_{\rm E}=0$, while for $a<R$, we have, instead of $r_0<0$, $\theta = \pi$ and $t_{\rm E}=\frac{\pi}{r_s}$ following the coordinate transformation \eqref{eqn:poincaretorindler} with $r_s$, the horizon radius in Schwarzchild coordinate. Therefore, the constant-$K$ curve can be equivalently classified by $(r_0,\,\theta = 0)$ or $(r_0,\,\theta = \pi)$. To obtain a solution of spatial slice from the constant-$K$ curve, we identify the point on the spatial slice with the affine parameter $\lambda = 0$ with the point $(r_0,\,\theta = 0)$ or $(r_0,\,\theta = \pi)$. Given that the affine parameter $\lambda$ on the spatial slice ranges between $-\frac{L}{2}$ and $\frac{L}{2}$, this identification places the middle point of the spatial slice on the axis $\theta =0$ or $\theta=\pi$ in the Poincar\'e disk coordinate. In principle, we can identify the middle point of the spatial slice with any points on the constant-$K$ curve within the Poincar\'e disk, while we can always move the middle point back to the $\theta =0$ or $\theta=\pi$ axis via an isometry. The isometry will covariantly change the dilaton profile, which absorb this degree of freedom on the position of the middle point into the variants of the dilaton profile on the spatial slice.  Following this identification, we derive, in Appendix \ref{sec:generalKcurve}, the solution of the spatial slice with extrinsic curvature $K$ and the proper length $L$ as follows
\begin{equation}
 \label{eqn:Kcurve1}
     r_{\text{Sch}}(\lambda) = r_{\text{Sch},0}+\frac{(r_{\text{Sch},0}\mp K\sqrt{r_{\text{Sch},0}^2-r_s^2})(-1+\cosh{(\sqrt{1-K^2}\lambda)})}{1-K^2},\,(|K|<1)\,,
 \end{equation}
 with $\lambda$ in between $-\frac{L}{2}$ and $\frac{L}{2}$. The solution is written in the Schwarzchild coordinate for convenience, and can be converted to the Poincar\'e disk coordinate via the relation 
 \begin{equation}
 \label{eqn:curverlambda}
     r(\lambda) = \sqrt{\frac{r_{\text{Sch}}(\lambda)-r_s}{r_{\text{Sch}}(\lambda)+r_s}}\,,
 \end{equation}
 where the $r_s$-dependence (as an auxiliary variable) disappears in the end. The $\mp$ sign in the expression depends on whether we place the middle point on $\theta=0$ or $\theta=\pi$, respectively, and the angular coordinate $\theta(\lambda)$ of the spatial slice can be deduced from the curve equation 
 \begin{equation}
 \label{eqn:curvethetalambda}
     \cos{\theta(\lambda)} = \frac{a^2+r(\lambda)^2-R^2}{2 a r(\lambda)}\,.
 \end{equation}

~\par
There are three undetermined, independent parameters $C_0,C_1,C_2$ in the general solution of the dilaton field \eqref{eqn:dilatonsolution}, so we can fix the solution by imposing the dilaton values at three different points on the spatial slice. For instance, we can set the dilaton values at two endpoints and the middle point to be $\phi_l,\,\phi_r,\,\phi_0$ correspondingly.
By using the solution of the spatial slice \eqref{eqn:curverlambda} and \eqref{eqn:curvethetalambda}, we obtain three independent equations
\begin{equation}
\label{eqn:threeequationslr0}
    \Phi\left(r(\frac{L}{2}),\theta(\frac{L}{2})\right) = \phi_l\,,\quad \Phi\left(r(-\frac{L}{2}),\theta(-\frac{L}{2})\right) = \phi_r\,,\quad \Phi\left(r(0),\theta(0)\right) = \phi_0\,,
\end{equation}
which allow us to solve for the coefficients $C_0$, $C_1$ and $C_2$. Accordingly, we can find a suitable dilaton profile $\phi(\lambda)$ for the spatial slice in terms of ($K,L,r_0$)\footnote{$r_0$ seems to be an extra parameter besides the extrinsic curvature, which classifies the possible spatial data. However, the value of $r_0$ will be fixed by requiring the spatial slice to intersect with the spacetime boundary that satisfies the EOMs \eqref{eqn:JTspacetimebc} on-shell. In the end, we will find that $r_0$ is no longer a free parameter.} compatible with the dilaton values $\phi_l,\,\phi_r,\,\phi_0$ fixed at three locations.  
 
\paragraph{A minisuperspace ansatz}~\par
In the rest of this section it is useful to regard the constant-curvature family as a \emph{JT minisuperspace ansatz} for the spatial data. Off-shell, what we really keep fixed is the {\it proper length} together with {\it a restricted family of dilaton profiles} labeled by the same parameter that becomes the geometric extrinsic curvature $K$ on shell. In that sense, $K$ should be viewed as a convenient label for the minisuperspace family rather than as the claim that every off-shell configuration literally has constant extrinsic curvature. In Appendix \ref{sec:Cauchycalculation}, we perform calculations for arbitrary choices of boundary condition $(\phi_l,\,\phi_r,\,\phi_0)$ and derive the minisuperspace ansatz for $\phi(\lambda)$ explicitly.

In particular, we can ease the discussions by focusing on the symmetric HH saddle, which has an isotropic dilaton field $\Phi(r)$ on the Poincar\'e disk that depends only on the radial coordinate $r$, i.e. $C_1=C_2=0$ in \eqref{eqn:dilatonsolution}.  The subsector of the minisuperspace compatible with the symmetric HH saddle is the reflection-symmetric family described as follows:
\begin{equation}
\label{eq:JT-symmetric-bc}
    \phi_l=\phi_r\equiv \phi_*\,,
    \qquad
    \phi_*=\phi_0\,\frac{X^0_*}{X^0_0}
    \equiv C(r_0,K)\,\phi_0\,,
\end{equation}
with
\begin{equation}
\label{eq:JT-Crk-def}
    C(r_0,K)=\frac{X^0_*}{X^0_0}
    =1+\frac{(1\mp 2Kr_0+r_0^2)\bigl(\cosh(\sqrt{1-K^2}\,L/2)-1\bigr)}{(1-K^2)(1+r_0^2)}\,.
\end{equation}
The variables $X_*^0$ and $X_0^0$ stand for the $X^0 = \frac{1+r^2}{1-r^2}$ coordinate of the endpoints $r(\pm\frac{L}{2})$ and the middle point $r(0)$ of the spatial slice in the embedding coordinates \eqref{eqn:embeddingmap}. The $\mp$ sign in $C(r_0,\,K)$ has the same origin as the $\mp$ sign in the solution \eqref{eqn:Kcurve1} of the spatial slice, which depends on whether the middle point of the spatial slice is on $\theta=0$ or $\theta=\pi$ axis. Solving the equations \eqref{eqn:threeequationslr0} for $C_0, \,C_1, \,C_2$ leads to
\begin{equation}
\label{eq:JT-symmetric-C012}
    C_1=0\,,\qquad C_2=0\,,\qquad C_0=\frac{\phi_0}{X^0_0}\,,
\end{equation}
and the dilaton profile on the spatial slice reduces to the following form
\begin{equation}
\label{eq:JT-phi-profile-symmetric}
    \phi(\lambda)=\frac{\phi_0}{X^0_0}\,X^0(\lambda)\,,
\end{equation}
with $X^0(\lambda) = \frac{1+r^2(\lambda)}{1-r^2(\lambda)}$. This reflection-symmetric subsector of the minisuperspace keeps the dependence of the dilaton profile $\phi(\lambda)$ on $(L,K,r_0,\phi_0)$\footnote{Note that when $K=0$, the $r_0$-dependence inside $C(r_0,K)$ disappears.} explicit while eliminating the dependence on $\phi_l$ and $\phi_r$. In the following step, we solve the position of the boundary circle $\CB$ in terms of the coefficients $C_\mu$, and we will show that the symmetric HH saddle with isotropic dilaton field $\Phi(r)$ significantly simplifies the EOMs of the spacetime boundary \eqref{eqn:JTspacetimebc}. Under the assumption that the boundary circle $\CB$ is placed at the center of the Poincar\'e disk, we find that the only free parameter $C_0$ in the general solution of the dilaton field will be completely fixed by the tension $T$ and the mass parameter $\mu$ of the spacetime boundary.

\paragraph{Find the position of the boundary circle $\CB$}~\par
 Given the solution of curves with constant extrinsic curvature and the general solution \eqref{eqn:dilatonsolution} of dilaton field with fixed $C_\mu$ coefficients, we can explicitly find the position of the boundary circle $\CB$ within the Poincar\'e disk, which supports the spacetime boundary $\CB_-$, by solving the EOMs of the spacetime boundary. The action on the spacetime boundary 
 \begin{equation}
     \int_{\mathcal{B}_-}\sqrt{h}\, (\Phi (K-T)-\mu)
 \end{equation}
 admits the following Neumann boundary conditions (i.e. the EOMs from spacetime boundary variation)
 \begin{equation}
     K_{\mathcal{B}_-} = T\,,\qquad  \partial_n \Phi_{\mathcal{B}_-}-T \Phi_{\mathcal{B}_-} = \mu\,.
 \end{equation}
 The spacetime boundary action after imposing the above boundary conditions is proportional to the proper length of the boundary $L_{\mathcal{B}_-}$, where the coefficient $\mu$ is the mass parameter. Note that we have three different terms associated with three coefficients $(C_0,\,C_1,\,C_2)$ in the general  solution of the dilaton field \eqref{eqn:dilatonsolution}. Here we compute the contribution of each term to the spacetime boundary EOM separately. When computing the extrinsic curvature and the normal derivative, we need to choose a set of normal vectors for the spacetime boundary different from the ones listed in \eqref{eqn:normalvector}, which point towards the center of the boundary circle, since the outward normal vector to the spacetime boundary is the opposite of \eqref{eqn:normalvector}. This new set of normal vectors, as commented before, point away from the center of the boundary circle $\CB$, and lead to the extrinsic curvature $K_{\CB_-}=T>1$, where 
 \begin{equation}
 \label{eqn:KBformula}
     K_{\mathcal{B}_-} = \frac{1-a_{\CB}^2+R_{\CB}^2}{2R_{\CB}},
 \end{equation}
 with $(a_{\CB},\,R_{\CB})$, two parameters that classify the boundary circle with constant extrinsic curvature.

 For the first term in the dilaton solution
 \begin{equation}
     \Phi(r,\theta) = C_0 \frac{1+r^2}{1-r^2}\,,
 \end{equation}
 its normal derivative is 
 \begin{equation}
 \label{eqn:normalderivativeC0}
     \partial_n\Phi = -C_0 \frac{a_\CB^2-R_\CB^2-r^2}{(1-r^2)R_{\CB}}\,.
 \end{equation}
 Plugging the formula of $K_{\CB_-}$ in terms of $a_{\CB}$ and $R_{\CB}$  from \eqref{eqn:KBformula} into the expression, we have
 \begin{equation}
     \Phi K_{\mathcal{B}_-} = C_0\frac{(1+r^2)(1-a_\CB^2+R_\CB^2)}{(1-r^2)2R_\CB}\,,
 \end{equation}
 where the expression of $K_{\mathcal{B}_-}$ differs from equation \eqref{eqn:KaRformula} by a sign because of the flip of the normal vector's direction.

 Then we consider the terms that depend on $\theta$ coordinate of the Poincar\'e disk
 \begin{equation}
     \Phi(r,\theta) = \frac{2r}{1-r^2}(C_1\sin{\theta}+C_2\cos{\theta})\,.
 \end{equation}
 The normal derivatives of these terms are
 \begin{equation}
     \partial_n\Phi = \frac{1-a_\CB^2+R_\CB^2}{2R_\CB}C_1\frac{2r}{1-r^2}\sin{\theta}+\frac{1-a_\CB^2+R_\CB^2}{2R_\CB}C_2\frac{2r}{1-r^2}\cos{\theta}-\frac{a_\CB C_2}{R_\CB}.
 \end{equation}
The first two terms are exactly $K_{\mathcal{B}_-}\Phi$. Summing these contributions together, the overall EOM of the spacetime boundary $\CB_-$ is then
\begin{equation}
    K_{\mathcal{B}_-}=T\,,\quad\mu = -C_0(\frac{1}{R_\CB}-K_{\mathcal{B}_-})-\frac{a_\CB C_2}{R_\CB}\,.
\end{equation}
Note that this EOM does not depend on the coefficient $C_1$, which is sensitive to the difference between $\phi_l$ and $\phi_r$, and when the boundary circle $\CB$, where the spacetime boundary lives on, is at the origin, i.e. $a_\CB=0$, the $C_2$-dependent term also vanishes. If we impose the fixed-mass $\mu$ and fixed-extrinsic-curvature $K_{\mathcal{B}_-}=T$ boundary condition on the spacetime boundary $\mathcal{B}_-$, the spacetime boundary will be completely fixed without any free parameters. We can solve the parameter $a_{\mathcal{B}}$ and $R_{\mathcal{B}}$ in terms of $T$ and $\mu$ 
\begin{equation*}
    a_{\mathcal{B}} = \pm\sqrt{1-2T R_\CB+R_\CB^2}\,,\quad R_{\mathcal{B}} = \frac{C_0(-\mu+T C_0)+C_2^2T\pm\sqrt{C_2^2(-\mu+(C_2^2-C_0^2)(T^2-1))}}{(-\mu+TC_0)^2-C_2^2} \,.
\end{equation*}
Specializing this boundary problem to the symmetric subsector \eqref{eq:JT-symmetric-C012} with an isotropic dilaton field, gives an actual simplification of the geometry. Since $C_2=0$, the $\CB_-$'s equation of motion reduces to $\mu=-C_0(\frac1{R_{\mathcal{B}}}-T)$, so the boundary circle radius is fixed by
 \begin{equation}
 \label{eq:JT-RA-symmetric}
     R_{\mathcal{B}}=\frac{C_0}{T C_0-\mu}
     =\frac{\phi_0}{T\phi_0-X^0_0\mu}\,.
 \end{equation}
 In particular, if we further require the boundary circle $\CB$ to sit at the center of the Poincar\'e disk, i.e. $a_{\mathcal{B}}=0$, we single out one value 
 \begin{equation}
     C_0=\frac{-\mu}{\sqrt{T^2-1}}\,,
 \end{equation}
 which determines the dilaton solution globally
 \begin{equation}
 \label{eqn:Symmetric-dilaton}
     \Phi(r,\theta)= \frac{-\mu}{\sqrt{T^2-1}}\frac{1+r^2}{1-r^2}\,.
 \end{equation}
 The positivity of the dilaton field requires $\mu<0$. We regard this configuration as the ``maximally-symmetric AdS$_2$ bubble''. In the remainder of this section, we focus on evaluating the HH wave function in the maximally symmetric AdS$_2$ bubble.

\subsection{HH wave function in maximally-symmetric \texorpdfstring{AdS$_2$}{AdS2} bubble}

The maximally-symmetric bubble in AdS$_2$ is a disk at the center of the Poincar\'e disk with a radius of $R_{\mathcal{B}}=T-\sqrt{T^2-1}$, where the radius $R_{\CB}$ is measured from the origin to the boundary circle $\CB$ by the flat metric on $\mathbb{R}^2$. 
The dilaton value is constant on the boundary circle $\CB$ of the bubble
\begin{equation}
    \Phi_{\mathcal{B}} = \frac{- T\mu}{T^2-1}.
\end{equation}
Since the spatial slice $\Sigma$ intersects the boundary circle $\CB$, for this symmetric bubble, the endpoint values of the dilaton field $\phi_l,\phi_r$ must be identical to $\Phi_{\mathcal{B}}$. By applying \eqref{eq:JT-Crk-def} to \eqref{eq:JT-symmetric-bc}, we can solve for the middle-point dilaton value $\phi_0$ in terms of the parameter $r_0$. In the meantime, as we fix the global dilaton solution \eqref{eqn:Symmetric-dilaton}, we have the other relation between $\phi_0$ and $r_0$, which allows us to obtain an equation for $r_0$
\begin{equation}
    \label{eqn:r0equation}
    \frac{-\mu}{\sqrt{T^2-1}}\frac{1+r_0^2}{1-r_0^2} = \frac{-T\mu}{C(r_0,K)(T^2-1)}\,.
\end{equation}
In what follows, we first evaluate the HH wave function for a geodesic spatial slice, where the result is particularly simple and helps build intuition, before turning to the more general family of spatial slices with constant $K$.

\subsubsection{Geodesic spatial slice}
When the spatial slice $\Sigma$ is a geodesic, the equation of $r_0$ is significantly simplified as follows
\begin{equation}
\label{eqn:geodesicr0equation}
    \frac{r_{\text{Sch},0}}{r_s}=\frac{1+r_0^2}{1-r_0^2} = \frac{T}{\sqrt{T^2-1}\cosh{\frac{L}{2}}}\,.
\end{equation}
Given that $\frac{1+r_0^2}{1-r_0^2}\ge 1$, we have 
\begin{align}
\label{eqn:lcrit0}
    L\leq L_{\text{crit}}^0 = 2\arctanh{\frac{1}{T}}
\end{align}
in order to have a real saddle. The equation of the geodesic curve labeled by length $L$ is then
\begin{equation}
    \frac{r_{\text{Sch}}(\lambda)}{r_s} = \frac{T}{\sqrt{T^2-1}}\frac{\cosh{\lambda}}{\cosh{\frac{L}{2}}}\,,
\end{equation}
with the dilaton profile
\begin{equation}
\label{eqn:geodesicdilaton}
    \phi(\lambda) = \frac{-\mu T}{T^2-1}\frac{\cosh{\lambda}}{\cosh{\frac{L}{2}}}\,.
\end{equation}
For $\phi_l=\phi(\frac{L}{2})$ and $\phi_r=\phi(-\frac{L}{2})$, we have $\phi_l=\phi_r=\Phi_{\mathcal{B}}$ as expected. Since the extrinsic curvature vanishes on the geodesic spatial slice, the boundary action on the spatial slice also vanishes. The on-shell action contributing to the HH wave function consists of only the corner term and the action on the spacetime boundary $\CB_-$. 

\begin{figure}[ht]
    \centering   \includegraphics[width=11cm]{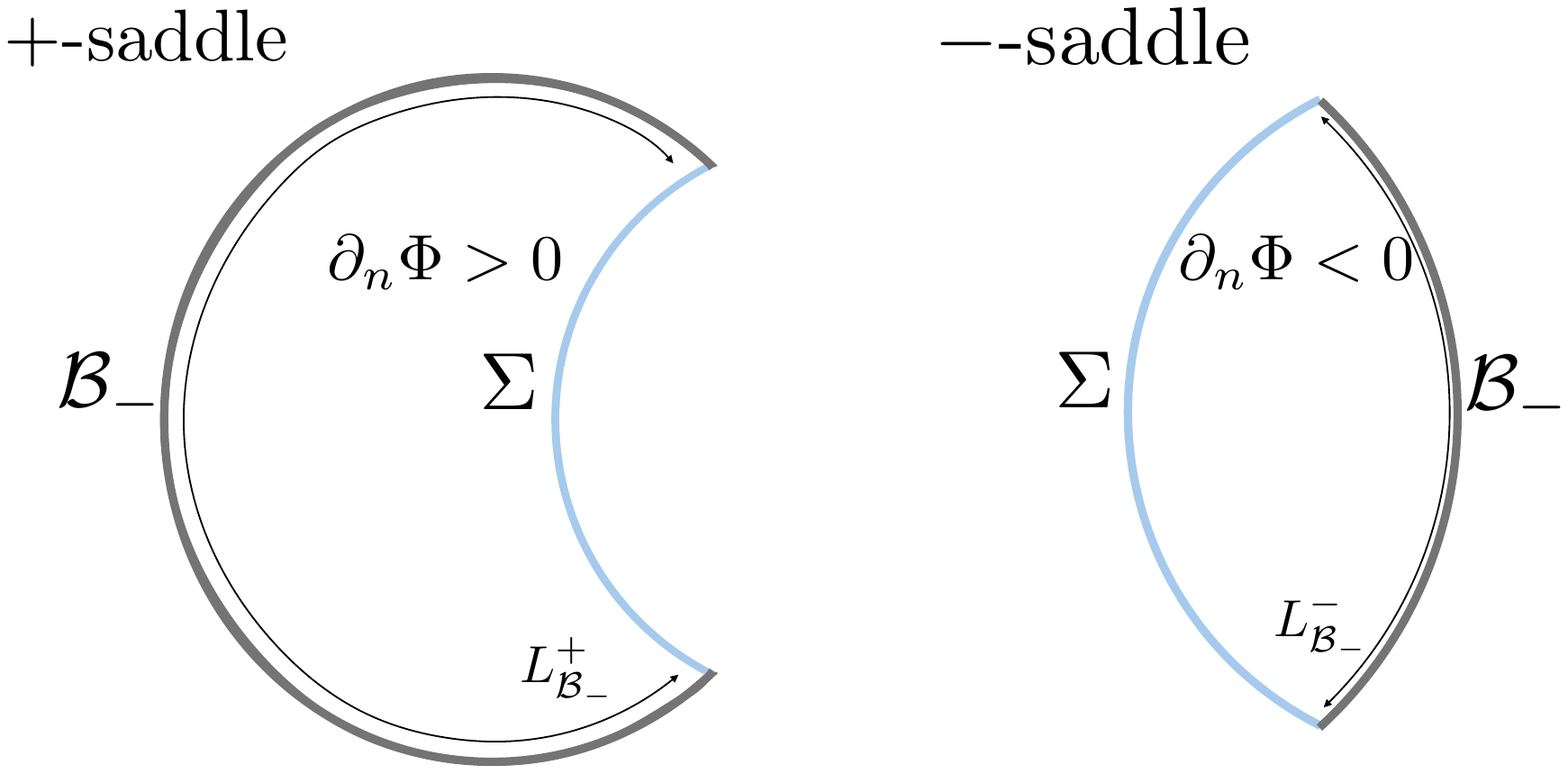}
    \caption{The graph shows that the geodesic spatial slice $\Sigma$ (Light blue curve) splits the geometry into two parts labeled as $+$-saddle and $-$-saddle. The spacetime boundary $\mathcal{B}_-$ (Light gray curve) is represented by a major arc in the $+$-saddle and by a minor arc in the $-$-saddle. The other reason for the $\pm$-notation of saddles comes from the sign of $\partial_n\Phi$ indicated in the graph. As $K$ is canonically conjugate to $\Phi$, $\partial_n \Phi$ is conjugate to the induced metric $\gamma$ on $\Sigma$. Although the geodesic spatial slice has $K_{\Sigma}=0$, we can distinguish the two saddles by $\partial_n\Phi$ that is nonzero.}  \label{fig:geodesiccauchy}
\end{figure}

The spatial slice $\Sigma$ cuts the AdS$_2$ bubble into two parts, both of which are extrema of the JT action \eqref{eqn:JTaction}. Note that the directions of the outward normal vector to the spatial slice are opposite on each side, which in principle reverses the sign of $K$. 
We refer to the saddle with positive $K$ on $\Sigma$ as ``$+$''-saddle and the one with negative $K$ as ``$-$''-saddle. For $K=0$, we use ``$+$''-saddle to label the geometry beyond the half of the bubble, and use ``$-$''-saddle for the other part as demonstrated in figure \ref{fig:geodesiccauchy}.
Since the ``$+$''-saddle and ``$-$''-saddle are complements to each other with respect to the whole symmetric bubble, the on-shell actions of both saddles sum to be the action on the bubble boundary $\mathcal{B}$ 
\begin{equation}
    I_+^{\text{on-shell}}+I_-^{\text{on-shell}}=I_{\mathcal{B}} = \frac{2\pi\mu}{\sqrt{T^2-1}}\,,
\end{equation} which is a constant fixed by parameters $T$ and $\mu$. Note the $\pm$ signs used here to distinguish saddles should not be confused with the subscript ``$-$'' used in $\CB_{-}$, which indicates $\CB_-$ is associated with the ket HH state.

The boundary action on $\mathcal{B}_-$ is proportional to the proper length, which depends on the angular coordinates $\theta$ of the endpoints of the spatial slice in the Poincar\'e coordinate. 
Provided the solution of the endpoint $\theta_*$ in \eqref{eq:JT-theta-star}, we have
\begin{equation}
\label{eqn:thetastarE}
    \theta_* = \arccos{\sqrt{T^2-(T^2-1)\cosh^2{\frac{L}{2}}}}\,.
\end{equation}
The proper circumference of the circle $\CB$ is $\frac{2\pi}{\sqrt{T^2-1}}$, so the proper length of arc $L_{\mathcal{B}_-}^{\pm}$ is 
\begin{equation}
    L_{\mathcal{B}_-}^- = \frac{2}{\sqrt{T^2-1}}\arccos{\sqrt{T^2-(T^2-1)\cosh^2{\frac{L}{2}}}} = \frac{2\pi}{\sqrt{T^2-1}}-L_{\mathcal{B}_-}^{+}\,,
\end{equation}
where $L_{\mathcal{B}_-}^\pm$ are depicted in figure \ref{fig:geodesiccauchy}.
We then have
\begin{equation}
    I_{\mathcal{B}_-}=\mu L_{\mathcal{B}_-}\,.
\end{equation}
The corner term can be computed by from the angle between normal vectors $n_{\mathcal{B}_-}$ and $n_\Sigma$ as 
\begin{equation}
\label{eqn:theta12}
    \Theta_{1,2}^-= \pi-\Theta_{1,2}^+ = \arccos{\frac{\sqrt{T^2-(T^2-1)\cosh^2{\frac{L}{2}}}}{\cosh{\frac{L}{2}}}}\,.
\end{equation}
The on-shell actions of ``$\pm$''-saddles are 
\begin{equation}
    -I_{\rm E}(L)=\frac{S_0}{2}+\begin{cases}
        -\frac{2\mu}{\sqrt{T^2-1}}\left(\pi-\theta_*\right)+\frac{\mu T}{T^2-1}\left(\pi-\Theta_1^--\Theta_2^-\right)& +\,\text{-saddle}\\
        -\frac{2\mu}{\sqrt{T^2-1}}\theta_*-\frac{\mu T}{T^2-1}\left(\pi-\Theta_1^--\Theta_2^-\right)&-\,\text{-saddle}
    \end{cases}
\end{equation}
Monotonicity of the two actions are different. ``$+$''-saddle leads to a wave function peaking at $L=L_{\text{crit}}^0=2\arctanh{\frac{1}{T}}$, where the on-shell geometry is exactly half of AdS$_2$ bubble. 
\begin{figure}[ht]
    \centering
\includegraphics[width=7cm]{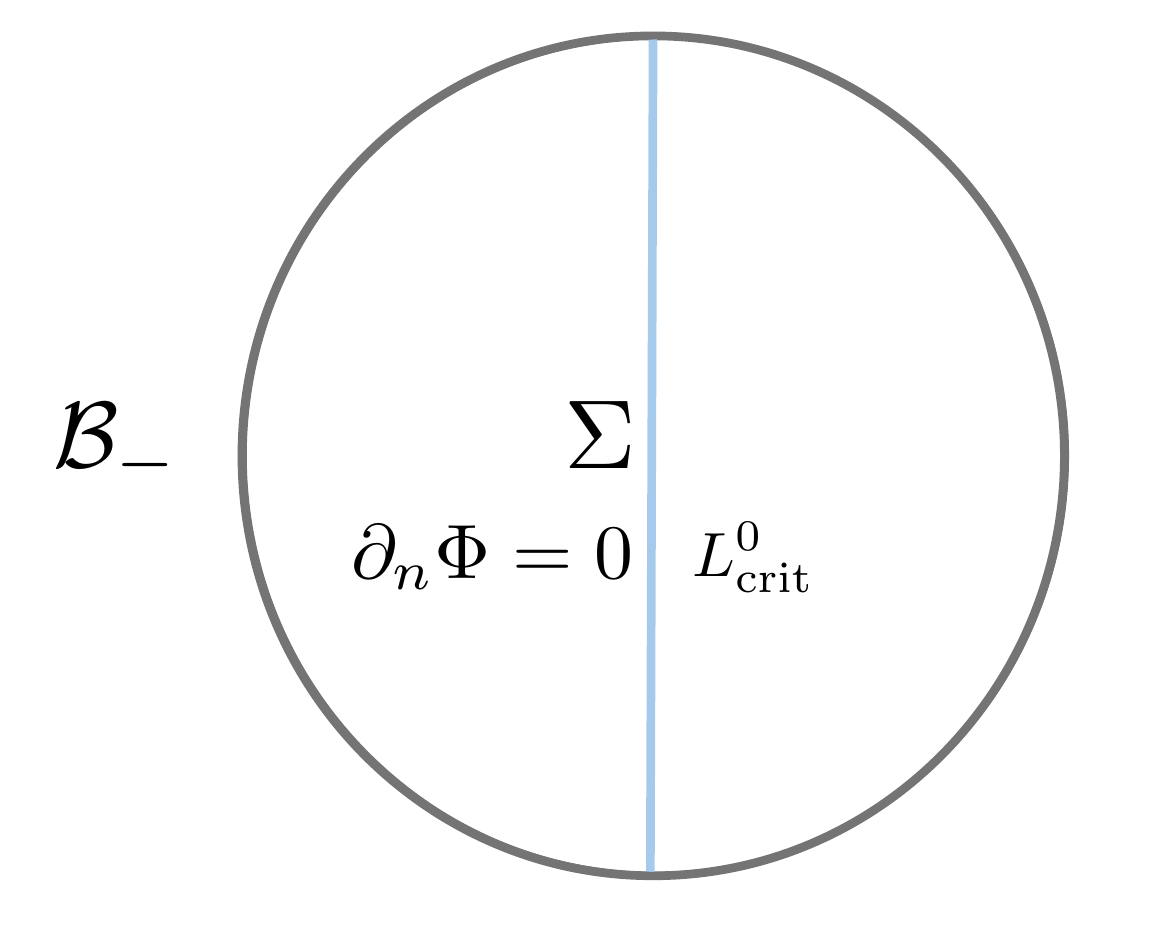}
    \caption{The HH wave function of geodesic spatial slice evaluated on the ``$+$''-saddle peaks at $L=L_{\text{crit}}^0$  with the normal derivative $\partial_n \Phi$ that vanishes everywhere on $\Sigma$.}
   \label{fig:geocritical}
\end{figure}
Instead, the action of the ``$-$''-saddle is monotonically decreasing with a maximum at $L=0$.

\paragraph{Which saddle should we pick up between them?}~\par 
As also commented in the section \ref{sec:AdS3}, the semiclassical calculation of HH wave function in dS gravity \cite{HH83} and in AdS$_3$ with fully-gravitating spacetime boundary, requires a choice of saddle. The choice of saddle depends on the prescription of the steepest-descent contour, which imposes the positivity requirement on the quantities that are canonically conjugate to the minisuperspace parameters. In dS gravity, the relevant quantity is the extrinsic curvature $K_{\Sigma}$ on the spatial slice, which is conjugate to the radius of the spatial slice sphere. In the AdS$_3$ discussion from section \ref{sec:AdS3}, we additionally require that  $K_{C}$, the extrinsic curvature of the intersection circle $C$ between the spatial slice and the spacetime boundary, to be positive, which are conjugate to the minisuperspace parameter $L$, the length of $C$. In the present JT calculation, besides the extrinsic curvature $K_\Sigma$, which is conjugate to the dilaton profile in this case, the contour choice for $\partial_n \Phi$, as the canonical conjugate of the induced metric, should also be considered. For geodesic spatial slice, the extrinsic curvature $K_\Sigma$ always equals zero, while the values of  $\partial_n \Phi$ are different between two saddles. If we adopt the same prescription used in \cite{HH83} and section \ref{sec:AdS3} and impose the positivity requirement on $\partial_n \Phi$, the ``$+$''-saddle can be singled out. Even though the ``$-$''-saddle has smaller Euclidean action, the steepest-descent contour prescription indicates that the HH wave function should be approximated by ``$+$''-saddle instead. Indeed, the semiclassical wave function evaluated at the ``$+$''-saddle mimics the monotonic growth of the HH wave function in dS gravity and in AdS$_3$ (figure \ref{fig:AdS3action}) with respect to the size of the spatial slice and peaks at the critical value $L_{\rm crit}$. 

\paragraph{A comparison to the AdS$_3$ case}~\par  
One might doubt why we expect that the semiclassical HH wave function evaluated in this section to behave qualitatively the same as the one calculated in section \ref{sec:AdS3}. If one stares at the chosen ``+''-saddle in this section, they will find it is different from the $s_{\rm small}$ saddle adopted in the previous section in the following way. ``$+$''-saddle covers more than half of the 1D boundary circle $\CB$, while $s_{\rm small}$ covers less than half of the 2D boundary sphere $\CB$. However, as we showed that the choice of these two saddles follows the same contour argument, there is no apparent contradiction. If we consider the JT calculation from a dimension-reduction point of view, the saddle point geometry we construct from a 2D disk can be lifted to part of the 3D solid torus $D_{2}\times S^1$. The spatial slice $\Sigma$ will be lifted to an annulus stretching between the boundary torus. The parameters $K_{\Sigma}$ and $\partial_n \Phi$ in JT language then becomes part of the extrinsic curvature of the spatial slice annulus embedded into the 3D solid torus, which is guaranteed to be positive for ``$+$''-saddle. Given that the 3D solid torus is topologically distinctive from the 3D hyperbolic balls considered in the previous section, there is no direct analogy we can draw between these two cases, except that we use the same contour prescription for saddle selection. Another question we can ask is if there is a counterpart of $K_C$ in JT calculation. We still consider the 3D lifting of the ``+''-saddle. From 3D perspective, the intersection between $\Sigma$ and $\CB_-$ is a union of two circles (boundary of the spatial slice annulus) on the boundary torus that wrap around the $S^1$ direction. Provided that the boundary torus has a flat induced metric on-shell, the corresponding $K_C$ for these two circles are zero. Therefore, this quantity does not play a role in saddle selection.

\subsubsection{Spatial slice with generic \texorpdfstring{$K$}{K}}
\begin{figure}[t]
    \centering
    \includegraphics[width=8cm]{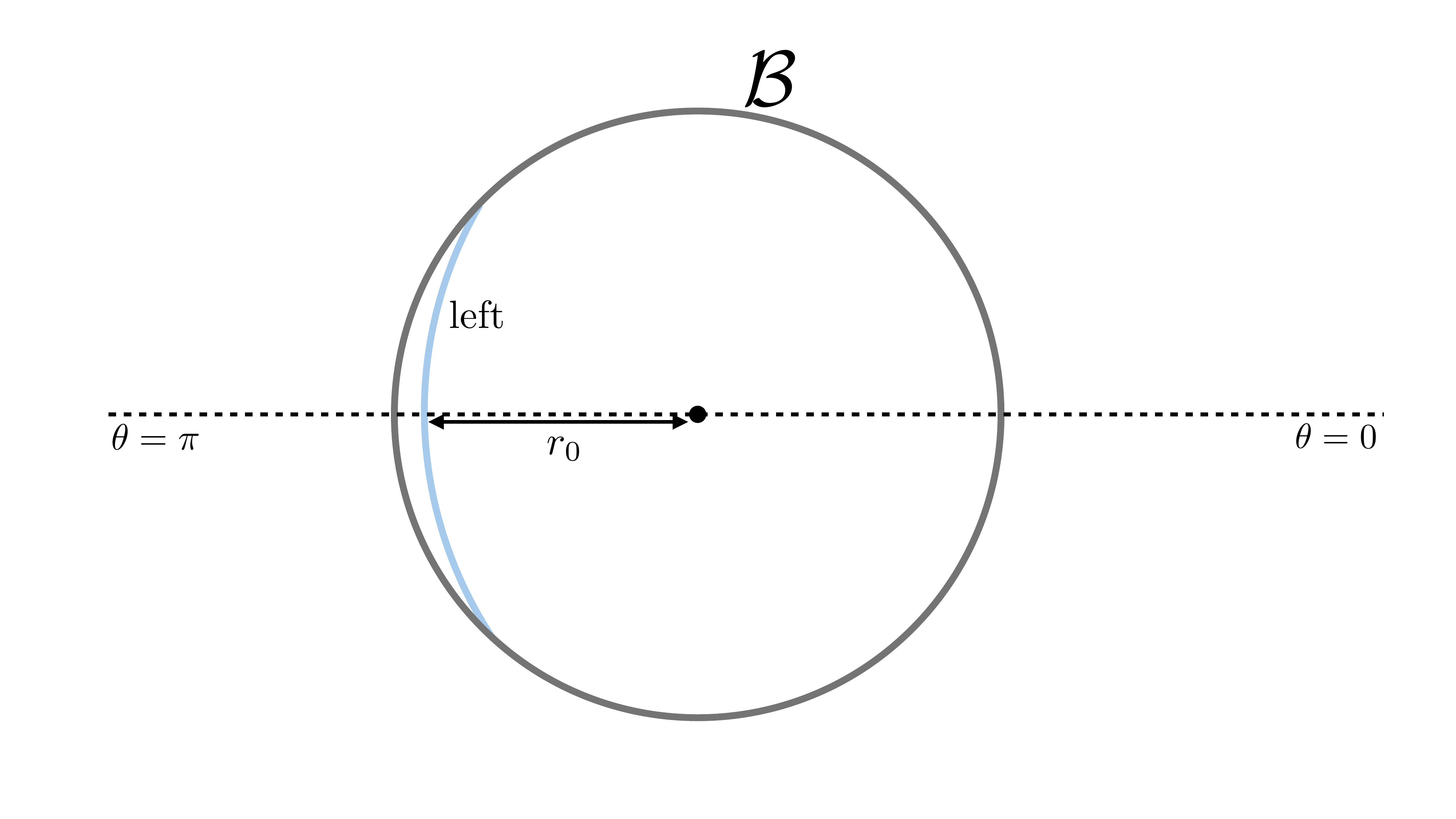}
    \includegraphics[width=8cm]{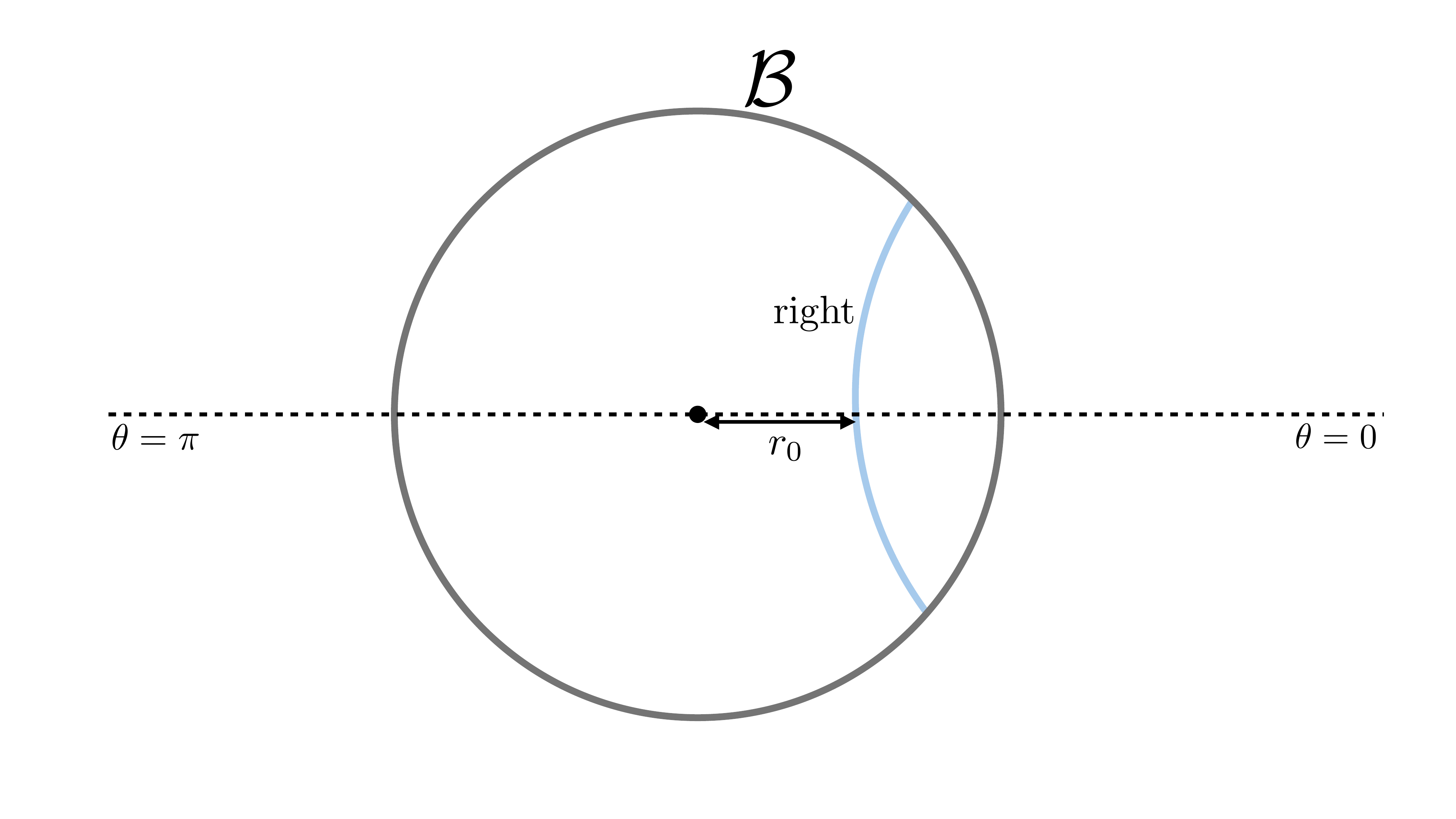}
    \caption{The ``$\mp$'' sign in the $C(r_0,K)$ factor of the equation \eqref{eqn:r0equation} of $r_0$ is sensitive to the relative position of the spatial slices (light blue) to the origin where the boundary circle $\CB$ is centered. For the spatial slice that intersects the $\theta= \pi$ axis on the left side of the origin, we choose the ``$+$'' sign in $C(r_0,K)$, and for the spatial slice that intersects the $\theta= 0$ axis on the right side of the origin, we choose the ``$-$'' sign. The ``left'' and ``right'' notions will be used in the following discussion of this section for the same purpose.}
    \label{fig:JTHHleftright}
\end{figure}

For generic $K$, we can also solve the parameter $r_0$ via equation \eqref{eqn:r0equation} to fix the position of the spatial slice $\Sigma$ in the AdS$_2$ bubble surrounded by $\CB$. However, there are some subtleties. One subtlety comes from the ``$\mp$'' sign in $C(r_0,K)$ \eqref{eq:JT-Crk-def}, which distinguishes the spatial slices intersecting the $\theta=0$ axis on the right side ($-$) of the origin and those intersecting the $\theta = \pi$ axis on the left side ($+$) of the origin (see figure \ref{fig:JTHHleftright}).
Another subtlety is from the quadratic nature of equation \eqref{eqn:r0equation}, which in principle has multiple roots, and we need to check if these roots are valid solutions of $r_0$. The most general solution of $r_0$ is given as follows
\begin{equation}
\label{eqn:r0solutionK}
    r_0 = \mp\frac{K(T^2-1)(x-1)\pm\sqrt{(1-K^2)(T^2-1)(x^2-K^2-T^2(x^2-1))}}{-(T^2-1)x-(1-K^2)T\sqrt{T^2-1}+K^2(T^2-1)}\,,
\end{equation}
where $x= \cosh{\left(\sqrt{1-K^2}\frac{L}{2}\right)}$. The $\mp$ sign is the same $\mp$ sign that appears in the expression of $C(r_0,K)$ \eqref{eq:JT-Crk-def} and the $\pm$ is from the double roots of the quadratic equation. To justify the validity of the roots, we recall that the $r_0$, defined by the absolute value $|a-R|$ (figure \ref{fig:PoincarediskConstantK}) in terms of two parameters $a$ and $R$ classifying the constant-extrinsic-curvature curves, must be a positive number, and for the two scenarios that the curves intersect the $\theta=0$ or $\theta =\pi$ axis on different sides of the origin, we require one of the following inequalities to be satisfied
\begin{equation}
\label{eqn:r0Kinequ}
    \begin{cases}
        r_0>K & \text{(right-side)}\\
        r_0<-K & \text{(left-side)}\,,
    \end{cases}
\end{equation}
which originates from the constraint that the radius $R_\Sigma$ of the circle (constant-$K$ circle in figure \ref{fig:PoincarediskConstantK})
\begin{equation}
    R_\Sigma = \begin{cases}
        \frac{1-r_0^2}{2(r_0-K)} & \theta = 0\,(\text{right-side})\\
        -\frac{1-r_0^2}{2(r_0+K)} & \theta=\pi\,(\text{left-side})
    \end{cases} \quad (0\leq r_0<1)
\end{equation}
as a function of $r_0$ and $K$, where the spatial slice $\Sigma$ lives, must also be a positive number. 

\begin{figure}[ht]
    \centering
    \includegraphics[width=6cm]{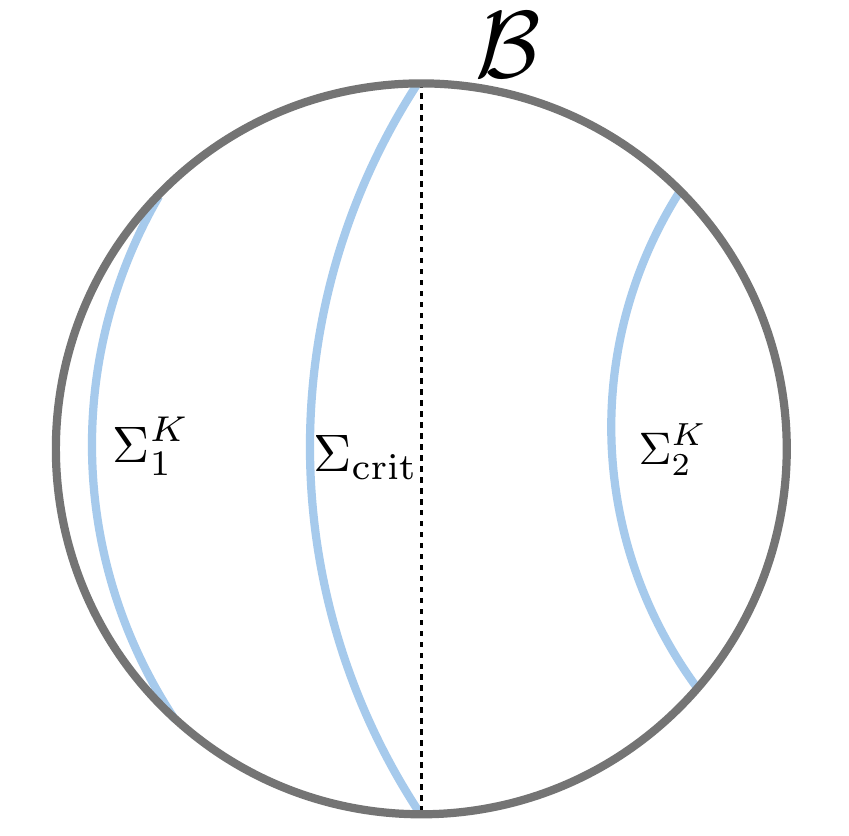}
    \caption{For generic $K$, there are two spatial slices depicted in light blue labeled by $\Sigma_1^K$ and $\Sigma_2^K$ with the same proper length $L$ and the same absolute value of the extrinsic curvature $|K|$. The two slices merge together into $\Sigma_{\text{crit}}$ at $L=L_{\text{crit}}^{|K|}$. The spatial slice $\Sigma_{\text{crit}}$ at criticality stretches between two antipodal points on $\CB$, and the corresponding spacetime boundary $\mathcal{B}_-$ is exactly the half circle.}
    \label{fig:generic_K}
\end{figure}

For fixed $0<|K|<1$, the real saddle exists for $L\leq L_{\text{crit}}^{|K|} = \frac{2}{\sqrt{1-K^2}}\arctanh{\frac{\sqrt{1-K^2}}{\sqrt{T^2-K^2}}}$, which reduces to $L_{\text{crit}}^0$ in \eqref{eqn:lcrit0} at $K=0$. 
For the proper length $L$ below the critical value, there are two positive roots and two negative roots in \eqref{eqn:r0solutionK}. Since these two positive roots will be either less than or greater than $|K|$, they will always satisfy one of the above inequalities \eqref{eqn:r0Kinequ}. Different from the $K=0$ case, we obtain two spatial slices inside the bubble with the same proper length $L$ for fixed nonzero $|K|$. At the critical proper length $L_{\text{crit}}^{|K|}$, the two slices merge into one as shown in figure \ref{fig:generic_K}.

When we take the critical value of the proper length, the parameter $r_0$ is given as
\begin{equation}
\label{eqn:r0criticalK}
    r_0 = -\frac{(T-\sqrt{T^2-1})(T-\sqrt{T^2-K^2})}{K}\,.
\end{equation}
The numerator of the expression \eqref{eqn:r0criticalK} is positive and real given that $|K|<1<T$. In order to get a positive $r_0 =|a-R|$, we have to take $K<0$. In this case, the intersection point lies on the $\theta =\pi$ axis on the left side of the origin provided that $r_0<-K=|K|$. The corresponding dilaton profile on this critical slice is 
\begin{equation}
    \phi(\lambda)^{|K|} = \frac{\mu\left(K^2\sqrt{T^2-1}-\sqrt{T^2-K^2}\cosh{(\sqrt{1-K^2}\lambda)}\right)}{(1-K^2)T\sqrt{T^2-1}}
\end{equation}
It is straightforward to evaluate the wave function at the critical proper length. In this case, the action on the spatial slice does not vanish. Integrating the dilaton profile over the whole spatial slice, we obtain the following action
\begin{equation}
    I_{\Sigma}^-(L_{\text{crit}}^{|K|}) =|K|\int_{-L_{\text{crit}}^{|K|}/2}^{L_{\text{crit}}^{|K|}/2}\,\phi(\lambda)\,d\lambda = -\frac{2\mu|K|}{T} \left(\frac{\sqrt{T^2-K^2}}{(1-K^2)(T^2-1)}-\frac{K^2\arctanh{\left(\frac{\sqrt{1-K^2}}{\sqrt{T^2-K^2}}\right)}}{(1-K^2)^{3/2}}\right),
\end{equation}
where the ``$+$''-saddle action is related by $I_{\Sigma}^+=-I_{\Sigma}^-$. We can also solve the angle coordinate of the endpoints $\theta_* = \frac{\pi}{2}$, which leads to an identical action on the spacetime boundary to the geodesic case
\begin{equation}
    I_{\mathcal{B}_-}^-(L_{\text{crit}}^{|K|})= I_{\mathcal{B}_-}^+(L_{\text{crit}}^{|K|}) = \frac{\pi\mu}{\sqrt{T^2-1}}.
\end{equation}
Finally, the intersection angles $\Theta_{1,2}$ also admit a simple expression in terms of $K$ and $T$ as 
\begin{equation}
    \Theta_{1,2}^-(L_{\text{crit}}^{|K|}) = \arccos{\frac{|K|}{T}} = \pi-\Theta_{1,2}^+(L_{\text{crit}}^{|K|})\,.
\end{equation}
The on-shell actions of ``$\pm$''-saddles for generic $|K|$ at the critical proper length are 
\begin{equation}
    -I_{\rm E}(L_{\text{crit}}^{|K|}) = -I_{\rm E}(L_{\text{crit}}^{0})- I_{\Sigma}^{\pm}(L_{\text{crit}}^{|K|})-\frac{\mu T}{T^2-1}(\pi-\Theta_1^{\pm}(L_{\text{crit}}^{|K|})-\Theta_2^{\pm}(L_{\text{crit}}^{|K|}))\,.
\end{equation}
By varying the absolute value $|K|$, the on-shell action has extrema at $K=0$.

The evaluation of the wave function for generic $L$ does not lead to a concise expression. Therefore, we perform a numerical analysis. Before starting calculating the wave function, we want to comment on the contour prescription and saddle point choice. 

\paragraph{Which saddle point should we choose?}~\par
Given the two positive roots of $r_0$ and the corresponding ``$\pm$''-saddles for each root, for fixed $|K|$, there are four different configurations, which mimic the four saddles $s_{\rm small}$, $s_{\rm large}$, $s_{\rm small}'$, $s_{\rm large}'$ in the AdS$_3$ discussion. However, there is one significant difference. In JT calculations, the two $r_0$ roots lead to different dilaton profiles on the spatial slice $\Sigma$, while in AdS$_3$ discussions, all four saddles have the same boundary data on the spatial slice $\Sigma$. Therefore, in the following discussion, we should select saddles for the two roots of $r_0$ separately. The steepest-descent contour argument used in the geodesic slice calculation also applies to the generic $K$ case. It selects the saddles with positive $K_\Sigma$ and positive $\partial_n\Phi$. The sign of the normal derivative $\partial_n \Phi$ is sensitive to the relative position of the spatial slice to the origin, the center of the symmetric bubble.
Recall that in equation \eqref{eqn:normalderivativeC0} we compute the normal derivative of the isotropic dilaton field, which shows dependences on $(a_\Sigma,\,R_\Sigma)$. For the spatial slice $\Sigma$ that intersects the axis from the left side of the origin, the radius $R_\Sigma$ is greater than $a_\Sigma$, which implies that $\partial_n\Phi$ over the whole slice has the same sign as $K$. The contour prescription chooses the ``$+$''-saddle, i.e. $K>0$ and $\partial_n \Phi>0$ everywhere on $\Sigma$. For the spatial slice $\Sigma$ that intersects from the right side, the radius $R_\Sigma$ is smaller than $a_{\Sigma}$. There are two potential subtleties that obstruct a consistent contour prescription.

\begin{figure}[t]
    \centering
    \includegraphics[width=12cm]{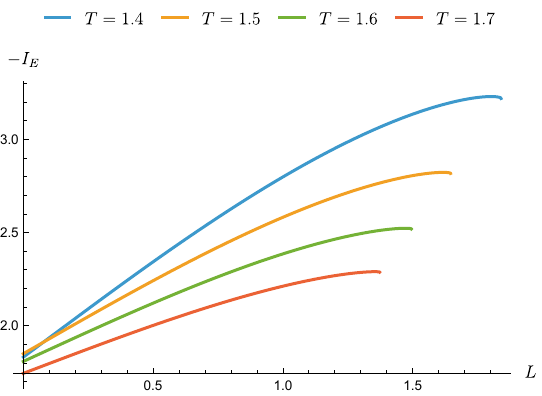}
    \caption{The "$+$"-saddle on-shell Euclidean action for $T=1.4,\,1.5,\,1.6,\,1.7$ is plotted as a function of the proper length $L$. In the above plot, we have subtracted the topological contribution $\frac{S_0}{2}$ off the action.}
    \label{fig:euclideanaction}
\end{figure}

\begin{itemize}
    \item Certain slice $\Sigma$ might have different signs of $\partial_n\Phi$ at different points. This happens when $r_0$ is close to $0$.
    \item The sign of $K_\Sigma$ is not guaranteed to be the same as $\partial_n \Phi$. 
\end{itemize}
To avoid these issues, we only consider the spatial slice that intersects from the left side of the origin in the following numerical calculation. 
We choose $T=\{1.4,\,1.5,\,1.6,\,1.7\},\,|K|=0.5,\,\mu = -1$ and plot the on-shell action calculated at the ``$+$''-saddle from $L=0$ to $L=L_{\text{crit}}^{0.5}$ in figure \ref{fig:euclideanaction}.
Although we always start from $L=0$ in the plot, the $L\rightarrow0$ limit of the spatial slice might not necessarily lie on the left side of the origin, since we have the constraint $r_0<-K$. When $T < \frac{1+K^2}{2|K|}$, the value of $r_0(L=0)$ will be greater than $-K$. It indicates that the spatial slice will instead intersect the axis from the right side of the origin and potentially suffer from the two subtleties mentioned above. 
Moreover, the action at $L=0$ is not monotonically decreasing with respect to $T$. We can actually explicitly write down the action at $L=0$
\begin{equation}
    -I_{\rm E}(L=0) = -\mu\left(\frac{2\pi}{\sqrt{T^2-1}}-\frac{\pi T}{T^2-1}\right)\,,
\end{equation}
which takes the maximum value at $T =\sqrt{1+\frac{2}{3}\sqrt{3}} \approx 1.468 $. The semiclassical HH wave function for nonzero $K$ has a different behavior near $L_{\text{crit}}^{|K|}$. Instead of peaking at the critical value of $L$ as the geodesic wave function does, the wave function for generic $K$ peaks at a specific $T$-dependent value of $L$ smaller than $L_{\text{crit}}^{|K|}$.

\subsubsection{Complex saddle}
From the equation \eqref{eqn:geodesicr0equation} of $r_0$, we find that in order for $r_0$ to be real, the proper length of the spatial slice must be smaller than a critical value $L_{\rm crit}^0$. One may then ask whether an on-shell geometry still exists when the proper length exceeds this bound. In this case, we can glue the Euclidean disk geometry to a Lorentzian AdS$_2$ strip to obtain an on-shell configuration, and the resulting on-shell action becomes complex. 

The motivation for this complex saddle comes from the observation that $L>L_{\text{crit}}$ renders the expressions inside the square roots in Eqs.~\eqref{eqn:thetastarE} and \eqref{eqn:theta12} negative, leading to an analytic continuation of the on-shell action. 

To understand how to glue the Euclidean geometry to the Lorentzian strip, we consider the global Lorentzian AdS$_2$ coordinates.
\begin{equation}
    \label{eqn:Lorentzianmetric}
    ds^2 = -(1+\rho^2)dt^2+\frac{d\rho^2}{(1+\rho^2)}\,,
\end{equation}
where we already set the AdS radius to be $1$ and we use $\rho$ as our spatial coordinate to distinguish with $r$ coordinate in the Euclidean geometry. The coordinate $\rho$ extends in both direction to the infinity, which represents the two asymptotic boundary in the global AdS$_2$. This coordinate can be transformed into the embedding coordinate as
\begin{equation}
    \label{eqn:Lorentzianembedding}
    X^0 = \sqrt{1+\rho^2}\cos{t}\,,\quad X^1 = \sqrt{1+\rho^2}\sin{t}\,,\quad X^2 =\rho\,. 
\end{equation}
The global Lorentzian AdS$_2$ is a hyperbola inside $(1+2)$-dimensional Minkowski spacetime with $-(X^0)^2-(X^1)^2+(X^2)^2=-1$. When we analytically continue our symmetric AdS$_2$ bubble to the Lorentzian spacetime, the spacetime boundary is no longer a single circle, but two timelike curves. Provided that the spacetime boundary of the bubble in the Euclidean geometry is a curve with the constant $X^0= \frac{T}{\sqrt{T^2-1}}$, we can find the trajectory of the spacetime boundary in the Lorentzian geometry by solving
\begin{equation}
    X^0 = \sqrt{1+\rho^2}\cos{t}=\frac{T}{\sqrt{T^2-1}}\,,
\end{equation}
which gives the following parametrized curve
\begin{equation}
    \rho(t) = \pm \frac{\sqrt{T^2\sin^2{t}+\cos^2{t}}}{\sqrt{T^2-1}\cos{t}}\,.
\end{equation}
When the global time $t = \pm \frac{\pi}{2}$, the spacetime boundary reaches the asymptotic boundary of the global AdS$_2$. The solution of the dilaton field can also be directly extended to the Lorentzian spacetime as 
\begin{equation}
    \Phi(\rho,t) = \frac{-\mu}{\sqrt{T^2-1}}\sqrt{1+\rho^2}\cos{t}\,.
\end{equation}

\begin{figure}[t]
    \centering
    \includegraphics[width=9cm]{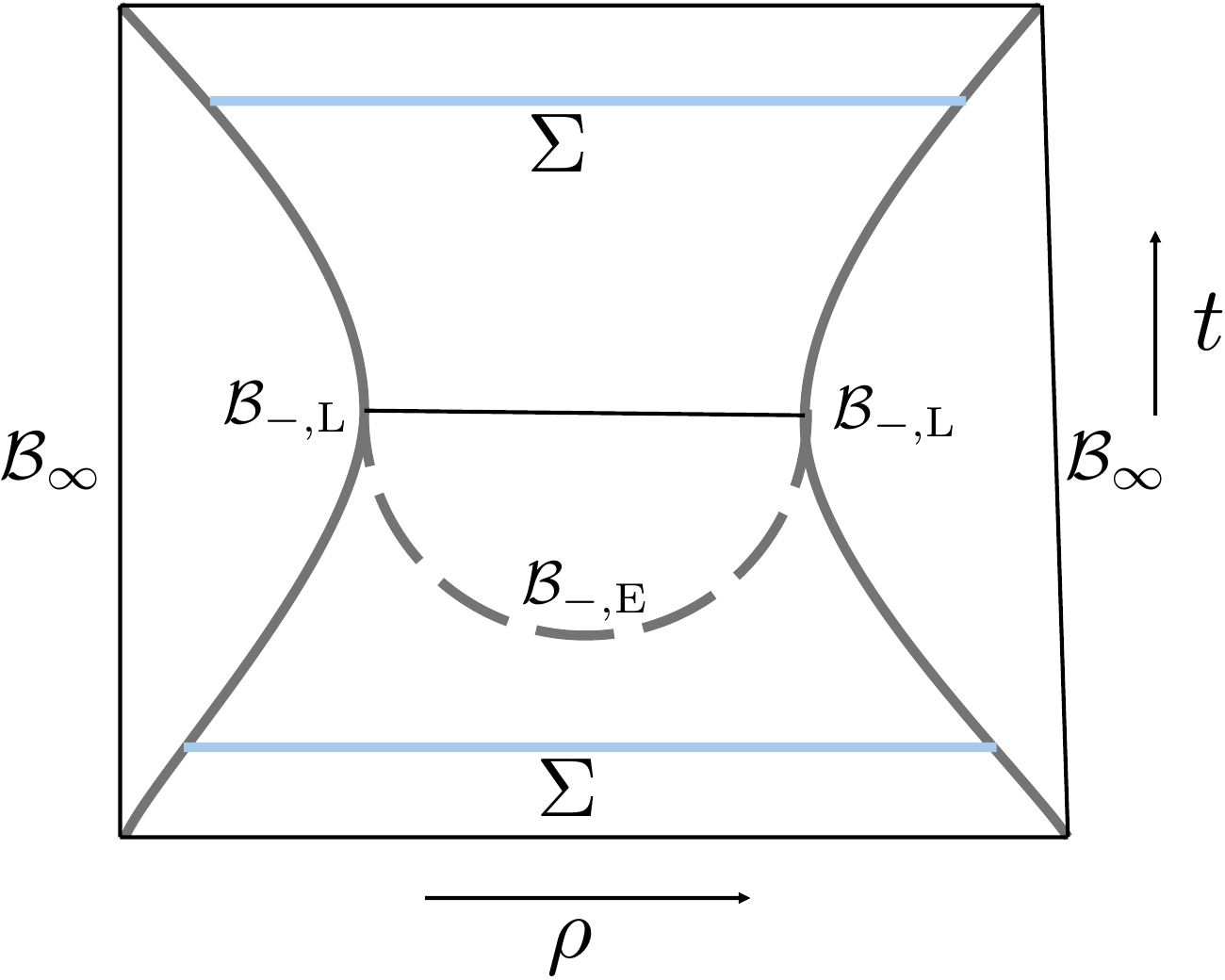}
    \caption{The picture shows the two branches of the complex saddle for the evaluation of HH wave function on geodesic spatial slice with $L>L_{\text{crit}}^0$. $\mathcal{B}_\infty$ represents the two asymptotic boundaries of global Lorentzian AdS$_2$. The dashed line shows the Euclidean part of the spacetime boundary $\CB_{-,\rm E}$ and the solid line for the Lorentzian part $\CB_{-,\rm L}$.}
    \label{fig:geodesiccomplex}
\end{figure}

For geodesic spatial slice, the dilaton profile at the critical length is 
\begin{equation}
    \phi_{\text{crit}}^0(\lambda) = \frac{-\mu}{\sqrt{T^2-1}}\cosh{\lambda}\,.
\end{equation}
In the meantime, the dilaton profile on the $t=0$ slice is $\Phi(\rho,0)=\frac{-\mu}{\sqrt{T^2-1}}\sqrt{1+\rho^2}$. Note that the constant-$t$ slice in Lorentzian AdS$_2$ is a spacelike geodesic with $\rho(\lambda) = \sinh{\lambda}$. This exactly matches the dilaton profile on the $t=0$ slice with $\phi^0_{\text{crit}}$, which suggests the gluing of the Euclidean half-disk to the Lorentzian AdS$_2$ along the $t=0$ slice. Furthermore, the geodesic slice with $L>L_{\text{crit}}^0$ can be associated to the constant-$t$ curve with
\begin{equation}
    t(L) = \pm\arccos{\frac{T}{\sqrt{T^2-1}\cosh{\frac{L}{2}}}}\,,
\end{equation}
which has the expected dilaton profile \eqref{eqn:geodesicdilaton} on it. The $\pm$ signs indicate that there exist two branches of solutions. Figure \ref{fig:geodesiccomplex} provides a demonstration of the complex geometry made of the gluing between the Euclidean half disk and the Lorentzian strip.
The corresponding on-shell actions of these two branches are
\begin{equation}
\label{eqn:JTcomplexaction}
    I(L^0) =I_{\rm E}(L_{\text{crit}}^0)+ i\frac{2\mu}{\sqrt{T^2-1}}\arcsinh{\left(T\tan{(t(L))}\right)}- i \frac{2\mu T}{T^2-1}\arcsinh{\left(\sqrt{T^2-1}\sin{(t(L))}\right)}\,.
\end{equation}
Different from the real saddle, the two branches of the complex saddle both have $\partial_n\Phi >0$, so the contour prescription, that imposes the positivity requirement on $\partial_n\Phi$, does not single out one saddle. 
This is similar to the discussion in dS and in AdS$_3$ (section \ref{sec:AdS3}), where the semiclassical wave function evaluated on the complex saddles is the sum of two branches. The above complex saddles can be generalized for the spatial slice with nonzero $K$ by analytically continuing the solutions of $r_0$. Since the analytic expression of the wave function for generic $K$ is lengthy, we omit the closed-form expression here, but the on-shell action of the complex saddle can be calculated numerically as the real saddle is.

\section{Alternative interpretations of the same computation} \label{sec:other_interpretations}

So far, we have focused on the fully gravitational HH wave function $\Psi_{\rm HH}$ on open spatial slices, defined by the GPI in \eqref{eq:open_HH}, and examined it in two representative settings: AdS$_3$ Einstein gravity and AdS$_2$ JT gravity. More generally, however, the GPI appearing on the right-hand side of \eqref{eq:open_HH} admits several alternative interpretations. In this section, we briefly discuss a number of these interpretations.

\subsection{The no-boundary wave function for a spatial subregion}
In our analysis above, we have interpreted $\Sigma$ as a spatial slice with a boundary, and $\mathcal{B}_-$ as a spacetime boundary, and computed the following distribution 
\begin{align}\label{eq:open_HH_recap}
    {\Psi_{\rm HH}[\gamma_{ab}] \equiv  
    \int_{\CB_-} \CD h_{ij}
    \int_{\partial \CM_- = \Sigma \cup \CB_-} \mathcal{D}g_{\mu\nu} ~e^{-I_{\rm E}[g_{\mu\nu}] }= \int_{\partial \CM_- \supset \Sigma} \mathcal{D}g_{\mu\nu} ~e^{-I_{\rm E}[g_{\mu\nu}] }}~, 
\end{align}
as the HH wave function. The bulk $\mathcal{M}_-$ is interpreted as a Euclidean history of the whole spatial slice, and $\CB_-$ is interpreted as a Euclidean history of the spatial boundary. Accordingly, the integral over $(\CB_-, h_{ij})$ is interpreted as summing over the history of the spatial boundary. 

\begin{figure}[ht]
    \centering
    \includegraphics[width=15cm]{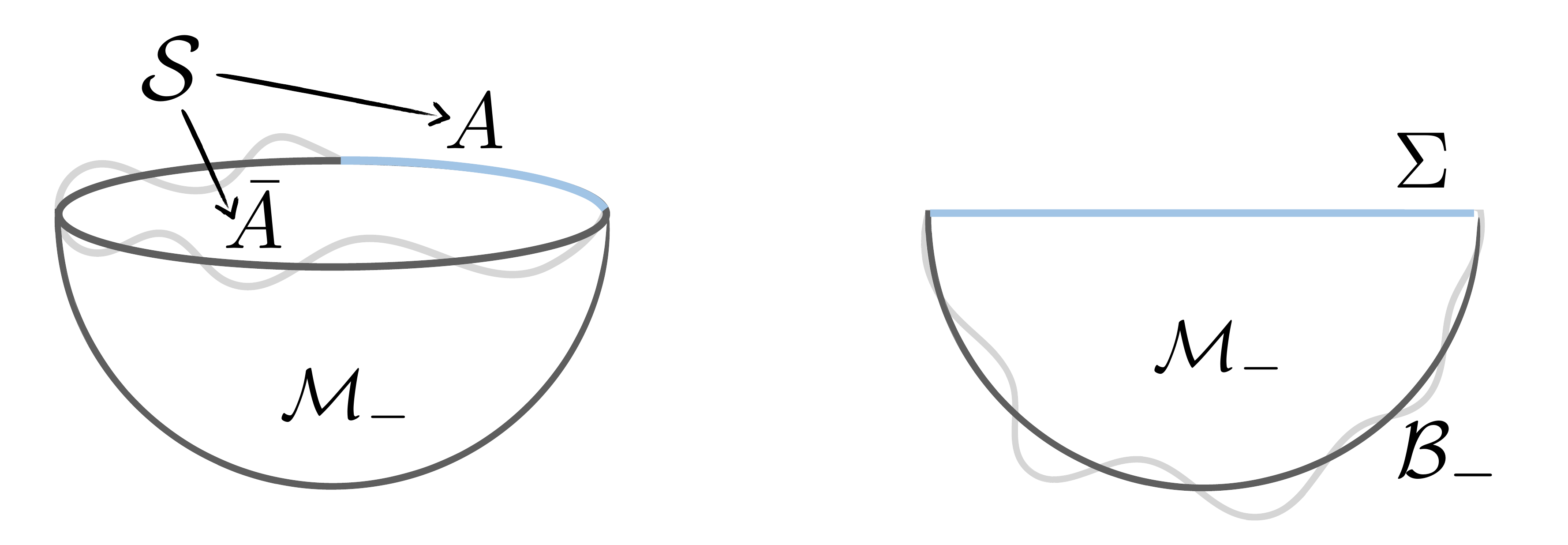}
    \caption{The setup of \eqref{eq:alternative_interpretation}, the no-boundary wave function for a spatial subregion (left), and its comparison to that of \eqref{eq:open_HH}, the HH wave function for an open spatial slice. The blue regions are fixed and the wavy boundaries are summed over in the GPI. 
    }
    \label{fig:HH_subregion}
\end{figure}

However, note that we can regard $\Sigma$ as a spatial subregion embedded in a closed spatial slice as well. For clearness, let us change the names for the elements involved in the computation above as, 
\begin{align}
    \Sigma \rightarrow A, \qquad \CB_- \rightarrow \bar{A},  \qquad A\cup \bar{A} = \mathcal{S}, 
\end{align}
and consider the same computation,
\begin{align}\label{eq:alternative_interpretation}
    {\Psi_{\rm HH}[\gamma_{ab}] \equiv  
    \int_{\bar{A}} \CD h_{ij}
    \int_{\partial \CM_- = \CS } \mathcal{D}g_{\mu\nu} ~e^{-I_{\rm E}[g_{\mu\nu}] }}~.  
\end{align}
This setup is sketched in figure \ref{fig:HH_subregion}
Here, we can interpret $\CS$ as a closed spatial slice, and $\CM_-$ as a Euclidean history of it. Then we divide it into a subregion $A$ and its complement $\bar{A}$. Now if we did not sum over $\bar{A}$, then it would be nothing but just the standard HH wave function for a closed spatial slice $\CS$. On the top of this, an additional integral sums over all possible ambient regions $\bar{A}$ outside of $A$, and hence presents a generalization to the HH {\it wave function} (not density matrix) for spatial subregions. 

Some readers may find it a little bit strange to associate a wave function but not a density matrix (See \cite{Page86,AKM20,DQSY20,ILM24,DW25} for different versions of density matrices akin to the HH wave function) to a spatial subregion. There are at least two ways to think about this. 

A relatively radical interpretation is the following. In gravitational theories, there is a sharp distinction between ${\rm cl}(A)=A\cup \partial A$ and ${\rm int}(A)=A\backslash \partial A$, because the information outside of $A$ should be holographically encoded on $\partial A$. Consequently, ${\rm cl}(A)$ should be associated with a pure state, and thus with a wave function, rather than with a mixed state described by a density matrix \cite{Wei26}. By contrast, ${\rm int}(A)$ should be associated with a mixed state. While it remains unclear whether \eqref{eq:alternative_interpretation} is the correct description of the pure state on ${\rm cl}(A)$, there is nothing {\it a priori} unnatural about assigning a pure state to a spatial subregion.

A more conservative interpretation is the following. Adopting a more conventional perspective, let us consider a gravitational density matrix of the Ivo-Li-Maldacena (ILM) type \cite{ILM24}. Even within this framework, the GPI in \eqref{eq:alternative_interpretation} constitutes an important sector of the ILM no-boundary density matrix once an end-of-the-world brane is included in the action. In fact, in the 2D case, this sector gives the leading contribution from the viewpoint of Euler-characteristic counting: it is described by a double-disk topology, whereas gluing the bra $\bar{A}$ to the ket $\bar{A}$ yields a single-disk topology.

Let us also comment on some subtle differences when comparing \eqref{eq:open_HH_recap} and \eqref{eq:alternative_interpretation}. Although they have the same form as we discussed above, they may bear some detailed differences in the action and the integral contour. 
First, in computing \eqref{eq:open_HH_recap} in sections \ref{sec:AdS3} and \ref{sec:JT-noboundary}, we included a corner term that vanishes at an angle of $\pi/2$ at the interface between $\Sigma$ and $\CB_-$, and we did not introduce a tension term on $\Sigma$, as these were treated as distinct boundary components. However, under the interpretation in \eqref{eq:alternative_interpretation}, where $A$ and $\bar{A}$ are regarded as lying on the same spatial slice, it is more natural to assign them the same tension term. In addition, the corner term would more naturally be taken to vanish at an angle of $\pi$ rather than $\pi/2$. Second, if one regards the Lorentzian path integral as more fundamental than the Euclidean one, then the GPIs in \eqref{eq:open_HH_recap} and \eqref{eq:alternative_interpretation} should be evaluated along different contours.

\subsection{A subregion generalization of Cauchy slice holography}

In the GPI formulation of AdS/CFT \cite{Witten98}, one fixes the metric $h_{ab}$ of the asymptotic boundary $(\mathcal{B},\gamma_{ab})$, sums over all the bulk configurations $\mathcal{M}$ such that $\partial \CM = \CB$, and identifies this as the CFT partition function defined on $\CB$: 
\begin{align}\label{eq:GKPW}
    Z_{\rm CFT}[\gamma_{ab}] = \int_{\partial \CM = \CB} \mathcal{D} g_{\mu\nu}~e^{-I_{\rm AdS}[g_{\mu\nu}]}\,.
\end{align}
By comparing \eqref{eq:GKPW} and \eqref{eq:original_HH}, it is clear that the same GPI can also be interpreted as a HH wave function on closed spatial slices in a gravity with negative cosmological constant. 
\begin{align}
    Z_{\rm CFT}[\gamma_{ab}] = \Psi_{\rm HH}[\gamma_{ab}]\,. 
\end{align}
This interpretation first appears in \cite{Maldacena02} and is also taken in e.g. \cite{Freidel08,IKTV20}.
This suggests, instead of accommodating a holographic theory on a spatial boundary, one can also accommodate a holographic theory on a spatial slice, where the time direction is emergent from the RG flow. One systematic realization of this idea is the so-called Cauchy slice holography \cite{AKW22}. While the work \cite{AKW22} mostly focuses on investigating the property of a $T^2$ deformed CFT partition function as a Wheeler-DeWitt wave function on a semiclassical back ground, it is natural to associate the GPI
\begin{align}\label{eq:partially_frozen_HH_recap}
    \Psi_{\rm HH}^{\CB_-}[\gamma_{ij}] = \int_{\partial \CM_- = \Sigma \cup \CB_-} \mathcal{D}g_{\mu\nu} ~e^{-I_{\rm AdS}[g_{\mu\nu}] },  
\end{align}
to it, considering its similarity to the setup in \cite{CKP20}. In this case, we can regard the spacetime boundary $\CB$ as a place to accommodate a seed CFT, and $\Psi_{\rm HH}^{\CB_-}$ as a QFT partition function flowed from it. 

From this point of view, the HH wave function in \eqref{eq:alternative_interpretation} defined on an open spatial slice can be regarded as a way to generalize the Cauchy slice holography.
In particular, as we have discussed in the last subsection, $\Psi_{\rm HH}[\gamma_{ab}]$ can be regarded as a HH wave function defined on a spatial subregion, and therefore accommodate a holographic dual of the 3D gravity to the spatial subregion.

\section{Norm of the HH wave function in AdS at one loop}
\label{sec:One_loop_norm_GPI}

\begin{figure}
    \centering
    \includegraphics[width=0.9\linewidth]{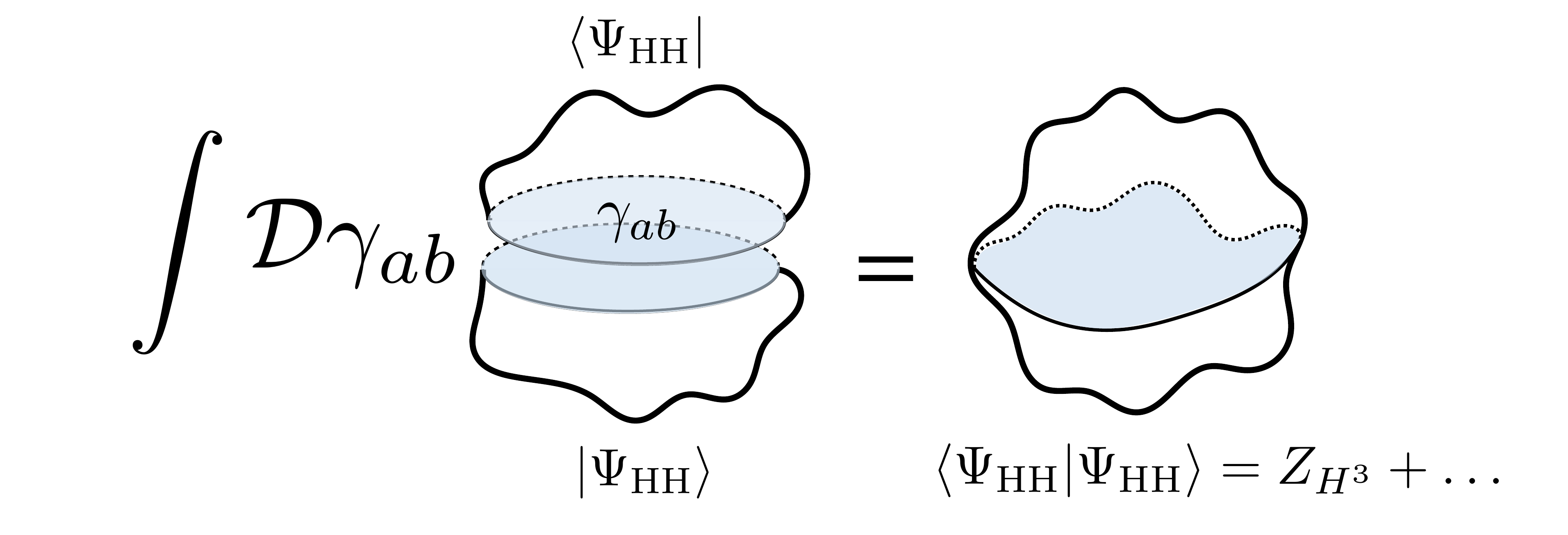}
    \caption{We consider the norm of the no boundary wavefunciton in the right panel, where we integrate over the metric fixed on the slice modulo diffeomorphisms. The norm of the Hartle-Hawking wavefunction is therefore an integration over all the degrees of freedom of the theory and amounts to compute the partition function on AdS$_D$ with Neumann boundary \cite{CM08}.}
    \label{fig:norm_and_partition_function}
\end{figure}

In this section, we compute the hyperbolic ball partition function with a gravitating spacetime boundary, which can be considered as the norm of the HH wave function $\Psi_{\rm HH}(\gamma_{ab})$ defined by \eqref{eq:open_HH} under the conventional interpretation. We work in AdS$_D$ with arbitrary dimension $D\geq 4$\footnote{The 3D case is qualitatively different since gravity is topological, and the would-be effective gravity on the spacetime boundary $S^2$ is special.}. We then derive a one-loop effective action on the dynamical spacetime boundary $\mathcal{B}$, treating the boundary gravitons fully off-shell while requiring the perturbed bulk metric to satisfy linearized Einstein’s equations. Finally, we evaluate the phase coming from the one-loop correction, which amounts to determining the number of negative modes of the effective action. The setup in this section is sketched in figure \ref{fig:norm_and_partition_function}, in particular we explain why computing the norm of the Hartle-Hawking wavefunction is equivalent to compute the partition function on EAdS.

The one-loop correction in AdS$_D$ when Dirichlet boundary conditions are imposed at the conformal boundary, which is standard in the AdS/CFT context , is reviewed in appendix \ref{appendix:1loop_without_boundary_AdSd}; see also\cite{GMY08,Suzuki2021,ST25,Sun2020,GKT14} and is known not to suffer from a phase factor problem.

\subsection{The Semi-classical approximation}
We shall first introduce the Euclidean action for Einstein gravity in $D$ dimensions with a negative cosmological constant $\Lambda$ and a dynamical spacetime boundary $\mathcal{B}$ with tension $T$:
\begin{equation}
    I_{\rm E}=-\frac{1}{16\pi G_N}\int_{\mathcal M}\!\sqrt g\,(R-2 \Lambda)
    -\frac{1}{8\pi G_N}\int_{\mathcal{B}}\!\sqrt h\,(K_{\mathcal{B}}-T)\,.
\label{eq:EuclActionFull_arbitrary_dimension}
\end{equation}
Einstein’s equations with a negative cosmological constant admit a maximally symmetric solution given by $D$-dimensional hyperbolic space. We choose the cosmological constant $\Lambda=-\frac{(D-1)(D-2)}{2}$ so that the AdS radius is set to unity. In these units, the hyperbolic ball metric on AdS$_D$ takes the form:
\begin{equation}
ds^2 = du^2 + \sinh^2 u~ d\Omega_{D-1}^2 \,.
\label{eq:Hdpolar}
\end{equation}
Similar to \cite{CM08}, the boundary is set free and hence fully gravitized. In particular it means that we do not impose Dirichlet boundary condition on the spacetime boundary, but rather sum over all possible configurations of $\CB$ in the GPI. In the semiclassical approximation, this means that 
\begin{equation}
    K_{ij}|_{\mathcal{B}}-(K_{\mathcal{B}}-T) h_{ij}=0\,,
    \label{eq:braneEOM_TildePi0}
\end{equation}
which is the equation of motion for the spacetime boundary obtained by varying the induced metric on $\mathcal{B}$. 

Note $\mathcal{B}$ has intrinsic dimension $D-1$.  Tracing over \eqref{eq:braneEOM_TildePi0} shows that $\mathcal{B}$ is totally umbilic:
\begin{equation}
K_{ij}|_{\mathcal{B}}=\kappa\,h_{ij}\,,
\qquad
\kappa=\frac{T}{D-2}\,,
\qquad
K_{\mathcal{B}}=\frac{D-1}{D-2}\,T\,.
\label{eq:umbilic_general_d}
\end{equation}
If we take $T = D-2$, then the tension term is identical to the usual holographic counterterm at the conformal boundary, this is regime considered in \cite{CM08} and where holographic renormalization is needed. For simplicity, we take $T>(D-2)$.  In this case, the Euclidean spacetime boundary $\CB$ is compact. In geodesic polar coordinates on $H^D$ \eqref{eq:Hdpolar}, the boundary sits at fixed radius $u=u_0$ with $\coth u_0=\frac{T}{D-2}$. The on-shell action reads:

 \begin{align}
 \label{eq:on-shell_action}
     I_{\rm E}^{\text{on-shell}}&=\frac{\text{Area}\left(S^{D-1}\right)}{8\pi G_N}\left((D-1)\int_{u=0}^{u_0}du~\sinh^{D-1}u-\frac{T}{D-2}\sinh^{D-1} u_0 \right) \\
     &=\frac{1}{8\pi G_N}\left((D-1)\text{Vol}(H^{D})-\frac{T}{D-2}\text{Area}(\mathcal{B}) \right)=-(D-2)\frac{\text{Area}\left(S^{D-1}\right)}{8\pi G_N}\int_{u=0}^{u_0}du\,\sinh^{D-3}u\,,\nonumber
 \end{align}
 where $\text{Area}\left(S^{D-1}\right)$ is the area of the unit $(D-1)$-sphere, and where we used $\frac{T}{D-2}=\coth u_0$ and an integral expression for the second term. The tree-level contribution to the norm of the HH wave function in the saddle-point approximation is then:
\begin{equation}
    |\Psi|^2\approx\exp(-I_{\rm E}^{\,\text{on-shell}})\,.
    \label{eq:semiclassical_answer}
\end{equation}

The rest of this section is devoted to computing the one-loop correction to this semiclassical answer, with a particular focus on the number of negative modes.

\subsection{Off-shell perturbation of the boundary sphere }
In the computation of the full one-loop partition function in arbitrary dimension $D$, it is difficult to disentangle the contributions of bulk and boundary gravitons. We formulate this problem in Appendix \ref{app:gauge_fixing}, where we also explain the gauge-fixing procedure in the presence of dynamical spacetime boundary, following \cite{Barvinsky05,Barvinsky06}, but solving the full problem remains beyond the scope of this paper.

\begin{figure}[ht]
    \centering
    \includegraphics[width=15cm]{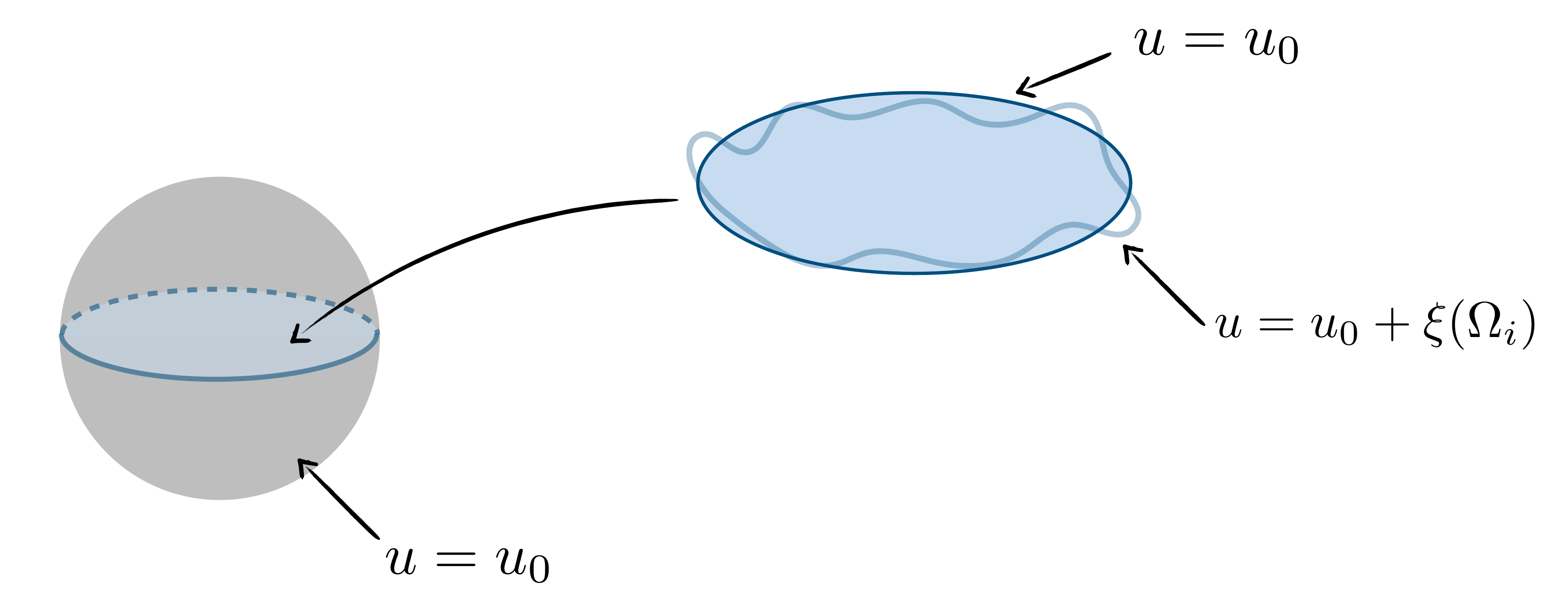}
    \caption{In pale blue we show the unperturbed spacetime boundary $\CB$ at $u=u_0$ in $H^d$. We consider the most general perturbations of the boundary metric $h_{ij}$ and of the boundary position $\xi(\Omega_i)$. The deformation of the pale blue slice is illustrated in blue, although it is only represented on the light blue equator, the perturbation is understood to extend over the entire sphere. The approximation we use is that bulk gravitons are on shell (represented in red): they solve the linear Einstein equations with off-shell boundary conditions. Of course the perturbation is that of the full $u=u_0$ surface, and not only that of the pale green equator. }
    \label{fig:effective_computation_norm}
\end{figure}

In this section, we adopt a simplified approach. We consider the most general off-shell metric perturbations on the spacetime boundary $\mathcal{B}$, while requiring the bulk gravitons to satisfy the linearized equations of motion, this approximation is explained in figure
\ref{fig:effective_computation_norm}. Our goal is to derive an effective one-loop action for these boundary gravitons while restricting the analysis to modes which satisfy the bulk equations with smooth interior geometries. We then use this effective action to evaluate the phase of the partition function. To construct the effective action, and in particular to solve the linearized equations around the saddle $H^D$, we follow \cite{HHR00,HHR01}, where a related problem was analyzed in $D=5$ in a different context. We find that, within this approximation, the phase of the partition function coincides with that obtained from Einstein gravity on the sphere $S^{D-1}$.

The boundary metric at $u=u_0$ is that of the round sphere of dimension $D-1$ that we perturb according to:
\begin{equation}
    ds^2=\sinh^2 u_0 d \Omega_{D-1}^2~=~H_{ij}dx^idx^j=\sinh^2 u_0 \hat{H}_{ij}dx^idx^j \,,\quad g_{ij}=\sinh^2 u_0 \hat{H}_{ij}+h_{ij}\,.
\end{equation}
We call $H_{ij}$ the unperturbed metric on the boundary $\mathcal{B}$: a sphere of radius $\sinh u_0$, $\hat{H}_{ij}$ the metric on the unit sphere, and the perturbations around $H_{ij}$ are called $h_{ij}$.
The most general perturbation on the boundary can be decomposed as:
\begin{equation}
    \label{eq:decomposition_boundary_metric}h_{ij}(x^i)=\phi_{ij}(x^i)+\hat{\nabla}_{(i}\kappa_{j)}(x^i)+\hat{\nabla}_{i}\hat{\nabla}_{j}\sigma(x^i)+\hat{H}_{ij}\psi(x^i)\,,
\end{equation}
where $\phi_{ij}$ is traceless and transverse, $\kappa_j$ is transverse and where $\hat{\nabla}$ are unit sphere covariant derivatives.  Such a decomposition is unique up to conformal killing vectors, indeed a transformation of the form $\sigma\rightarrow\sigma+ \chi\,,\psi\rightarrow\psi +\epsilon \chi$ can leave the boundary metric invariant if:
\begin{equation}
    \hat{\nabla}_{i}\hat{\nabla}_j\chi+\epsilon~\hat{H}_{ij}\chi=0\,.
    \label{eq:conformal_killing_vec}
\end{equation}
This is exactly the equation for sphere conformal Killing vectors, and it requires solutions with $\epsilon=1$. Indeed, tracing this equation leads to:
\begin{equation}
    \hat{\nabla}^2\chi+(D-1)\epsilon\chi=0\,.
\end{equation}
This is the problem of finding the eigenvalues of the scalar Laplacian on $S^{D-1}$; these are the well-known spherical harmonics:
\begin{equation}
    \hat{\nabla}^2Y^{(l)}+l(l+D-2)Y^{(l)}=0\,.
\end{equation}
Hence \eqref{eq:conformal_killing_vec} has a solution only if $\epsilon=1$; this corresponds to the $l=1$ modes. We will come back to this conformal-mode problem later. 

Under a general diffeomorphism $\chi_{i}$ on the boundary, which can be decomposed into a transverse vector and a scalar as $\chi_{i}=\tilde{\chi}_{i}+\partial_{i}\chi$, the coefficients transform as:
\begin{equation}
    \kappa_i\rightarrow\kappa_i+\tilde{\chi}_i\,,\quad\sigma\rightarrow\sigma+\chi\,,\quad\phi_{ij}\rightarrow\phi_{ij}\,,\quad\psi\rightarrow\psi\,,
\end{equation}
so that $\phi_{ij}$ and $\psi$ are invariant under boundary diffeomorphisms. Such perturbations do not satisfy the boundary equations of motion. Nevertheless, we impose that they solve the bulk Einstein’s equations, subject to the boundary condition $h_{ij}(x,u_0)=h_{ij}(x)$. This is the only approximation made in this section. The most general metric perturbation can then be written as:
\begin{equation}
ds^2=\left(du^2+\sinh^2 u~d \Omega_{D-1}^2\right)+ N^2 du^2+2N_i dx^idu + h_{ij} dx^idx^j,.
\end{equation}
We adopt a parametrization analogous to the ADM formalism, in which the geometry is foliated by spheres and the radial coordinate plays the role of time; it is notably useful for holographic renormalization \cite{ElvangHad16}. The shift vector $N_i$ can be decomposed as $N_i=\hat{N}_i+\partial_i n$, where $\hat{N}_i$ is transverse.
Similarly, $h_{ij}$ admits a decomposition analogous to \eqref{eq:decomposition_boundary_metric}. The difference is that the corresponding fields are no longer restricted to the boundary, but are promoted to bulk fields:
\begin{equation}
 h_{ij}(u,x^i)=\phi_{ij}(u,x^i)+\hat{\nabla}_{(i}\kappa_{j)}(u,x^i)+\hat{\nabla}_{i}\hat{\nabla}_{j}\sigma(u,x^i)+\hat{H}_{ij}\psi(u,x^i)\,.
\end{equation}
 Of course the decomposition we introduced is not gauge invariant, since under an infinitesimal diffeomorphism generated by the vector field $\xi^{\mu}$, the metric transforms as:
\begin{equation}
    g_{\mu\nu}\rightarrow g_{\mu\nu}+\nabla_{\mu}\xi_{\nu}+\nabla_{\nu}\xi_{\mu}\,.
\end{equation}
A general diffeomorphism can be decomposed as
$\xi_{\mu}=\xi_u \delta^{u}_{\mu}+(\hat{\xi}_{i}+\partial_i \xi)\delta^{i}_{\mu}$,
where $\hat{\xi}_{i}$ is transverse and transforms as a vector under the $SO(D)$ isometries, while $\xi_u$ and $\xi$ transform as scalars. The corresponding gauge-invariant combinations (independent of $D$) are\cite{HHR00,HHR01}:
\begin{align}
\label{eq:def_gauge_gnvariant_quantities}
    F_1~=~&N^2-\partial_{u}\left(\frac{\psi}{\sinh u\cosh u}\right)\,,\\
    F_2~=~&n-\frac{1}{2}\partial_u\sigma-\frac{\psi}{2\sinh u\cosh u}+\coth u\sigma\,,\\
    V_i~=~&\hat{N}_i-\partial_u\kappa_i+2 \coth u\kappa_i\,.
\end{align}
Our next task is to linearize the Einstein's equations given by:
\begin{equation}
    R_{\mu\nu}-\frac{1}{2} R g_{\mu\nu}=-\Lambda g_{\mu\nu}\,.
\end{equation}
Around the background solution \eqref{eq:Hdpolar}, one obtains:
\begin{equation}
    \frac{1}{2} \left(\nabla_{(\mu}\nabla^{\lambda}\delta g_{\nu)\lambda}-\nabla^2 \delta(g_{\mu\nu})-\nabla_{\mu}\nabla_{\nu}\delta g\right)=\frac{2\Lambda }{(D-1)(D-2)}g_{\mu\nu}\delta g-\frac{2\Lambda }{(D-1)(D-2)} \delta g_{\mu\nu}\,.
\end{equation}
Using $\Lambda=-\frac{(D-1)(D-2)}{2 \ell^2}$ to write the metric as a function of the AdS length $\ell=1$. The linearized Einstein's equations take the simple form:
\begin{equation}
\label{eq:linearized_einstein_equation}
    \frac{1}{2} \left(\nabla_{(\mu}\nabla^{\lambda}\delta g_{\nu)\lambda}-\nabla^2 \delta(g_{\mu\nu})-\nabla_{\mu}\nabla_{\nu}\delta g\right)=-\frac{1 }{\ell^2}g_{\mu\nu}\delta g+\frac{1}{\ell^2} \delta g_{\mu\nu}\,.
\end{equation}
The next subsection is devoted to solving these linearized Einstein's equations, decomposed into scalar, vector, and tensor sectors under the action of $SO(D)$. The tensor part is left to Appendix.\ref{appendix:solving_for_the_tensor_mdoes} following \cite{HHR00}. Solving the tensor modes is not of great interest since these modes are non dynamical in the effective action for gravity on the spacetime boundary $\CB$.

\subsection{Solving for the scalar and vector parts}
\label{sec:solving_the_bulk_equations}
The linearized Einstein's equations are gauge invariant and can be expressed in terms of the gauge-invariant variables introduced in \eqref{eq:def_gauge_gnvariant_quantities}. Taking advantage of foliating AdS$_{D}$ by $(D-1)$-spheres, we further decompose these equations into scalar, vector, and tensor sectors under the action of $SO(D)$. The $uu$ component of the Einstein's equations yields a scalar equation:
\begin{equation}
    \hat{\nabla}^2F_1-2\partial_{u}\left(\hat{\nabla}^2F_2\right)-(D-1)\cosh u~\sinh u\partial_{u}F_1-2(D-1)\sinh^2 u~F_1=0\,.
\end{equation}
The $u i$ equation can be decomposed in a scalar equation and a vector one:
\begin{equation}
    \hat{\nabla}^2V_i=-(D-2)V_i\,,\quad \partial_i\left(\cosh u~\sinh uF_1-2F_2\right)=0\,.
    \label{eq:vec_and_scalar_part_rhoj}
\end{equation}
Similarly, the $ij$ equation leads to a tensor equation (the analysis of this equation can be found in Appendix.\ref{appendix:solving_for_the_tensor_mdoes}). The $ij$ linearized Einstein's equations have a scalar equation:
\begin{align}
    0=&\hat{\nabla}_{i}\hat{\nabla}_j\left(-F_1+2\partial_{u}F_2+2\coth u (D-3) F_2 \right)\\
    &+\hat{H}_{ij}\left(\cosh u~\sinh u\partial_{u} F_1+2\coth u\hat{\nabla}^2F_2+\left(2(D-1)\cosh^2 u-2\right)F_1\right)\,,
\end{align}
and a vector equation:
\begin{equation}
    \left(\partial_{u}+(D-3)\coth u\right) \hat{\nabla}_{(i}V_{j)}=0\,.
    \label{eq:vec_part_ij}
\end{equation}

Solving first for the vector mode using \eqref{eq:vec_part_ij}, we obtain: 
\begin{equation}
    \hat{\nabla}_{(i}V_{j)}(u,x^i)=\left(\frac{\sinh u_0}{\sinh u}\right)^{D-3}\hat{\nabla}_{(i}V_{j)}(u_0,x^i)\,.
\end{equation}
Regularity imposes $\hat{\nabla}_{(i}V_{j)}=0$, so that by combining with \eqref{eq:vec_and_scalar_part_rhoj}, we obtain $V_i=0$. The scalar equations can be written as:
\begin{align}
    &\hat{\nabla}^2F_1+(D-1)F_1=0\,,\\
    &\left(\cosh u~\sinh u\partial_{u}+(D-1)\cosh^2 u-2\right)\left(\hat{\nabla}_i\hat{\nabla}_j+\hat{H}_{ij}\right)F_1=0\,.
\end{align}
This can be solved first as a function of $u$, and we obtain:
\begin{equation}
    \left(\hat{\nabla}_i\hat{\nabla}_j+\hat{H}_{ij}\right)F_1(u,x^i)=\frac{\sinh^{D-1} u_0\tanh^2 u~}{\sinh^{D-1} u\tanh^2 u_0}\left(\hat{\nabla}_i\hat{\nabla}_j+\hat{H}_{ij}\right)F_1(u_0,x^i)\,.
\end{equation}

This is singular at $u=0$, so we need to impose: $\left(\hat{\nabla}_i\hat{\nabla}_j+\hat{H}_{ij}\right)F_1(u,x^i)=0$. This equation is just the equation for conformal killing vectors, and is ambiguous in our mode decomposition. It is consistent to send $F_1=0$, and then we immediately obtain that $F_2$ is an arbitrary function of $u$, which is pure gauge and not interesting in our analysis.

In summary, solving the scalar and vector sectors, together with the requirement of smoothness in the bulk, implies that the following gauge-invariant quantities vanish:
\begin{align}
    F_1=F_2=V_i=0\,.
\end{align}
Since the bulk metric perturbations are on-shell, we are free to choose a convenient gauge. We adopt Gaussian normal coordinates, imposing $N = N_i = 0$. In this gauge, the remaining fields can be readily solved in terms of their boundary values. The most general solution takes the form:
\begin{equation}
    \sigma(x,u)\!=\!a(x)\sinh^2 u~+2 b(x)\sinh u\cosh u,~\kappa_i(x,u)\!=\!\kappa_i(x) \sinh^2 u~\,,~\psi(x,u)\!=\!2 b(x)\sinh u\cosh u\,.
\end{equation}

For some function $a(x),\,b(x)$ that should be matched with boundary data. In general there is also an extra scalar $\xi(x)$: the displacement of the boundary. Namely , because of the perturbation of the metric the boundary sits at $u=u_0+\xi(x^i)$, but we end up with three scalars to fix $a(x),\,b(x),\,\xi(x)$ in terms of two boundary scalar perturbations $\psi(x),\,\sigma(x)$. This tension is resolved because we should further require regularity in the interior, which amounts to imposing that the perturbations remain finite as $u = 0$. The only constraint comes from smoothness of the trace of the perturbation $H^{ij}h_{ij}$ which is gauge invariant, it imposes that $b(x)=0$. The full relation between the boundary data $\sigma(x),\,\psi(x)$ and $a(x),\,\xi(x)$ can be obtained by expanding the perturbation of the metric around $u_0$ and evaluate to first order in $\xi$.  In particular the metric perturbation expands in the bulk through:
\begin{equation}
    \sigma(x,u)=\sigma(x)\sinh^2 u~\,,\quad \kappa_i(x,u)=\kappa_i(x) \sinh^2 u~\,,\quad \psi(x,u)=2\xi(x) \sinh u_0\cosh u_0\,.
\end{equation}
Note that the coefficient for $\psi$ does not depend on $u$. This is the approach chosen in \cite{HHR00,HHR01}. We now adopt a different perspective. Starting from the configuration with $a(x)=0$ and general $\xi$, we perform a change of coordinates such that the brane is brought back to $\tilde{u}=u_0$, while maintaining Gaussian normal form for the metric in the bulk. Under a general diffeomorphism generated by $\beta^{\mu}(x,y)$, the metric transforms as:
\begin{equation}
    h_{\mu\nu}\rightarrow h_{\mu\nu}+\nabla_{(\mu}\beta_{\nu)}\,.
\end{equation}
Accordingly, 
\begin{align}
    N&\rightarrow N+2\partial_{u}\beta_{u}\\
    N_i&\rightarrow N_i+\partial_{u}\beta_i+\partial_i\beta_{u}-2\beta_i\coth u\\
    h_{ij}&\rightarrow h_{ij} +\hat{\nabla}_{(i}\beta_{j)}-2\Gamma^{u}_{ij}\beta_{u}\,.
\end{align}
We perform a coordinate transformation such that the brane is located at a fixed position, which requires $\beta_{u}=\xi$. The condition of Gaussian normal coordinates imposes that the lapse $N$ remains trivial, and therefore $\beta_{u}$ must be independent of $u$.
Requiring in addition that $N_i=0$ determines the spatial components of the diffeomorphism to be
$\beta_i=\cosh u~\sinh u\partial_i\xi+\sinh^2 u~\tilde{\beta}_i$,
where $\tilde{\beta}_i$ corresponds to a residual gauge transformation acting on the boundary.
With these conditions, the resulting metric perturbation on the sphere takes the form:
\begin{equation}
\label{eq:metric_perturbation_final}
    h_{ij}=\phi_{ij}(x,u)+\sinh^2 u~\hat{\nabla}_{(i}\kappa_{j)}(x)+\sinh^2 u~\hat{\nabla}_{i}\hat{\nabla}_{j}\sigma(x)+2\sinh u\cosh u\left(\hat{\nabla}_{i}\hat{\nabla}_{j}\xi+\hat{H}_{ij}\xi\right)\,.
\end{equation}
Here the bending mode becomes an independent field on the boundary, of course now $H^{ij}h_{ij}$ is not smooth at $u=0$, but we know that this is a coordinate artifact coming from having fixed the boundary at $u=u_0$. We have made the $u$ dependence explicit to distinguish what can be gauged away from what cannot.

\subsection{The effective action and the phase of the GPI}
Let us now compute the effective action for the boundary metric $h_{ij}(x)$. The bulk gravitons are on shell up to second order, so Einstein's equations are satisfied and the second-order perturbation of the bulk terms vanishes exactly. The boundary satisfies the equations of motion to first nontrivial order. The total action, perturbed to second order, reduces to:
\begin{align}
    I_{\rm E}+I_{\mathcal{B}}=&I_{\rm E}^{\text{on-shell}} -\frac{1}{16\pi G_N}\int_{\mathcal{B}}d^{D-1}x\sqrt{H}\delta\left(\left(K_{ij}|_{\mathcal{B}}-K_{\mathcal{B}} H_{ij}+T H_{ij}\right)\right)\delta(h^{ij})\,,
\end{align}
where $I_{\rm E}^{\text{on-shell}}$ was given in \eqref{eq:on-shell_action}. The variation of $K_{ij}|_{\mathcal{B}}-(K_{\mathcal{B}}-T)H_{ij}$ is given by
\begin{align}
    &\delta\left(K_{ij}|_{\mathcal{B}}-(K_{\mathcal{B}}-T)H_{ij}\right) \nonumber\\
    =&-\left(\hat{\nabla}_{i}\hat{\nabla}_{j}\xi+\hat{H}_{ij}~\xi\right)+\hat{H}_{ij}\left(\hat{\nabla}^{2}\xi+(D-1)\right)+\frac{1}{2}g^{uu}\partial_{u}(\phi_{ij})-\coth u~\phi_{ij}\,.
\end{align}
We finally obtain the one-loop effective action for the boundary modes, with the normal displacement $\xi$ and traceless-transverse two-tensor $\phi_{ij}$:
\begin{align}
    I=I_{\rm E}^{\text{on-shell}}&-\frac{1}{16\pi G_N}\int_{\mathcal{B}}d^{D-1}x~\sinh^{D-5}(u_0)\left(D-2\right)(-2\sinh u_0\cosh u_0~\xi)\left(\hat{\nabla}^2+(D-1)\right)\xi \nonumber\\
    &-\frac{1}{16\pi G_N}\int_{\mathcal{B}}d^{D-1}x~\sinh^{D-5}(u_0)\left(\frac{1}{2}\partial_{u}\phi_{ij}-\coth u_0~\phi_{ij}\right)(-\phi^{ij})\,.
\end{align}
Indices are raised and lowered with $\hat{H}_{ij}$. We now include the contribution from the gauge-fixing term.
The gauge-fixing condition involves only derivatives tangential to the boundary, and is therefore insensitive to the $u$-dependence. In order to make contact with the existing literature which evaluates the phase of the one loop partition function in dS \cite{Polchinski88,ST25,CSTY25}, we shall decompose the metric according to:
\begin{equation}
    h_{ij}=\phi_{ij}+\bar{\nabla}_{(i}\bar{\kappa}_{j)}+\bar{\nabla}_{i}\bar{\nabla}_{j}\sigma-\frac{1}{(D-1)}H_{ij}\bar{\nabla}^2\sigma+\frac{1}{(D-1)}H_{ij}h\,,
\end{equation}
where the trace $h$ of the metric perturbation with respect to $H_{ij}$ and can be written in terms of coefficients of the decomposition \eqref{eq:metric_perturbation_final}:
\begin{equation}
    h=2(D-1)\coth u~\xi+\bar{\nabla}^2\sigma
\end{equation}
We denote by $\bar{\nabla}$ the covariant derivative on the sphere of radius $\sinh u_0$. The fact that only tangential derivatives enter at this stage is crucial, as it allows one to absorb the $\bar{\nabla}^2\xi$ dependence into $\sigma$ (that we also rescaled by a $\sinh(u_0)^2$). The gauge-invariant quantity becomes $\xi=\frac{h-\bar{\nabla}^2\sigma}{2(D-1)\coth u_0}$. The action can be written as
\begin{align}
    I_{\rm E}=&-\frac{1}{16\pi G_N}\int_{\mathcal{B}}d^{D-1}x\sqrt{\hat{H}}\frac{(D-2)\sinh^{D-1}(u)}{2(D-1)^2\coth u}\left(h-\bar{\nabla}^2\sigma\right)\left(-\bar{\nabla}^2-\frac{(D-1)}{\sinh^2 u~}\right)\left(h-\bar{\nabla}^2\sigma\right)\nonumber\\
    &+I_{\rm E}^{\text{on-shell}}-\frac{1}{16\pi G_N}\int_{\mathcal{B}}d^{D-1}x~\sinh^{D-5}(u)\left(\frac{1}{2}\partial_{u}\phi_{ij}-\coth u\phi_{ij}\right)(-\phi^{ij})\,.
\end{align}
in terms of these quantities. The second term is non dynamical, hence the path integral over $\phi_{ij}$ leads to convergent gaussian integrals over the coefficients of its decomposition into spherical harmonics. We shall now only consider the first term (which is the only dynamical one), it reads
\begin{equation*}
    I_{\rm E}^{\text{eff}}=-\frac{1}{16\pi G_N^{{\rm eff}\,(D-1)} }\int_{\mathcal{B}}d^{D-1}x\sqrt{H}\frac{(D-3)(D-2)}{4(D-1)^2}\left(h-\bar{\nabla}^2\sigma\right)\left(-\bar{\nabla}^2-\frac{(D-1)}{\sinh^2 u~}\right)\left(h-\bar{\nabla}^2\sigma\right)\,.
\end{equation*}
where we introduced the effective Newton constant on the boundary $G_N^{{\rm eff}\,(D-1)}=\frac{(D-3)\coth u}{2}G_N$.  To compute the phase of the gravitational path integral and as mentioned in Appendix.\ref{app:gauge_fixing}, we shall add the gauge fixing term:
\begin{equation}
    I_{\text{gauge fixing}}=\frac{1}{32G_N^{{\rm eff}\,(D-1)}}\int_{\mathcal{B}} d^{D-1}\sqrt{H}\left(\bar{\nabla}^{i}h_{ij}-\frac{1}{2}\bar{\nabla}h\right)^2
\end{equation}
The computation reduces to that of the standard partition function on the sphere. The essential difference is that the transverse traceless modes are non-dynamical in the present setting. 
Restricting to the dynamical degrees of freedom, the resulting quadratic effective action in this approximation takes the form:
\begin{align}
\label{eq:gauge_fixed_action_at_one_loop}
    I_{\rm E}+I&_{\text{gauge fixing}}=\frac{1}{16 \pi G_N^{{\rm eff}\,(D-1)}}\int_{\mathcal{B}} d^{D-1}x\sqrt{H}\left(\frac{-(D-3)}{8(D-1)}\right)h\left(-\bar{\nabla}^2-2 \frac{(D-2)}{l_{\text{eff}}^2}\right)h\\
    &+\frac{1}{16 \pi G_N^{{\rm eff}\,(D-1)}}\int_{\mathcal{B}} d^{D-1}x\sqrt{H}\frac{(D-2)}{D}\sigma(-\bar{\nabla}^2)\left(-\bar{\nabla}^2-\frac{(D-1)}{l_{\text{eff}^2}}\right)\left(-\bar{\nabla}^2-2 \frac{(D-2)}{l_{\text{eff}}^2}\right)\sigma \nonumber\\
    &+\frac{1}{32 \pi G_N^{{\rm eff}\,(D-1)}}\int_{\mathcal{B}}d^{D-1}x\sqrt{H}\kappa^{i}\left(-\bar{\nabla}^2-\frac{D-2}{l_{\text{eff}}^2}\right)\kappa_i
\end{align}
where $l_{\text{eff}}^2=\sinh^2 u_0$ is the radius of the sphere. In addition to the modes written in \eqref{eq:gauge_fixed_action_at_one_loop}, one should also add the $bc$ ghosts system which do not lead to any negative modes, see Appendix \ref{app:gauge_fixing} for the ghost, and \cite{ST25} for a proof that they don't lead to negative modes. It is well known that the trace mode $h$ has a wrong-sign kinetic term. Our effective action suffer from the same problem which can be remedied by an ultra-local field redefinition $h' = i h$. The negative modes only comes from the trace part of the action \cite{Polchinski88}. The spectrum of the differential operator involved for the trace part:
\begin{equation}
    \left(-\bar{\nabla}^2-2\frac{(D-2)}{l_{\text{eff}^2}}\right)h=0\,,
\end{equation}
have eigenfunctions which are simply the spherical harmonics $Y^{(l,m)}$ with eigenvalues: $l(l+D-2)-2(D-2)$. The $l=0$ and $l=1$ modes have negative eigenvalues, there are $D+1$ such negative eigenvalues and therefore the phase of the norm of the HH wavefunction is $i^{D+1}$ at one loop.

We have shown that when the boundary is fully gravitized and Neumann boundary conditions are imposed, the norm suffers from the same phase factor as dS$_{D-1}$ (the sphere on the boundary). 
\begin{align}\label{eq:1-loop_phase_AdS}
    \boxed{ Z^{\text{1-loop}}_{\rm hyperbolic\,ball} = (\mp i)^{D+1}\times \left(\text{positive number}\right)}\,,
\end{align}
While this computation was performed in the approximation in which bulk gravitons are on shell, we expect the phase factor to remain the same in the full one-loop computation. Indeed, the full off-shell computation with Dirichlet boundary conditions leads to no phase at all;
\begin{align}\label{eq:1-loop_phase_AdS_DBC}
    \boxed{Z^{\text{1-loop}}_{\rm hyperbolic\,ball\,DBC} =  \left(\text{positive number}\right)}\,,
\end{align}
therefore, we expect any possible phase to arise from “boundary modes,” which are captured in our approximation.

\section{One-loop correction of dS gravity with fixed equator}
\label{sec:paper2-fixed-equator}

The appearance of the one-loop phase \cite{Polchinski88} in the fully gravitational PI of dS gravity has been a longstanding problem in tension with the state-counting interpretation of the sphere partition function for decades, and has been extensively addressed recently \cite{Maldacena24,ST25,IMS25,CJ25,CSTY25,IT26,GS26}. 
As discussed in some of these literature \cite{Maldacena24,ST25,CJ25}, it is also in tension with interpreting the sphere partition function as the leading contribution of the norm of the HH wave function. 

It was suggested that the one-loop phase in the sphere partition function may be removed either by introducing an observer to the path integral \cite{Maldacena24,CSTY25} or by considering different gravity models\cite{GS26}, while our AdS analysis in the previous section provides a new perspective to this problem. The AdS analysis in the previous section shows that, in a fully gravitational setup, the hyperbolic ball partition function, which was interpreted as the norm computation, receives a nontrivial phase contribution from the one-loop correction, whereas in the frozen-boundary setup (the standard AdS/CFT setting) it does not. This suggests that the presence or absence of such a phase may depend on whether the path integral is fully gravitational or partially frozen. 

This observation motivates us to reconsider the one-loop correction to the sphere partition function in Einstein gravity within a partially frozen GPI. As the simplest example, we study the case in which the equator of the sphere is frozen. 
As we show below, this partially frozen sphere partition function does not acquire a nontrivial phase from the one-loop correction, similar to the AdS with the Dirichlet boundary condition case.

\subsection{Review on the sphere partition function and its negative modes}
\label{sec:sphere_partition_function}

\begin{figure}[ht]
    \centering
    \includegraphics[width=13.5cm]{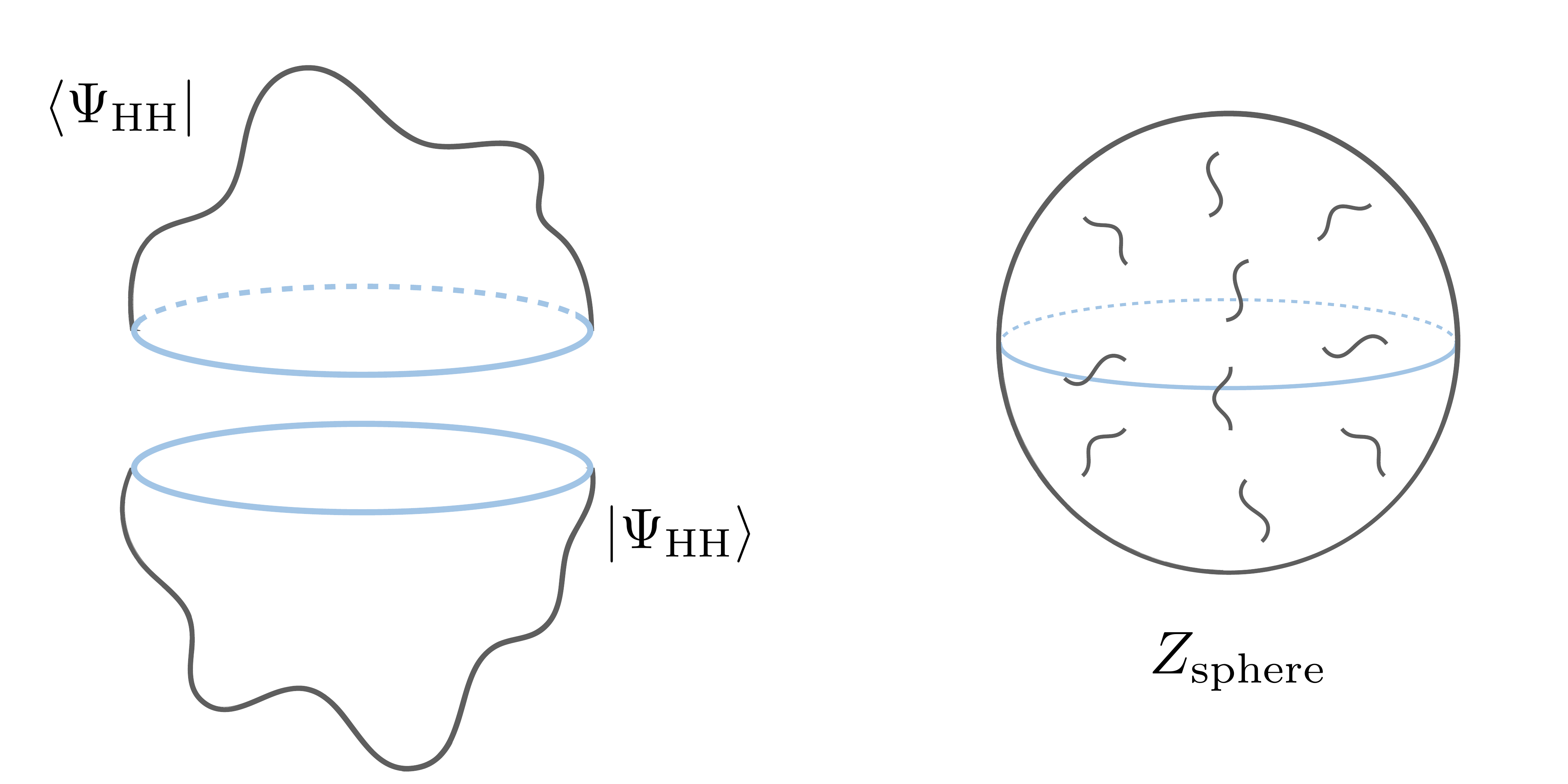}
    \caption{The sphere partition function is usually considered to be the leading contribution of the norm of the HH no-boundary state in dS gravity (shown in the left panel), which is in tension with its nontrivial phase. 
    The $(D-1)$-dimensional spatial slice shown in blue color specifies the data to evaluate the HH no-boundary wave function. 
    In the fully gravitational GPI, we expand the metric around the spherical saddle (shown in the right panel) and integrate out all possible metric fluctuations (demonstrated by wavy lines) over the whole sphere. The one-loop correction leads to a nontrivial phase. 
    }
    \label{fig:fullygravitational}
\end{figure}

Before moving to a partially frozen GPI calculation, we briefly review the phase problem of the sphere partition function in pure dS gravity from the fully gravitational GPI (see figure \ref{fig:fullygravitational} for a demonstration of the setup). 

We begin with Euclidean Einstein gravity on the round sphere $S^D$,
\begin{equation}
    I_{\rm EH} =-\frac{1}{16\pi G_N} \int_{S^D} d^Dx\, \sqrt{g}\, (R-2\Lambda)\,,
    \qquad
    \Lambda = \frac{(D-1)(D-2)}{2}\,,
\end{equation}
where we set the dS radius to 1. 
The saddle-point geometry is the unit sphere, with tree-level action
\begin{equation}
\label{eqn:sphereonshell}
    I_0 = -\frac{\mathrm{Vol}(S^{D-2})}{4G_N} = -\frac{\mathrm{A}_c}{4G_N}\,.
\end{equation}
To compute the one-loop contribution, we consider the metric perturbation
\begin{equation}
\label{eq:metric_perturbation}
    g_{\mu\nu}=\hat g_{\mu\nu}+\sqrt{32\pi G_N}\,h_{\mu\nu}
\end{equation}
about the round sphere metric $\hat{g}_{\mu\nu}$ and fix de Donder gauge by inserting the following gauge-fixing term
\begin{equation}
    I_{\text{gauge-fixing}}
    = \frac{1}{2}\int d^Dx\, \sqrt{\hat{g}}\,
    \left(\hat{\nabla}^\mu h_{\mu\nu}-\frac{1}{2}\hat{\nabla}_\nu h\right)^2\,.
\end{equation}
The corresponding ghost term is
\begin{equation}
    I_{\text{ghost}} = \frac{1}{2}\int d^Dx\,\sqrt{\hat{g}}\, b_\mu(-\hat{\nabla}^2-(D-1))c^\mu\,.
\end{equation}
Decomposing the metric perturbation into its traceless $\phi_{\mu\nu}$ and trace parts $\phi$ gives the standard quadratic action
\begin{equation}
\label{eqn:quadraticaction}
    I_{\text{EH}}+I_{\text{gauge-fixing}}
    = \frac{1}{2}\int d^Dx\,\sqrt{\hat{g}}\,
    \left(
        \phi_{\mu\nu}(-\hat{\nabla}^2+2)\phi^{\mu\nu}
        -\frac{1}{2}D(D-2)\phi(-\hat{\nabla}^2-2(D-1))\phi\right)\,.
\end{equation}
The traceless tensor sector is positive.  The trace sector has the wrong overall sign, and after the Gibbons--Hawking--Perry contour rotation \cite{GHP78} $\phi\rightarrow \pm i\hat{\phi}$\footnote{The overall $\pm$ sign of the contour rotation is undetermined at this stage and will be addressed at the end of this section.}, one is left with finitely many negative modes of the scalar operator
\begin{equation}
    -\hat{\nabla}^2-2(D-1).
\end{equation}
Its scalar harmonic eigenvalues are
\begin{equation}
    \lambda_l=l(l+D-1)-2(D-1)\quad (l\in\mathbb{Z})\,,
\end{equation}
so the $l=0$ mode and the $l=1$ multiplet are negative for $D\ge 3$ with degeneracy $1$ and $D+1$, respectively.  This is the familiar source of the $(\mp i)^{D+2}$ phase in the one-loop sphere partition function\cite{Polchinski88,ADLS20},  
\begin{align}\label{eq:sphere_phase_fully_gravitational}
    \boxed{ Z^{\text{1-loop}}_{\rm sphere} = (\mp i)^{D+2}\times \left(\text{positive number}\right)}\,.
\end{align}
The ghost sector has both negative modes and zero modes, but the zero modes are just the $\mathrm{SO}(D+1)$ isometries and are removed by dividing by the group volume. The negative mode in the ghost sector can be resolved by considering the appropriate Faddeev-Popov measure.

\subsection{Partially frozen sphere partition function with a fixed equator}

\begin{figure}[ht]
    \centering
    \includegraphics[width=13.5cm]{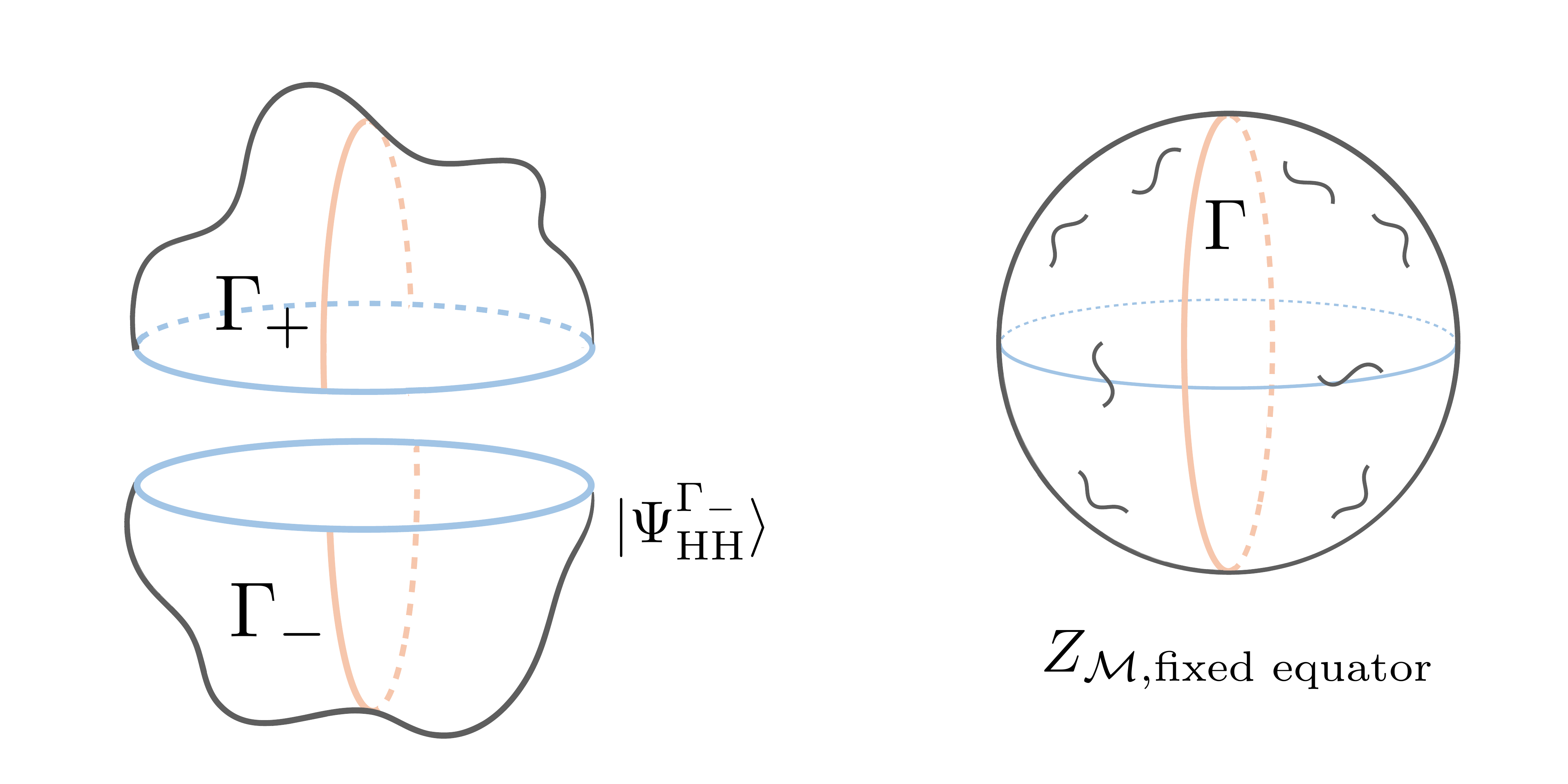}
    \caption{In the partially frozen GPI, we consider a sphere partition function with a fixed equator, as shown in the right panel. The metric on the codimension-1 surface $\Gamma$ (orange) is frozen to be an $S^{D-1}$ with the same dS radius for simplicity. The metric fluctuctaions (wavy lines) are required to vanish on $\Gamma$. This partially frozen sphere partition function may be interpreted as the leading contribution of the norm of a partially frozen HH state $\ket{\Psi_{\rm HH}^{\Gamma_-}}$, where $\Gamma_- \sim \Gamma/Z_2$ is frozen in its GPI preparation, as shown in the lower half of the left panel. The upper half of the left panel shows its bra counterpart, whose GPI preparation also includes a frozen region $\Gamma_+ \sim \Gamma_- \sim \Gamma/Z_2$. Note that $\Gamma$ wraps around the Euclidean temporal direction. }
    \label{fig:partially_frozen}
\end{figure}

Motivated by the comparison between the 1-loop correction of the hyperbolic ball partition function with a gravitating boundary \eqref{eq:1-loop_phase_AdS} and that with the Dirichlet boundary condition \eqref{eq:1-loop_phase_AdS_DBC}, we would like to consider a partially frozen sphere partition function as an analogue of the hyperbolic ball partition function with DBC in dS. 

As the simplest example, we consider the sphere partition function with a fixed equator. As shown in figure \ref{fig:partially_frozen}, such a fixed-equator sphere partition function may be interpreted as the norm of a partially frozen HH wave function. 

More precisely, we consider the path integral on manifold $\CM$ consisting two parts $\CM = \CM_L \cup \CM_R$, where $\CM_L$ and $\CM_R$ are glued together along the co-dimension 1 interface $\Gamma = \partial \CM_L = \partial \CM_R$, and the induced metric on $\Gamma$ is frozen to be a $(D-1)$-dimensional sphere whose radius is the same as the dS radius in $D$-dimensions. 

The gravitational action on $\CM$ is
\begin{equation}
\begin{aligned}
    I =& -\frac{1}{16\pi G_N} \int_{\CM_L\cup\CM_R} d^Dx\, \sqrt{g}\, (R-2\Lambda)\\&-\frac{1}{8\pi G_N}\left(\int_{\partial \CM_L}\,d^{D-1}x\,\sqrt{H_L}\,K_L+\int_{\partial \CM_R}\,d^{D-1}x\,\sqrt{H_R}\,K_R\right)\,,
\end{aligned}
\end{equation}
where we introduce two GHY terms and fix the induced metric $H_{ij}|_{\partial \CM_L} = H_{ij}|_{\partial \CM_R}$ on the  interface $\Gamma$. The gluing condition identifies the induced metric on both sides. In our case, the induced metric on the interface $\Gamma$ is fixed to be 
\begin{equation}\label{eq:Gamma_metric}
    ds^2_{\Gamma} = d\Omega_{D-1}^2\,.
\end{equation}
where the radius of $\Gamma$ is set to $1$ in accordance with the dS radius $R_{\text{dS}}=1$. 

The saddle geometry on $\CM$ is the $D$-dimensional unit sphere with the interface $\Gamma$ placed at the equator. The on-shell action in this case will be identical to the action \eqref{eqn:sphereonshell} obtained in the fully gravitational path integral, while the one-loop contribution to the partition function will be different. Let us proceed to the 1-loop analysis in the following.

\paragraph{1-loop analysis on half spheres}~\par
\begin{figure}[ht]
    \centering
    \includegraphics[width=13cm]{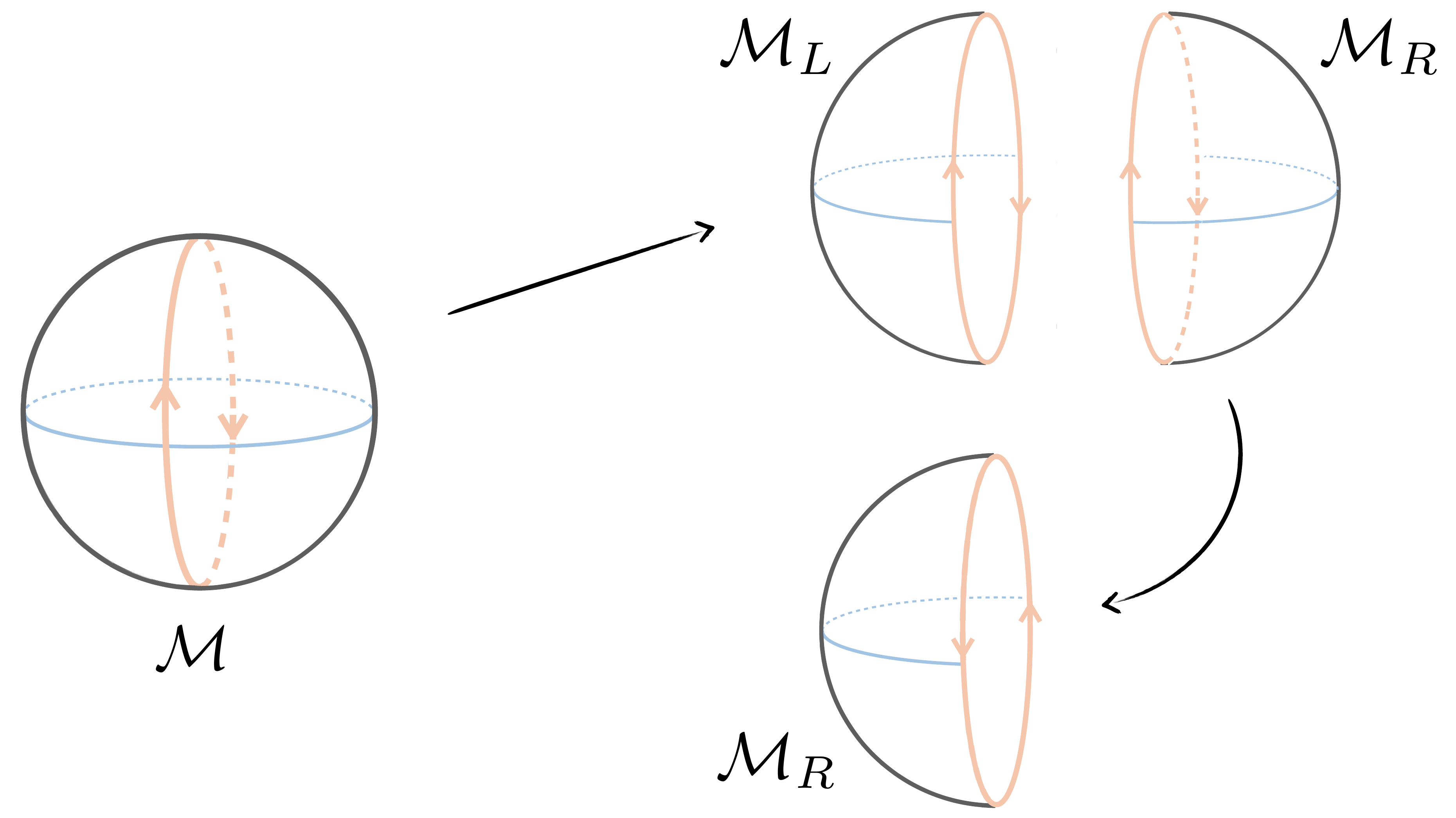}
    \caption{Splitting the partially frozen sphere (left) into two half-spheres (right upper) is convenient for evaluating the partition function. It is important to note that the two half-spheres $\CM_L$ and $\CM_R$ wraps the Euclidean temporal direction $\tau$ in opposite orientations, which is easy to see when flipping $\mathcal{M}_R$ in the sketch (right lower).}
    \label{fig:split_two_hemispheres}
\end{figure}

In practice, it is convenient to firstly consider the case of half-spheres, and then glue two of them together, as shown in figure \ref{fig:split_two_hemispheres}. 

More precisely, we consider the path integral on the $D$-dimensional manifold $\CM_L$ with a $(D-1)$-dimensional boundary $\partial \CM_L = \Gamma$, with the action given by 
\begin{align}
    I^L =& -\frac{1}{16\pi G_N} \int_{\CM_L} d^Dx\, \sqrt{g}\, (R-2\Lambda)-\frac{1}{8\pi G_N}\int_{\partial \CM_L = \Gamma}\,d^{D-1}x\,\sqrt{H_L}\,K_L.
\end{align}
Fixing the metric at $\CM_L =\Gamma$ to be \eqref{eq:Gamma_metric}, the saddle is exactly the hemisphere. The equatorial extrinsic curvature vanishes, so the on-shell action is exactly half the action of the full sphere. 

The setup of the one-loop calculation will be similar to the fully gravitational one with the metric perturbation given in \eqref{eq:metric_perturbation}. Subject to the fixed boundary induced metric, the metric perturbation around the saddle must vanish on the boundary
\begin{equation}
    g_{\mu\nu}|_{\partial \CM_L} = \hat{g}_{\mu\nu}|_{\partial\CM_L}=H_{ij}\,,\quad h_{\mu\nu}|_{\partial \CM_L} = 0\,.
\end{equation}
We again decompose the metric perturbation into the traceless part $\phi_{\mu\nu}$ and the trace part $\phi$, and the above boundary condition simply imposes that $\phi_{\mu\nu}|_{\partial \CM_L}=0$ and $\phi|_{\partial \CM_L} =0$ on the equator. Since all eigenfunctions of the Laplacian on the hemisphere can be extended over the whole sphere,  by imposing the boundary condition on the equator, only the odd spherical harmonics survive. The constant $l=0$ mode is therefore removed immediately from the one-loop determinant.

For the $l=1$ multiplet, we can read out the eigenfunction by embedding the sphere in $\mathbb R^{D+1}$ as
\begin{equation}
    x_1^2+\cdots+x_{D+1}^2=1
\end{equation}
and placing the equator at $x_{D+1}=0$.  The general $l=1$ scalar harmonic \cite{RO84} can be expressed as
\begin{equation}
    \phi(x)=C_\mu x^\mu\,.
\end{equation}
Requiring $\phi|_{x_{D+1}=0}=0$ forces all coefficients to vanish except $C_{D+1}$. Thus, exactly one member of the $l=1$ multiplet survives under the fixed-equator boundary condition. The fixed-equator boundary condition effectively projects out $(D+1)$ negative modes from the trace sector and finally leads to a one-loop partition function with the following behavior on the hemisphere,
\begin{align}
    Z^{\text{1-loop}}_{\CM_L,\text{fixed-equator}} = \mp i\times \left(\text{positive number}\right)\,.
\end{align}
There is a $\mp i$ phase that does not depend on dimension $D$.

\paragraph{Gluing the two half spheres with consistent time directions}~\par

Let us then go back to the full sphere by gluing two fluctuating hemispheres together. We proceed in a way such that the difference from the fully gravitational analysis in section \ref{sec:sphere_partition_function} is transparent. 

On the full sphere, we expand the metric around the saddle-point geometry
\begin{equation}
    g_{\mu\nu}^{L/R}=\hat{g}_{\mu\nu}^{L/R}+\sqrt{32\pi G_N}h_{\mu\nu}^{L/R}\,,
\end{equation}
where the superscript $L/R$ stands for $\CM_L$ and $\CM_R$. The metric perturbations on both sides can be decomposed into the traceless part $\phi_{\mu\nu}^{L/R}$ and the trace part $\phi^{L/R}$. The boundary condition imposed on the interface $\Gamma$ requires that $\phi^{L/R}_{\mu\nu}|_\Gamma$ and $\phi^{L/R}|_\Gamma$ to vanish. The action quadratic in the metric perturbation is then
\begin{equation}
    I^{L/R} = \frac{1}{2}\int_{\CM_{L/R}}\,d^Dx\,\sqrt{\hat{g}^{L/R}}\,  \left(
        \phi_{\mu\nu}^{L/R}(-\hat{\nabla}^{2}+2)\phi^{L/R,\mu\nu}
        -\frac{1}{2}D(D-2)\phi^{L/R}(-\hat{\nabla}^{2}-2(D-1))\phi^{L/R}\right)\,.
\end{equation}
Compared with the quadratic action \eqref{eqn:quadraticaction} in the fully gravitational GPI, the action here is spanned by modes in $(L,R)$ pairs, so the one-loop correction to the partition function is the product of the contributions from $L$ and $R$ modes, i.e. 
\begin{equation}
    Z^{1-\text{loop}}_{\CM,\text{fixed-equator}} = Z^{1-\text{loop}}_{\CM_L,\text{fixed-equator}}Z^{1-\text{loop}}_{\CM_R,\text{fixed-equator}}\,.
\end{equation}
Each factor of the product contains a $\mp i$ phase with an undetermined overall sign, as commented before. 

As a result, a naive guess of the one-loop corrections to the sphere partition function with fixed equator would be 
\begin{equation}
    Z^{1-\text{loop}}_{\CM,\text{fixed-equator}} \overset{?}{=} (\pm 1)\times(\text{positive number})\,,
\end{equation}
which is manifestly real and the overall sign depends on the prescriptions of $\phi^\pm$ contour rotation. This result is, however, obtained when treating the choices of contours in $\CM_L$ and $\CM_R$ completely independently. Let us think more about this point. 

In \cite{Maldacena24}, a preferable contour prescription is suggested by analytically continuing the Newton's constant to
\begin{equation}
    \frac{1}{G_N}\rightarrow \frac{1}{|G_N|}(1-i\epsilon),
\end{equation}
which generates an infinitesimal imaginary part for $G_N$ and avoids the large $t$ growth in Lorentzian evolution. Once we choose this specific continuation of $G_N$, the contour rotation $\phi\rightarrow -i \hat{\phi}$ is selected. If we flip the sign of the $i\epsilon$-prescription above, the other direction of contour rotation will be the preferable one instead. This suggests us to consider if we should choose the opposite $i\epsilon$-prescriptions for $\CM_L$ and $\CM_R$. 

Indeed, this is the case. As shown in figure \ref{fig:split_two_hemispheres}, the interface $\Gamma$ on $\CM_L$ and $\CM_R$ wrap the Euclidean temporal direction $\tau$ in opposite ways. Therefore, when we analytically continue the Euclidean time $\tau$ to the Lorentzian time, different $i\epsilon$-prescriptions should be taken on each side. This choice of time directions will make the phase factors in $ Z^{1-\text{loop}}_{\CM_L,\text{fixed-equator}}$ and $ Z^{1-\text{loop}}_{\CM_R,\text{fixed-equator}}$ off by a sign, which leads to a positive definite one-loop partition function
\begin{equation}
    \boxed{Z^{1-\text{loop}}_{\CM,\text{fixed-equator}} = (-i)_{L} (i)_{R} \times (\text{positive number})= +1\times (\text{positive number})}\,.
\end{equation}
In this way, the partial frozen sphere partition function studied here receives no phase contribution from the 1-loop correction, in contrast to the fully gravitational one in \eqref{eq:sphere_phase_fully_gravitational}.

\section{Conclusion and discussion}
\label{sec:conclusion}

In this paper, we revisited the HH wave function in AdS spacetime, where natural spatial slices $\Sigma$ are open. This structure necessitates a piece of spacetime boundary $\CB_-$, which leaves us with a choice of whether to sum over the spacetime boundary configuration $\CB_-$ or not. We argued that the fully gravitational HH wave function, in which $\CB_-$ is summed over, is more akin to the original HH proposal. On the other hand, the partially frozen HH wave function, in which $\CB_-$ is fixed, is the one most studied in the existing literature, due to its connection to the AdS/CFT correspondence. For this reason, we studied the fully gravitational HH wave function, including explicit examples in AdS$_3$ Einstein gravity and AdS$_2$ JT gravity, and the phase associated with the one-loop computation in AdS$_D$, and made a comparison with the more conventional partially frozen HH wave function. Based on observations from this comparison, we also studied the one-loop aspects of a partially frozen sphere partition function in dS Einstein gravity.

The first explicit example of the fully gravitational HH wave function was studied in section \ref{sec:AdS3}. We considered AdS$_3$ Einstein gravity coupled to a spacetime boundary with tension $T>1$, such that the brane configurations are compact on shell. 
We considered relevant saddle points within a minisuperspace treatment, where the spatial slice $\Sigma$ is a hyperbolic disk with a finite cutoff. We identified the domain in minisuperspace in which real Euclidean saddle points exist, and described the complex saddle point that appears outside this domain. Based on this, we evaluated the semiclassical HH wave function. In particular, a closed-form formula is available when we further restrict to the case where the hyperbolic radius on $\Sigma$ is identical to the AdS radius in the bulk. We also considered a second example in AdS$_2$ JT gravity in section \ref{sec:JT-noboundary} and performed an analysis parallel to that in AdS$_3$. The analysis turns out to be slightly more complicated in JT gravity due to the presence of the dilaton field. We considered a one-parameter family of dilaton profiles that admit constant extrinsic curvature when on shell. This can be understood as a minisuperspace approximation in JT gravity. We also discussed the issue of how a contour should be chosen, and how the probability distribution interpretation should be built on a phase space where diffeomorphisms are partially broken. We note that the HH wave function in dS gravity with timelike boundaries has been recently studied in \cite{BBLST26}. It would be interesting to understand the relations to our setups. 

In Section \ref{sec:other_interpretations}, we discussed two alternative interpretations of the GPI studied here: the no-boundary wave function for spatial subregions and subregion generalizations of Cauchy slice holography. Both are interesting directions for future research.

There is also a natural holographic reading of our computation. 
Let us take the AdS$_3$ case as an example. 
Before the spacetime boundary is allowed to gravitate, AdS$_3$ is dual to a holographic CFT$_2$ via the AdS/CFT correspondence. Allowing the boundary geometry to fluctuate then couples the holographic CFT to gravity. This is the so-called brane world holography \cite{RS99-1,RS99-2,SMS99,GT99,Gubser99}. More precisely, we expect that AdS$_3$ gravity with a gravitating boundary can be effectively described by a 2D Liouville gravity coupled to a large-$c$ holographic CFT \cite{ST22BCFT,NSS24}. In the lower-dimensional picture, the two minisuperspace variables $(L,A)$ characterize the 2D gravitational sector and the 2D matter sector, respectively. From this perspective, it would be interesting to understand the connection between our analysis and \cite{ABM24,AHK25}, where a dS$_2$ cosmology, realized by coupling Liouville gravity to a matter CFT, is considered.

As a comparison between the fully gravitational HH wave function and the partially frozen one in AdS, in section \ref{sec:One_loop_norm_GPI}, we studied the one-loop correction to the hyperbolic ball partition function in $D$ dimensions, which is interpreted as the leading contribution to the norm of the HH wave function. We found that the one with the gravitating boundary has a dimension-dependent phase $(\mp i)^{D+1}$ arising from the boundary modes, whereas that with the fixed Dirichlet boundary is a positive number, as expected from the AdS/CFT correspondence. This result is natural from the brane world perspective, since our setup can be regarded as a codimension-1 sphere partition function via brane world holography, and the phase is identical to that in dS Einstein gravity. However, we should note that the result here is still nontrivial, not only because we are performing a different analysis, but also because, even from the lower-dimensional perspective, the brane-localized theory is not pure Einstein gravity, but in general a higher derivative gravity coupled to a holographic CFT. Given the fact that the one-loop phase depends on the details of the gravitational theory, as emphasized in \cite{GS26}, the direct evaluation in the brane world EFT remains nontrivial. 

The comparison between the two different HH wave functions in AdS motivated us to consider a partially frozen HH wave function in dS gravity. 
For this reason, we studied the one-loop correction to a sphere partition function with a fixed equator in section \ref{sec:paper2-fixed-equator}, which is interpreted as the norm of a partially frozen HH wave function in dS. We found that the phase vanishes in a nontrivial way. In particular, the consistency of the time orientation along the frozen equator played a central role in canceling the phase. 

Our frozen spacetime subregion is similar in flavor to Maldacena's observer \cite{Maldacena2024-2}, in the sense that both break the time reparameterization invariance of the gravitational sector. However, they work in different ways. Maldacena's observer is a quantum clock coupled to gravity, and the Hamiltonian constraint acts on the composite system. Therefore, the Hamiltonian constraint seems to be broken if one only looks at the gravity sector. On the other hand, our treatment is more brutal: the Hamiltonian constraint is broken by the presence of the frozen subregion, in the same way that an AdS boundary breaks the global Hamiltonian constraint \cite{Wei25}. See also \cite{NU25,NU26,BC26}. It would be interesting to study whether generic mechanisms breaking the time reparameterization invariance of the gravitational sector can resolve the phase issue in computing physical observables.\footnote{We thank Daniel Jafferis for discussions on this point.} 

Although we have only studied a very simple example of the partially frozen dS partition function in this paper, we expect that freezing any spacetime subregion that respects a notion of time-reflection symmetry will lead to a similar phase-canceling result. We would like to address this question in future work. 

One lesson one may draw from the results in this paper is that physically sensible predictions may need to be extracted from a partially frozen GPI, rather than from the fully gravitational one. However, unlike the AdS case, in dS or other closed universes, there does not exist a canonical choice of the frozen spacetime subregion. Therefore, one needs to answer the following three questions in order to proceed along this line: 1. What is the physical interpretation of the partially frozen GPI? 2. Freezing different subregions leads to different theories. How are they related to each other? 3. What is the exact theory defined from a partially frozen GPI beyond the AdS/CFT case? In \cite{Wei25}, we have qualitatively argued that a partially frozen GPI may be interpreted as defining observer-dependent holography. We will make this precise and answer question 2 in an upcoming paper \cite{Wei26}. We would also like to address question 3 in future work.

\section*{Acknowledgements}
We are grateful to Jordan Cotler, Daniel Jafferis, Quentin Lamouret, Tadashi Takayanagi and Joaquin Turiaci for useful discussions. GA is supported by the Fannie and John Hertz Foundation and NSF GRFP. ZW is supported by the Society of Fellows at Harvard University. MZ is supported by World Premier International Research Center Initiative (WPI Initiative), MEXT, Japan.

\appendix
\section{Analytic continuation of the Hartle--Hawking wave-function for \texorpdfstring{$L>L_{\text{crit}}(T)$}{l=1}}
\label{app:analytic_continuation_l=1}
In this subsection we explicitly perform the analytic continuation of the HH wave function in the restricted superspace $\ell=1$.
It is convenient to introduce the dimensionless boundary length
\begin{equation}
    y \;\equiv\; \frac{L}{2\pi},
\end{equation}
and to recall that $T=\coth u_0$ with $u_0>0$.
For $L>L_{\rm crit}(T)$ we have
\begin{equation}
    \kappa(L;T)\;\equiv\;\frac{\cosh u_0}{\sqrt{1+y^2}}\in(0,1),
    \qquad
    \chi(L;T)\;\equiv\;T\,\frac{y}{\sqrt{1+y^2}}\in(1,T).
    \label{eq:kappa-chi-def2}
\end{equation}
The first inequality is equivalent to $L>L_{\rm crit}(T)$ because $L_{\rm crit}(T)=2\pi\sinh u_0$ implies
$\sqrt{1+y^2}>\sqrt{1+\sinh^2u_0}=\cosh u_0$.
The second inequality follows from monotonicity of $y/\sqrt{1+y^2}$ and the fact that
$\chi(L_{\rm crit}(T);T)=1$.

\medskip
\noindent
\emph{Step 1: analytic continuation of $\arccosh$ and $\arcsin$.}
Since $\kappa\in(0,1)$, the inverse hyperbolic cosine is purely imaginary:
\begin{equation}
    \arccosh(\kappa)\;=\;\pm\, i\,\arccos(\kappa).
    \label{eq:acosh-cont2}
\end{equation}
To see this, define $\beta\equiv\arccos(\kappa)\in(0,\pi/2)$, so that $\cos\beta=\kappa$.
Then $\cosh(i\beta)=\cos\beta=\kappa$, i.e.\ $i\beta$ is a solution of $\cosh z=\kappa$.
On the principal branch of $\arccosh$ (continuously connected to the real axis from $\kappa>1$),
one takes $\arccosh(\kappa)=+\,i\beta$; the second choice $-i\beta$ corresponds to the complex-conjugate
saddle and will be reinstated at the end.

Similarly, for $\chi>1$ the inverse sine acquires an imaginary part.
A convenient analytic definition is
\begin{equation}
    \arcsin z \;\equiv\; -\,i\,\log\!\Big(i z+\sqrt{1-z^2}\Big),
\end{equation}
with the branch of the square root and logarithm fixed so that $\arcsin z\sim z$ as $z\to0$.
For real $\chi>1$, one has $\sqrt{1-\chi^2}= i\sqrt{\chi^2-1}$, and continuity across $\chi=1$
fixes
\begin{equation}
    \arcsin(\chi)\;=\;\frac{\pi}{2}\,\mp\, i\,\arccosh(\chi),
    \qquad (\chi>1),
    \label{eq:asin-cont2}
\end{equation}
since $\arcsin(1)=\pi/2$ and $\arccosh(1)=0$.
Here again the two signs correspond to the two complex-conjugate branches.

\medskip
\noindent
\emph{Step 2: complex on-shell action and its phase.}
Applying \eqref{eq:acosh-cont2}--\eqref{eq:asin-cont2} to \eqref{eq:IE-ell1-closed} yields, for $L>L_{\rm crit}(T)$,
a pair of complex saddle-point actions
\begin{equation}
    I_{\rm E}^{(\ell=1)}(L)\;=\; I_R^{(\ell=1)}(L)\ \pm\ i\,I_I^{(\ell=1)}(L),
    \label{eq:IE-ell1-complex-decomp2}
\end{equation}
with real and imaginary parts
\begin{equation}
\boxed{
\begin{aligned}
    I_R^{(\ell=1)}(L)
    &= -\frac{1}{4 G_N}\,u_0 \,,
    \\
    I_I^{(\ell=1)}(L)
    &= \frac{1}{4 G_N}\,\beta(L;T)\;-\;\frac{L}{8\pi G_N}\,\gamma(L;T)\,,
\end{aligned}}
\label{eq:IR-II-ell12}
\end{equation}
where we defined the two positive functions
\begin{equation}
    \beta(L;T)\;\equiv\;\arccos\!\Big(\kappa(L;T)\Big)
    =\arccos\!\Bigg(\frac{\cosh u_0}{\sqrt{1+(L/2\pi)^2}}\Bigg)\in(0,\tfrac{\pi}{2}),
    \label{eq:beta-def2}
\end{equation}
and
\begin{equation}
    \gamma(L;T)\;\equiv\;\arccosh\!\Big(\chi(L;T)\Big)
    =\arccosh\!\Bigg(T\,\frac{(L/2\pi)}{\sqrt{1+(L/2\pi)^2}}\Bigg)\in(0,\arccosh T).
    \label{eq:gamma-def2}
\end{equation}
At the transition point $L=L_{\rm crit}(T)$ we have $\kappa=\chi=1$ so $\beta=\gamma=0$ and the imaginary part vanishes,
in agreement with the existence of a real Euclidean saddle for $L\le L_{\rm crit}(T)$.

\section{Hartle--Hawking wave function for sphere topology}
The main text focuses on a HH wave function prepared on a 
spatial slice $\Sigma$ with boundary, in the presence of a spacetime boundary $\mathcal{B}$ whose boundary coincides with $\partial \Sigma$, in AdS$_D$ gravity. 
It is also useful to record the more standard ``closed'' version of the HH construction in AdS$_D$ gravity, where the final slice has the topology of $S^{D-1}$ and no spacetime boundary is present.
This case is closely related to the usual Euclidean AdS path integral corresponding to the CFT partition function on $S^{D-1}$ and has been discussed since early work on Euclidean preparations of AdS states \cite{Maldacena02}. 
This version of the HH wave function has been explicitly studied in 2D JT gravity in \cite{IKTV20}. In this appendix, we study it in 3D. 

\paragraph{Definition and minisuperspace ansatz}~\par
Let $\Sigma\simeq S^2$ be a closed two-manifold equipped with a metric $\gamma_{ab}$.  
The HH wave function we study is
\begin{equation}
    \Psi_{\rm HH}^{S^2}[\gamma]\equiv \int_{g|_{\Sigma}=\gamma}\frac{\mathcal D g}{\mathrm{Diff}}\,\exp\big[-I_{\rm E}[g]\big],
    \label{eq:HHsphereDef}
\end{equation}
where it is more convenient to write the Euclidean action with only bulk and GHY terms, without the spacetime boundary term and the corner term, 
\begin{equation}
    I_{\rm E}[g]= -\frac{1}{16 \pi G_N}\int_{\mathcal M}\!\sqrt g\,(R+2)
    -\frac{1}{8\pi G_N}\int_{\Sigma}\!\sqrt\gamma\,K_{\Sigma}.
    \label{eq:IEsphere}
\end{equation}
since $\Sigma$ is smooth and closed.

We consider the saddle point contribution under the following minisuperspace calculation. We take $\Sigma$ to be a round two-sphere of radius $a$,
\begin{equation}
    ds^2_{\Sigma}=a^2\,d\Omega_2^2,\qquad d\Omega_2^2=d\theta^2+\sin^2\theta\,d\phi^2.
\end{equation}
The symmetry suggests an $O(3)$-invariant bulk ansatz
\begin{equation}
    ds^2=d\chi^2+b(\chi)^2\,d\Omega_2^2,\qquad 0\le \chi\le \chi_0,
    \label{eq:sphereAnsatz}
\end{equation}
where $\chi=0$ is the place where the manifold caps off 
and $\chi=\chi_0$ is $\Sigma$.
Regularity at $\chi=0$ imposes
\begin{equation}
    b(0)=0,\qquad b'(0)=1,
\end{equation}
so that the geometry closes off smoothly.

\paragraph{Solving the equations of motion}~\par
For the metric \eqref{eq:sphereAnsatz} the nonzero components of the Ricci tensor are
\begin{equation}
    R_{\chi\chi}=-2\frac{b''}{b},\qquad
    R_{ij}=\Big(1-(b')^2-b\,b''\Big)\,\gamma^{S^2}_{ij},
\end{equation}
where $\gamma^{S^2}_{ij}$ is the unit-sphere metric.
The Einstein's equation $R_{\mu\nu}=-2g_{\mu\nu}$ then gives two ordinary differential equations.
The $\chi\chi$ component implies
\begin{equation}
    -2\frac{b''}{b}=-2\qquad\Longrightarrow\qquad b''=b.
    \label{eq:bppEq}
\end{equation}
The angular components give
\begin{equation}
    1-(b')^2-b\,b''=-2b^2.
\end{equation}
Using \eqref{eq:bppEq} this becomes the first integral
\begin{equation}
    (b')^2=1+b^2.
    \label{eq:bprimeEq}
\end{equation}
The unique solution of \eqref{eq:bppEq} and \eqref{eq:bprimeEq} obeying the regularity conditions is
\begin{equation}
    b(\chi)=\sinh\chi.
\end{equation}
Therefore the saddle is a hyperbolic three-ball (a region in $H^3$) with boundary at $\chi=\chi_0$.

It is also useful to make uniqueness completely explicit.  The general solution of \eqref{eq:bppEq} is
$b(\chi)=A\sinh\chi+B\cosh\chi$. Regularity at $\chi=0$ requires $b(0)=0$, which forces $B=0$.
Then $b'(0)=A=1$ fixes $A=1$.  Hence the regular $O(3)$-invariant Einstein filling is unique and is given by
$b(\chi)=\sinh\chi$.

Matching the induced metric at the boundary gives
\begin{equation}
    a=b(\chi_0)=\sinh\chi_0\qquad\Longleftrightarrow\qquad \chi_0=\operatorname{arcsinh}a.
    \label{eq:chi0froma}
\end{equation}

\paragraph{On-shell action and the semiclassical wave function}~\par 
Evaluating \eqref{eq:IEsphere} on the saddle gives an expression for the minisuperspace wave function. On shell $R=-6$, so the bulk term reduces to a volume:
\begin{equation}
    -\frac{1}{16 \pi G_N}\int_{\mathcal M}\!\sqrt g\,(R+2)=\frac{1}{4\pi G_N}\,\mathrm{Vol}(\mathcal M).
\end{equation}
For \eqref{eq:sphereAnsatz} with $b(\chi)=\sinh\chi$ the volume is
\begin{equation}
\begin{aligned}
    \mathrm{Vol}(\mathcal M)
    &=\int_0^{\chi_0}\!d\chi\,4\pi\,\sinh^2\chi
    =4\pi\int_0^{\chi_0}\!d\chi\,\frac{\cosh 2\chi-1}{2}
    =\pi\,\sinh 2\chi_0-2\pi\,\chi_0.
\end{aligned}
\label{eq:VolBall}
\end{equation}
The GHY term involves the extrinsic curvature of the surface $\chi=\chi_0$. We have
\begin{equation}
    K_{ab}|_{\Sigma}=\frac12\mathcal L_n\gamma_{ab}=\frac{b'}{b}\,\gamma_{ab}=\coth\chi_0\,\gamma_{ab},\qquad
    K_{\Sigma}=2\coth\chi_0.
\end{equation}
Since $\sqrt\gamma=b(\chi_0)^2\sqrt{\gamma_{S^2}}=\sinh^2\chi_0\sqrt{\gamma_{S^2}}$, we obtain
\begin{equation}
    -\frac{1}{8\pi G_N}\int_{\Sigma}\!\sqrt\gamma\,K_{\Sigma}
    =-\frac{1}{8\pi G_N}\,4\pi\,\sinh^2\chi_0\,(2\coth\chi_0)
    =-\frac{\pi}{2G_N}\,\sinh 2\chi_0.
    \label{eq:GHYSphere}
\end{equation}
Combining \eqref{eq:VolBall} and \eqref{eq:GHYSphere}, the on-shell Euclidean action is
\begin{equation}
\begin{aligned}
    I_{\rm E}^{S^2}(\chi_0)
    &=\frac{1}{4\pi G_N}\Big(\pi\sinh 2\chi_0-2\pi\chi_0\Big)-\frac{\pi}{2G_N}\sinh 2\chi_0
    =-\frac{\pi}{4\pi G_N}\,\sinh 2\chi_0-\frac{\pi}{2G_N}\,\chi_0.
\end{aligned}
\label{eq:IEsphereOnShell}
\end{equation}
Using \eqref{eq:chi0froma} and $\sinh 2\chi_0=2\sinh\chi_0\cosh\chi_0=2a\sqrt{1+a^2}$, we can write this as a function of the boundary radius:
\begin{equation}
    I_{\rm E}^{S^2}(a)=-\frac{\pi}{2G_N}\Big(a\sqrt{1+a^2}+\operatorname{arcsinh}a\Big).
    \label{eq:IEsphereOfa}
\end{equation}
The semiclassical HH wave function in this minisuperspace is therefore
\begin{equation}
    \Psi_{\rm HH}^{S^2}(a)\ \approx\ \exp\Big[-I_{\rm E}^{S^2}(a)\Big]
    =\exp\left[\frac{\pi}{2G_N}\Big(a\sqrt{1+a^2}+\operatorname{arcsinh}a\Big)\right].
\end{equation}
Note that, for large $a$, the action behaves as
\begin{equation}
    I_{\rm E}^{S^2}(a)= -\frac{\pi}{2G_N}\Big(a^2+\log(2a)+\frac12+\mathcal O(a^{-2})\Big),
\end{equation}
so the unrenormalized wave function grows rapidly with the boundary size.  
If one instead takes $a\to\infty$ while fixing only the \emph{conformal} class of the boundary metric (as in AdS/CFT), one should add the standard local counterterms on $\Sigma$ to define a renormalized action.

\section{Constant \texorpdfstring{$K$}{K} curve on the Poincar\'{e} disk}
\label{sec:generalKcurve}

The saddle-point geometry in Section \ref{sec:JT-noboundary} must be a subregion of the Euclidean 
AdS$_2$ space bounded by two curves $\Sigma$ and $\mathcal{B}_{-}$, and we focus on the subclass of curves with constant extrinsic curvature, so it is useful to work out the general solutions of constant-$K$ curves on the Poincar\'{e} disk. This is the purpose of this appendix.

It turns out to be convenient to use the Poincar\'e disk coordinate $(r,\theta)$ of the Euclidean AdS$_2$ with the metric \eqref{eqn:Poincaredisk metric}
\begin{equation}
    ds^2 = \frac{4(dr^2+r^2d\theta^2)}{(1-r^2)^2}\,,
\end{equation} where $r$ is between $0$ and $1$, and $\theta$ takes a value between $0$ and $2\pi$. The Poincar\'e disk can be embedded into the two-dimensional plane $(x,y) \in \mathbb{R}^2$ via the map $x=r\cos{\theta},\,y=r\sin{\theta}$ as a unit disk $(x^2+y^2\leq 1)$. Depending on the values of the extrinsic curvature $K$, these curves can be classified into the following categories:
 \begin{itemize}
     \item \textbf{Geodesics} ($K=0$): The geodesic lines on the Poincar\'{e} disk can be constructed from a circle in $\mathbb{R}^2$ that intersects the boundary of the Poincar\'e disk orthogonally. The arc of that circle within the Poincar\'e disk is a geodesic.
     
     \item \textbf{Hypercycles} ($0<|K|<1$): The hypercycles can be constructed similarly by intersecting a circle in $\mathbb{R}^2$ non-orthogonally with the boundary of the Poincar\'e disk. Note that the extrinsic curvature of the hypercycle is equal to $\pm \cos{\varphi}$ (the choice of the sign $\pm$ depends on whether the normal vector is outward or inward), where $\varphi$ is the intersection angle between the hypercycle curve and the boundary of the Poincar\'e disk in $\mathbb{R}^2$. 
     
     \item \textbf{Horocycles} ($|K|=1$): Horocycles are circles in $\mathbb{R}^2$ that are within the Poincar\'{e} disk and tangent to the disk boundary. They meet with the boundary at exactly one point except that the boundary of disk itself, as a special case of horocycle, also has $K=1$. 
     
     \item \textbf{Hyperbolic Circles} ($|K|>1$): Hyperbolic circles are circles in $\mathbb{R}^2$ that are completely contained in the Poincar\'e disk without any intersections with the boundary.
 \end{itemize}

 Since all constant-$K$ curves come from circles in the two-dimensional plane $\mathbb{R}^2$, 
 in the following discussion, we will use the two parameter $(a,R)$ to label the constant-$K$ curves. $a$ stands for the position of the center of the circle in $\mathbb{R}^2$ and $R$ is the corresponding radius. In principle, the position of the center should consist of two variables, while without loss of generality, we consider the case where the center lies on the axis $\theta = 0$. All other cases will be equivalent up to a rotation, which is an isometry\footnote{Notice that when we apply an isometric transformation, the dilaton field will transform covariantly. Explicitly, the rotation of the coordinate will also rotate the vector $(C^1,C^2)$ in the dilaton solution \eqref{eqn:dilatonsolution}.}. In the polar coordinate of $\mathbb{R}^2$, the outward\footnote{In Section \ref{sec:JTminsuperspaceansatzandmaximallybubble}'s discussion, we see that the listed normal vector points outward on the spatial slice, while it points inward on the spacetime boundary. To evaluate the on-shell action, we always select the normal vector pointing outward in order to be consistent with the convention of Gauss-Bonnet term in \eqref{eqn:JTaction}.} normal vector of the $(a,R)$ circle can be expressed in terms of $r$ and $\theta$ as follows
 \begin{equation}
 \label{eqn:normalvector}
     n^r = \frac{(a\cos{\theta}-r)(1-r^2)}{2R},\, n^\theta = \frac{-a\sin{\theta}(1-r^2)}{2rR},
 \end{equation}
 where $R = \sqrt{a^2+r^2-2ar\cos{\theta}}$. The normal vector above has been normalized with respect to the Poincar\'e disk metric \eqref{eqn:Poincaredisk metric}. The extrinsic curvature can be computed via the formula
 \begin{equation}
 \label{eqn:Kformula}
     K=\nabla_\mu n^\mu = \text{sgn}(r^2-1)\frac{1-a^2+R^2}{2R}\equiv\text{sgn}(r^2-1)\cos{\varphi}.
 \end{equation}
 Some observations: We only consider the arc inside the Poincar\'e disk, i.e. $r^2<1$. For $a>1$, the circle intersects the disk only if $a-1<R<a+1$. In this range, $K$ goes between $-1$ and $1$
 \begin{equation}
     \begin{cases}
         K>0 & a-1<R<\sqrt{a^2-1}\\
         K=0 & R=\sqrt{a^2-1}\\
         K<0 & \sqrt{a^2-1}<R<a+1
     \end{cases}.
 \end{equation}
For $a<1$, the radius of the circle ranges between $0$ and $a+1$. In particular, when the radius $R<1-a$, the circle is completely within the disk. The extrinsic curvature $K$ with respect to the normal vector \eqref{eqn:normalvector} is
\begin{equation}
    \begin{cases}
        K<-1 & 0<R<1-a\\
        K=-1 & R=1-a\\
        -1<K<0 & 1-a<R<1+a
    \end{cases}.
\end{equation}
 
 The equation of this curve can be summarized as
 \begin{equation}
 \label{eqn:curveequation1}
     \cos{\theta} = \frac{a^2+r^2-R^2}{2ar}.
 \end{equation}
 The evaluation of the proper length is simpler in the Schwarzschild (Rindler) coordinate \eqref{eqn:EuclideanSchwarzschild}. We can translate the above curve equation into the Schwarzschild coordinate via the coordinate transformation \eqref{eqn:poincaretorindler}
 \begin{equation}
 \label{eqn:curveequation2}
     \cos{r_s t_E} = \frac{a^2-R^2+\frac{r_{\text{Sch}}-r_s}{r_{\text{Sch}}+r_s}}{2 a\sqrt{\frac{r_{\text{Sch}}-r_s}{r_{\text{Sch}}+r_s}}}\,.
 \end{equation}
 Note that the constant-$K$ curve in the Schwarzchild (Rindler) coordinate is no longer circular.
 
 For fixed $K$,  the radius and the position of the circle in $\mathbb{R}^2$ can be expressed in terms of $K$ and $r_0$
 \begin{equation}
     R= \frac{1-r_0^2}{2(r_0-K)},\, a=R+r_0.
 \end{equation}
 $r_0$ is the distance from the origin of $\mathbb{R}^2$ to the closest point on the circle. Note that we only consider the case $a>R$. For the case $a<R$, we can simply flip the sign ahead of $r_0$ and redo the computation. Also, the change of the direction of the normal vector can be addressed by changing the sign of $K$. In Schwarzschild (Rindler) coordinate, we have $r_{\text{Sch},0} = r_s\frac{1+r_0^2}{1-r_0^2}$. The relation between $r_{\text{Sch}}$ and the affine parameter $\lambda$ is given by
 \begin{equation}
 \label{eqn:Kcurve}
     r_{\text{Sch}}(\lambda) = r_{\text{Sch},0}+\frac{(r_{\text{Sch},0}\mp K\sqrt{r_{\text{Sch},0}^2-r_s^2})(-1+\cosh{(\sqrt{1-K^2}\lambda)})}{1-K^2},\,(|K|<1)
 \end{equation}
 where $r_{\text{Sch}}(0)=r_{\text{Sch},0} = \sqrt{A^2+r_s^2}$. The sign $\mp$ comes from the distinction between two cases $a>R$ and $a<R$. When $K=0$, the above formula reduced to 
 \begin{equation}
     r_{\text{Sch}}(\lambda) = \sqrt{A^2+r_s^2}\cosh{\lambda},
 \end{equation}
 which agrees with the geodesic solution in the Euclidean AdS$_2$ space. The result can be simply generalized to $|K|\ge 1$ via analytical continuation. In the special case $|K| =1$, the expression becomes
 \begin{equation}
     r_{\text{Sch}}(\lambda) =r_{\text{Sch},0}+(r_{\text{Sch},0}\pm A)\frac{\lambda^2}{2}.
 \end{equation}
 Note that for $|K|\leq 1$, the affine parameter $\lambda$ can extend all the way to infinity. However, the hyperbolic $\cosh$-function becomes the cosine function for $|K|>1$, which suggests that the proper circumference of $|K|>1$ circle in the Poincar\'e coordinate is finite. In fact, all circles with the same $|K|>1$ have the the same proper circumference
 \begin{equation}
     \lambda_{\text{circ}} = \frac{2\pi}{\sqrt{K^2-1}},
 \end{equation}
 regardless of the position of the center.

\section{Minisuperspace ansatz for \texorpdfstring{$\phi(\lambda)$}{philambda} on \texorpdfstring{$\Sigma$}{Sigma}}
\label{sec:Cauchycalculation}
In Section \ref{sec:JT-noboundary}, we evaluate the HH wavefunction in terms of the proper length and the dilaton profile as the spatial data imposed on the spatial slice $\Sigma$. Although the proper length and the dilaton profile can be arbitrarily chosen, we restrict our discussion to a one-parameter family of the dilaton profile, which is labeled by the parameter that becomes the extrinsic curvature of the spatial slice in the on-shell geometry. 

This restricted family of the dilaton profile is regarded as the JT minisuperspace ansatz for $\phi(\lambda)$. In Section \ref{sec:JT-noboundary}, we have worked in this minisuperspace without explicitly writing down the expression of the dilaton profile, since it was not necessary.

For completeness, in this appendix, we fill this gap and aim to derive the explicit ansatz for $\phi(\lambda)$ in terms of the minisuperspace parameter $K$ (on-shell extrinsic curvature) step by step. 
First, we determine the endpoint coordinates of the length-$L$ spatial slice $\Sigma$ in terms of the geometrical parameters
 $(K,L,r_0)$ that pin down a unique constant-K curve introduced in Appendix \ref{sec:generalKcurve}, and then we solve for the coefficients $(C_0,C_1,C_2)$ in the dilaton solution \eqref{eqn:dilatonsolution} that reproduce the three boundary data
 $(\phi_l,\phi_r,\phi_0)$ of the dilaton profile defined in \eqref{eq:boundary_condition_dilaton}, which represent the dilaton values at the endpoints and the middle point of the spatial slice $\Sigma$ respectively. Finally, we write down the minisuperspace ansatz for $\phi(\lambda)$ for the constant-$K$, length-$L$ spatial slice $\Sigma$.

\subsubsection*{Endpoints of the symmetric length-$L$ segment}

We focus on the one--parameter family of constant--extrinsic--curvature curves (``hypercycles'') with $|K|<1$ described in Appendix \ref{sec:generalKcurve}, and we
parametrize the curve by its proper length $\lambda$ so that $\lambda=0$ is the point closest to the Poincar\'e disk origin.
In Schwarzschild (Rindler) coordinates this turning point is at
\begin{equation}
    r_{\rm Sch}(0)=r_{{\rm Sch},0}\,,\qquad t_E(0)=0\,,
\end{equation}
and by construction the curve is invariant under the reflection $\lambda\mapsto-\lambda$ combined with $t_E\mapsto-t_E$.
The radial coordinate along the curve is given in \eqref{eqn:Kcurve}:
\begin{equation}
\label{eq:JT-rSch-lambda}
    r_{\rm Sch}(\lambda)=r_{{\rm Sch},0}+\frac{\bigl(r_{{\rm Sch},0}\mp K\sqrt{r_{{\rm Sch},0}^2-r_s^2}\bigr)\bigl(\cosh(\alpha\lambda)-1\bigr)}{\alpha^2}\,,
    \qquad \alpha\equiv\sqrt{1-K^2}\,.
\end{equation}
For a finite spatial slice of proper length $L$ we restrict to $\lambda\in[-L/2,L/2]$.
Since \eqref{eq:JT-rSch-lambda} depends on $\lambda$ only through $\cosh(\alpha\lambda)$, the two endpoints have the same radial
coordinate,
\begin{equation}
\label{eq:JT-rSch-endpoints}
    r_{\rm Sch}\Bigl(\pm\frac{L}{2}\Bigr)=r_{{\rm Sch},*}
    \equiv r_{{\rm Sch},0}+\frac{\bigl(r_{{\rm Sch},0}\mp K\sqrt{r_{{\rm Sch},0}^2-r_s^2}\bigr)\bigl(\cosh(\alpha L/2)-1\bigr)}{\alpha^2}\,.
\end{equation}
The Euclidean times are opposite,
\begin{equation}
\label{eq:JT-tE-endpoints-sym}
    t_E\Bigl(\pm\frac{L}{2}\Bigr)=\pm t_{E,*}\,,
\end{equation}
and we now determine $t_{E,*}$ explicitly.

To pass to Poincar\'e disk coordinates we use
\begin{equation}
\label{eq:JT-Sch-to-disk}
    r_{\rm Sch}=r_s\,\frac{1+r^2}{1-r^2}\,,\qquad r^2=\frac{r_{\rm Sch}-r_s}{r_{\rm Sch}+r_s}\,,
    \qquad \theta=r_s t_E\,.
\end{equation}
In particular, the turning point $r_0\equiv r(0)$ and $r_{{\rm Sch},0}$ are related by
\begin{equation}
\label{eq:JT-r0-rSch0}
    r_0^2=\frac{r_{{\rm Sch},0}-r_s}{r_{{\rm Sch},0}+r_s}\,,\qquad
    \frac{r_{{\rm Sch},0}}{r_s}=\frac{1+r_0^2}{1-r_0^2}\,.
\end{equation}
Similarly, the endpoint radius is
\begin{equation}
\label{eq:JT-rstar-from-rSchstar}
    r_*^2\equiv r^2\Bigl(\frac{L}{2}\Bigr)=\frac{r_{{\rm Sch},*}-r_s}{r_{{\rm Sch},*}+r_s}.
\end{equation}
It is useful to eliminate $r_{{\rm Sch},0}$ in favor of $r_0$ using \eqref{eq:JT-r0-rSch0}.
Substituting \eqref{eq:JT-rSch-lambda} into \eqref{eq:JT-Sch-to-disk} and simplifying gives a closed form for the disk radius along
the curve:
\begin{equation}
\label{eq:JT-r-lambda-closed}
    r(\lambda)^2
    =\frac{2(\alpha r_0)^2+\bigl(1+r_0^2\mp 2Kr_0\bigr)\bigl(\cosh(\alpha\lambda)-1\bigr)}{2\alpha^2+\bigl(1+r_0^2\mp 2Kr_0\bigr)\bigl(\cosh(\alpha\lambda)-1\bigr)}.
\end{equation}
In particular,
\begin{equation}
\label{eq:JT-rstar}
    r_*^2
    =\frac{2(\alpha r_0)^2+\bigl(1+r_0^2\mp 2Kr_0\bigr)\bigl(\cosh(\alpha L/2)-1\bigr)}{2\alpha^2+\bigl(1+r_0^2\mp 2Kr_0\bigr)\bigl(\cosh(\alpha L/2)-1\bigr)}.
\end{equation}
This exhibits the expected limits: $r(0)=r_0$ and $r(\lambda)\to1$ as $|\lambda|\to\infty$.

The constant-$K$ curve is an arc of a Euclidean circle in the Poincar\'e disk.
Eliminating the circle parameters $(a,R)$ in favor of $(K,r_0)$ using
$R=(1-r_0^2)/(2(-K\pm r_0))$ and $a=R\pm r_0$,
its equation can be written directly as a relation between $r$ and $\theta$:
\begin{equation}
\label{eq:JT-cos-theta-of-r}
    \cos\theta
    =\frac{-K\,(r^2+r_0^2)\pm r_0\,(r^2+1)}{r\,(1\mp 2Kr_0+r_0^2)}\,.
\end{equation}
Along the symmetric segment we take $\theta(0)=0$ (or $\theta(0)=\pi$ for $a<R$) and choose the branch for which $\theta(\lambda)$ is an odd function with respect to $\theta=0$ or $\pi$, so that
$\theta(\pm L/2)=\pm\theta_*$ or $\theta(L/2)=2\pi-\theta(-L/2)=\theta_*$.  Evaluating \eqref{eq:JT-cos-theta-of-r} at $r=r_*$ gives
\begin{equation}
\label{eq:JT-theta-star}
    \theta_*\equiv\theta\Bigl(\frac{L}{2}\Bigr)
    =\arccos\!\left[\frac{-K\,(r_*^2+r_0^2)\pm r_0\,(r_*^2+1)}{r_*\,(1\mp 2Kr_0+r_0^2)}\right]\,,
    \qquad
    t_{E,*}=\frac{\theta_*}{r_s}\,.
\end{equation}
Equations \eqref{eq:JT-rSch-endpoints} and \eqref{eq:JT-tE-endpoints-sym}--\eqref{eq:JT-theta-star} pin down the relation
between the endpoint coordinates $(r_{\rm Sch},t_E)$ and the intrinsic parameters $(K,L,r_{{\rm Sch},0})$ (or equivalently $(K,L,r_0)$).

\subsubsection*{Solving for $C_0,C_1,C_2$ and the profile $\phi(\lambda)$}

On Euclidean AdS$_2$ in the Poincar\'e disk coordinate the general JT dilaton solution can be written as a linear function on the embedding
hyperboloid,
\begin{equation}
\label{eq:JT-dilaton-embedding}
    \Phi(X)=C_\mu X^\mu\,,
\end{equation}
with
\begin{equation}
\label{eq:JT-embedding-coords}
    X^0=\frac{1+r^2}{1-r^2}\,,\qquad
    X^1=\frac{2r}{1-r^2}\sin\theta\,,\qquad
    X^2=\frac{2r}{1-r^2}\cos\theta\,.
\end{equation}
Restricting to the spatial segment yields a one--dimensional profile
\begin{equation}
\label{eq:JT-phi-lambda-general}
    \phi(\lambda)=C_0\,X^0(\lambda)+C_1\,X^1(\lambda)+C_2\,X^2(\lambda)\,,
\end{equation}
where $r(\lambda)$ is given by \eqref{eq:JT-r-lambda-closed} and $\theta(\lambda)$ is fixed implicitly by
\eqref{eq:JT-cos-theta-of-r} (with the branch choice described above).

We now impose the three boundary conditions \eqref{eq:boundary_condition_dilaton} on the dilaton profile.
At $\lambda=0$ we have $(r,\theta)=(r_0,0)$, so $X^1(0)=0$ and
\begin{equation}
\label{eq:JT-bc-midpoint}
    \phi_0=C_0\,X^0_0+C_2\,X^2_0\,,
    \qquad
    X^0_0\equiv X^0(r_0)=\frac{1+r_0^2}{1-r_0^2}\,,\qquad
    X^2_0\equiv X^2(r_0,0)=\frac{2r_0}{1-r_0^2}\,.
\end{equation}
At the endpoints $\lambda=\pm L/2$ we have the same radius $r_*$ and opposite angles $\pm\theta_*$, hence
$X^0(\pm L/2)=X^0_*$ and $X^2(\pm L/2)=X^2_*$ while $X^1(\pm L/2)=\pm X^1_*$.  Explicitly,
\begin{equation}
\label{eq:JT-Xstar-defs}
    \begin{aligned}
        X^0_*\equiv X^0(r_*)&=\frac{1+r_*^2}{1-r_*^2},\\
        X^1_*\equiv X^1(r_*,\theta_*)&=\frac{2r_*}{1-r_*^2}\sin\theta_*,\\
        X^2_*\equiv X^2(r_*,\theta_*)&=\frac{2r_*}{1-r_*^2}\cos\theta_*.
    \end{aligned}
\end{equation}
The endpoint conditions \eqref{eq:boundary_condition_dilaton} become
\begin{equation}
\label{eq:JT-bc-endpoints}
    \phi_l=C_0\,X^0_*+C_1\,X^1_*+C_2\,X^2_*,\qquad
    \phi_r=C_0\,X^0_*-C_1\,X^1_*+C_2\,X^2_*.
\end{equation}
Subtracting and adding these equations gives immediately
\begin{equation}
\label{eq:JT-C1-solved}
    C_1=\frac{\phi_l-\phi_r}{2X^1_*},
    \qquad
    \frac{\phi_l+\phi_r}{2}=C_0\,X^0_*+C_2\,X^2_*.
\end{equation}
Together with \eqref{eq:JT-bc-midpoint}, this determines $(C_0,C_2)$ uniquely provided the determinant
$\Delta\equiv X^0_0X^2_*-X^0_*X^2_0$ is nonzero.  Solving the $2\times2$ system yields
\begin{equation}
\label{eq:JT-C0C2-solved}
    C_0=\frac{\phi_0 X^2_* - \frac{\phi_l+\phi_r}{2}X^2_0}{\Delta},
    \qquad
    C_2=\frac{\frac{\phi_l+\phi_r}{2}X^0_0-\phi_0 X^0_*}{\Delta},
    \qquad
    \Delta\equiv X^0_0X^2_*-X^0_*X^2_0.
\end{equation}
Equations \eqref{eq:JT-C1-solved} and \eqref{eq:JT-C0C2-solved} are the desired expressions for the three integration constants
$(C_0,C_1,C_2)$ in terms of the boundary data $(\phi_l,\phi_r,\phi_0)$ and the geometric parameters $(L,K,r_0)$.
The dependence on $(L,K,r_0)$ is entirely through $(r_*,\theta_*)$ determined in
\eqref{eq:JT-rstar} and \eqref{eq:JT-theta-star}.

Substituting back into \eqref{eq:JT-phi-lambda-general} gives the full dilaton profile along the segment.
It is convenient to present the result in a form that makes linearity in the boundary data manifest.
Define $X^\mu(\lambda)$ by \eqref{eq:JT-embedding-coords} evaluated on the curve, and denote $\phi_{\rm av}\equiv(\phi_l+\phi_r)/2$.
Then \eqref{eq:JT-C1-solved}--\eqref{eq:JT-C0C2-solved} imply
\begin{equation}
\label{eq:JT-phi-profile-final}
\boxed{\;
\begin{aligned}
    \phi(\lambda)
    &=\frac{\phi_l-\phi_r}{2X^1_*}\,X^1(\lambda)\\
    &\quad
    +\frac{\phi_0}{\Delta}\Bigl(X^2_*X^0(\lambda)-X^0_*X^2(\lambda)\Bigr)\\
    &\quad
    +\frac{\phi_{\rm av}}{\Delta}\Bigl(X^0_0X^2(\lambda)-X^2_0X^0(\lambda)\Bigr),\qquad 
    \Delta=X^0_0X^2_*-X^0_*X^2_0.
\end{aligned}
\;}
\end{equation}
This expression should be understood together with the explicit relations
\eqref{eq:JT-r-lambda-closed} and \eqref{eq:JT-cos-theta-of-r}, which determine $r(\lambda)$ and $\theta(\lambda)$ for fixed $(K,r_0)$.
The first term in \eqref{eq:JT-phi-profile-final} is odd under $\lambda\mapsto-\lambda$ and controls the antisymmetric difference
$\phi_l-\phi_r$, while the remaining terms are even and depend only on $(\phi_l+\phi_r,\phi_0)$, as expected from the reflection symmetry of the chosen interval.

\section{One-loop aspects of AdS}
In this appendix, we summarize some aspects of the one-loop contribution in AdS spacetime in addition to the analysis of section \ref{sec:One_loop_norm_GPI}. In the first subsection \ref{appendix:1loop_without_boundary_AdSd}, we show that the one-loop partition function in AdS with Dirichlet boundary conditions does not have negative modes at one loop. We then clarify what we mean by Neumann boundary conditions in \ref{app:NeumanBC_clarified}, before explaining how to gauge-fix the action with a dynamical spacetime boundary in \ref{app:gauge_fixing}. The last subsection, \ref{appendix:solving_for_the_tensor_mdoes}, is a technical note on how to solve the tensor sector of the linearized Einstein equations following \cite{HHR00}, this is a complement of the analysis of \ref{sec:solving_the_bulk_equations}.

\subsection{The one-loop correction with Dirichlet boundary condition}
\label{appendix:1loop_without_boundary_AdSd}

In this appendix, we briefly review the one-loop correction to the partition function in AdS$_D$ with Dirichlet boundary conditions at the conformal boundary\cite{GMY08,Suzuki2021,ST25,Sun2020,GKT14}. We particularly focus our attention on the number of negative modes and hence the phase of the one-loop partition function. We perturb the action up to second order around the metric of Euclidean AdS$_D$ that we denote $\hat{g}^{(0)}_{\mu\nu}$:
\begin{equation}
    g_{\mu\nu}=\hat g_{\mu\nu}^{(0)}+\sqrt{32\pi G_N}\,h_{\mu\nu}.
\end{equation}
We fix de Donder gauge by inserting the following gauge-fixing term:

\begin{equation}
    I_{\text{gauge-fixing}}
    = \frac{1}{2}\int d^Dx\, \sqrt{\hat{g}^{(0)}}\,
    \left(\hat{\nabla}^\mu h_{\mu\nu}-\frac{1}{2}\hat{\nabla}_\nu h\right)^2\,.
\end{equation}

Decomposing the metric perturbation into its traceless part $\phi_{\mu\nu}$ and trace part $h$ gives the standard quadratic action :
\begin{equation}
    I_{\text{E}}+I_{\text{gauge-fixing}}
    = \frac{1}{2}\int d^Dx\,\sqrt{g^{(0)}}\,
    \left(
        \phi_{\mu\nu}(-\hat{\nabla}^2+2)\phi^{\mu\nu}
        -\frac{1}{2}D(D-2)h(-\hat{\nabla}^2+2(D-1))h\right)\,.
\end{equation}
The trace part has the wrong sign for the kinetic term, an ultralocal redefinition of $h\rightarrow ih$ leads to the correct sign. We obtain the following action :
\begin{equation}
\label{eq:AdS_at_one_loop}
    S_{\text{EH}}+S_{\text{gauge-fixing}}
    = \frac{1}{2}\int d^Dx\,\sqrt{g^{(0)}}\,
    \left(
        \phi_{\mu\nu}(-\hat{\nabla}^2+2)\phi^{\mu\nu}
        +\frac{1}{2}D(D-2)h(-\hat{\nabla}^2+2(D-1))h\right)\,.
\end{equation}
The gauge fixing procedure leads to ghost terms:
\begin{equation}
    I_{\text{ghost}} = \frac{1}{2}\int d^Dx\,\sqrt{g^{(0)}}\, b_\mu(-\hat{\nabla}^2+(D-1))c^\mu\,.
\end{equation}

The goal is now to compute the number of negative modes in this action. As reviewed in \ref{sec:sphere_partition_function}, the negative modes come from the trace sector. The differential operator for dS (left) compare to AdS (right) is:
\begin{equation}
    -\hat{\nabla}_{{\rm sphere}}^2-2(D-1)h\,,\quad -\hat{\nabla}_{{\rm EAdS}}^2+2(D-1)h\,.
\end{equation}
The difference of sign in between the two expressions makes all the difference \cite{Sun2020}, dS has $D+2$ negative modes, while AdS has none. The traceless part of the one-loop effective action does not have negative modes in both cases. Therefore the one-loop partition function for AdS with Dirichlet boundary condition is positive and does not suffer from the phase issue.

\subsection{What people mean by ``Neumann boundary conditions"}
\label{app:NeumanBC_clarified}

In this subsection we make a clarification on what is meant by ``Neumann boundary conditions", since it can mean different things in different contexts. 
The Gibbons-Hawking-York term in \eqref{eq:EuclActionFull} is the boundary term that makes the Dirichlet problem (fixed induced metric $h_{ij}$) well posed. Varying \eqref{eq:EuclActionFull} with respect to $h_{ij}$ gives:

\begin{equation}
\label{eq:deltaI_Q_def}
\delta I_{\rm E}\big|_{\mathcal{B}}
=-\frac{1}{16\pi G_N}\int_{\mathcal{B}}\!\sqrt h\,\Big(K_{ij}|_{\mathcal{B}}-K_{\mathcal{B}} h_{ij}+T h_{ij}\Big)\,\delta h^{ij}\,.
\end{equation}
In order to satisfy the variational principle, besides setting the Dirichlet boundary condition $\delta h^{ij} = 0 $, we can as well impose the boundary EOM as the boundary condition 
\begin{align}\label{eq:NBC_appendixF}
    K_{ij}|_{\mathcal{B}}-K_{\mathcal{B}} h_{ij}+T h_{ij} = 0. 
\end{align}
This is what we mean by the Neumann boundary condition in the main text, and in most of the literature (see e.g. \cite{CM08, Takayanagi11}). 
However, it is important to note that 
this condition (as well as \eqref{eq:braneEOM_TildePi0}  is a boundary equation of motion but not a choice of boundary data, so it is not on the same footing with the Dirichlet boundary condition $\delta h^{ij} = 0 $. In the main text, we consider a fully gravitized boundary, which means we do not impose any boundary condition off shell, and on the spacetime boundary $\mathcal{B}$ the induced metric $h_{ij}$ is dynamical. The Neumann boundary condition \eqref{eq:NBC_appendixF} arises as a consequence of the saddle point approximation.

By contrast, if we think ``Neumann boundary condition" as something fixing the momentum conjugate to $h_{ij}$, 
then it should fix the Brown-York momentum $\Pi_{ij}$ (the momentum conjugate to $h_{ij}$) as boundary data $\delta \Pi_{ij} =0$. To implement this version of the Neumann boundary condition in the path integral, one needs to engineer the action such that the variational principle leads to \cite{KR16,Witten18} :
\begin{equation}
    \delta I_{\rm E}\big|_{\mathcal{B}}=-\frac{1}{16\pi G_N}\int_{\mathcal{B}}\sqrt{h}~ \delta(\Pi_{ij})h^{ij}\,,\quad \Pi_{ij}\;\equiv\;K_{ij}|_{\mathcal{B}}-K_{\mathcal{B}} h_{ij}+T h_{ij}\,.
    \label{eq:Strict_Neumann}
\end{equation}
These are Neumann boundary conditions in the strict sense because they impose that the extended Brown-York momentum $K_{ij}|_{\mathcal{B}}-K_{\mathcal{B}} h_{ij}+T h_{ij}$ is held fixed off-shell. By contrast, the metric on the boundary $\mathcal{B}$ is free. Having a variational principle which leads to \eqref{eq:Strict_Neumann} can be engineered through a different boundary term from \eqref{eq:EuclActionFull}, 
\begin{equation}
    I_{\mathcal{B}}=-\frac{(4-D)}{8\pi G_N}\int \sqrt{h}~ K_{\mathcal{B}}-\frac{1}{8\pi G_N}\int \sqrt{h}~ T\,.
\end{equation}
obtained by a gravitational Legendre transform \cite{KR16}. Note that we do not change ensemble, which means the integration is still over metrics and not over its canonical momentum: the Legendre transform is only for the boundary term.

\subsection{Gauge fixing with dynamical boundary}
\label{app:gauge_fixing}

In this subsection we explain the Fadeev-Popov procedure and gauge fixing of the Einstein-Hilbert action in the presence of a dynamical spacetime boundary. This question was raised and solved in \cite{Barvinsky05,Barvinsky06}, whose results we summarize. The main question is to understand which of the diffeomorphisms are gauge redundancy, and which ones are physical degrees of freedom. The answer is simple, the diffeomorphisms normal to $\mathcal{B}$ are real degrees of freedom since they move the boundary. By contrast, diffeomorphisms tangent to $\mathcal{B}$ remain redundancies of the spacetime boundary and hence must be fixed by an additional boundary gauge condition \cite{Barvinsky05,Barvinsky06}.

Let us consider the diffeomorphisms generated by the vector fields $\xi^{\mu}$ (bulk diffeo) and  $\chi^i$ (boundary diffeo) tangent to the spacetime boundary. The bulk metric $g$ and boundary metric $h$ transforms as:
\begin{equation}
\label{eq:diffeo_bulk_and_boundary}
    \delta_\xi g_{\mu\nu}=\nabla_{(\mu}\xi_{\nu)}\,,\quad  \delta_\chi h_{ij}=D_{(i}\chi_{j)}\,,
\end{equation}
where $\nabla_\mu$ is the covariant derivative in the bulk while $D_i$ is the derivative tangential to spacetime boundary. The condition that diffeomorphisms normal to the boundary vanish at the boundary $\mathcal{B}$ and the consistency of the gauge transformation leads to:
\begin{equation}
    \xi^{\perp}|_{\mathcal{B}}=0\,,\quad \xi^i|_\mathcal{B}=\chi^i\,.
\end{equation}
We should therefore gauge-fix the Einstein-Hilbert action with this constraint. It was shown in \cite{Barvinsky06} that we should independently gauge fix the diffeomorphisms generated by the two vector fields $\xi$ and $\chi$ introduced in \eqref{eq:diffeo_bulk_and_boundary}. We choose the de-Donder gauge, for both the bulk and boundary metrics:
\begin{equation}
\nabla^{\mu}g_{\mu\nu}-\frac{1}{2}\nabla_{\nu}g=0,\quad D^{i}h_{ij}-\frac{1}{2}D_{j}h=0\,,
\end{equation}
with the subtlety that Dirichlet boundary conditions should be imposed for the bulk diffeomorphisms on the spacetime boundary (all of the diffeomorphisms and not only the normal ones). This is of great importance for the ghost degrees of freedom. The associated Faddeev–Popov procedure leads to the functional integral:
\begin{equation}
\int_{\xi_{\mu}\overset{\partial M}{=}0} d\xi d\chi ~\delta( \nabla^{\mu}g_{\mu\nu}-\frac{1}{2}\nabla_{\nu}g)~\delta(D^{i}h_{ij}-\frac{1}{2}D_{j}h)=\frac{1}{\det_{D}\left(\nabla_{\alpha}\nabla^{\alpha} g_{\mu\nu}+R_{\mu\nu} \right)}\frac{1}{\det\left(D_{a}D^{a} h_{ij}+R_{ij}|_{\mathcal{B}} \right)}\,,
\end{equation}
where the subscript $D$ on $\det_{D}$ emphasizes that the determinant is taken with Dirichlet boundary conditions. The gauge-fixed gravitational path integral follows directly by adding the following term:

\begin{align}
1&=\int_{\xi_{\mu}\overset{\partial M}{=}0} d\xi d\chi ~\delta( \nabla^{\mu}g_{\mu\nu}-\frac{1}{2}\nabla_{\nu}g)~\delta(\nabla^{i}h_{ij}-\frac{1}{2}\nabla_{j}h)\det_{D}\left(\nabla_{\alpha}\nabla^{\alpha} g_{\mu\nu}+R_{\mu\nu} \right)\det\left(D_{a}D^{a} h_{ij}+R_{ij}|_{\mathcal{B}} \right) \nonumber \\
&=\int_{\xi_{\mu}\overset{\partial M}{=}0} d\xi d\chi~dc_D~ db_D~d\kappa ~d\beta   ~e^{\int dx^{d+1}\sqrt{g}~ \left(\nabla^{\mu}g_{\mu\nu}-\frac{1}{2}\nabla_{\nu}g \right)\left(\nabla^{\alpha}g_{\alpha}^{~\nu}-\frac{1}{2}\nabla^{\nu}g \right)}\\
&\hspace{5.5cm} e^{\int dx^d~\sqrt{h}~ \left(D^{~q}h_{q i}-\frac{1}{2}D_{i}h \right)\left(D^{j}h_{j}^{~i}-\frac{1}{2}D^{i}h \right)} \nonumber\\
&\hspace{6.5cm} e^{\int d^{d+1}x b^{\mu}\left(\nabla_{\alpha}\nabla^{\alpha} g_{\mu\nu}+R_{\mu\nu} \right)c^{\nu}}e^{\int d^{d}x \beta^{i}\left(D_{a}D^{a} h_{ij}+R_{ij}|_{\mathcal{B}}\right)\kappa^{i}}\,.\nonumber
\end{align}
Here the $b,c$ ghosts are subject to Dirichlet boundary conditions, and the $\beta,\kappa$ ghost is a $b,c$ ghost system on the spacetime boundary.

In this subsection we have shown how to gauge-fix the GPI to compute the full one-loop correction to the norm of the HH wavefunction. In Section.\ref{sec:One_loop_norm_GPI}, we do not solve the full one-loop problem, the boundary gravitons are off-shell while the bulk ones are kept on-shell. Therefore, the only gauge fixing condition we need to impose is the boundary one.

\subsection{Solving for the tensor modes}

\label{appendix:solving_for_the_tensor_mdoes}

In this subsection we solve for the tensor modes of equation \eqref{eq:linearized_einstein_equation} in Section.\ref{sec:solving_the_bulk_equations}; this was already done in \cite{HHR00} in arbitrary dimensions, so we shall be brief. For the tensor part of the $ij$ component of Einstein's equations, we can focus on:
\begin{equation}
    \nabla^2h_{\mu\nu}=-2h_{\mu\nu}\,.
\end{equation}
Decomposing this equation leads to:
\begin{equation}
    \partial_{uu}\phi_{ij}+((D-1)-4)\coth u\partial_{u}\phi_{ij}- 2\left((D-1)-2\right)\left(\coth^2 u\right)\phi_{ij}+\frac{1}{\sinh^2 u~}\hat{\nabla}^2\phi_{ij}=0\,.
\end{equation}
Since $\phi_{ij}(x)$ is fixed at the boundary, it can therefore be decomposed (in a Fourier-like decomposition) in a basis of transverse and traceless tensor spherical harmonics $Y_{ij}^{(l)}$, where obviously $l$ hides the usual degeneracies of each representation. We characterize $Y_{ij}^{(l)}$ by its eigenvalue:
\begin{equation}
    \nabla^{2}Y_{ij}^{(l)}=(2-l(l+(D-1)-1)Y_{ij}^{(l)}\,.
\end{equation}
The boundary transverse traceless part of the metric can be solved accordingly:
\begin{equation}
    \phi_{ij}(u,x)=\sum_l a_l Y_{ij}^{(l)}(x)\,,\quad a_l=\int d^{D-1}x\sqrt{\hat{H}}\phi^{ij}(x)Y_{ij}^{(l)}(x)\,.
\end{equation}

To solve Einstein's equations, each mode acquires an independent $u$ dependence, and we are left with the decomposition:
\begin{equation}
    \phi_{ij}(x)=\sum_l a_l f_l(u)Y_{ij}^{(l)}(x)\,,\quad a_l=\int d^{D-1}x\sqrt{\hat{H}}\phi^{ij}(x)Y_{ij}^{(l)}(x)\,.
\end{equation}
With boundary conditions $f_l(u=u_0)=1$. The solution is given in term of an hypergeometric function F:
\begin{equation}
    f_l(u)=\frac{\sinh^{l+2} u}{\cosh^l u}F\left(\frac{l}{2},\frac{l+1}{2},l+\frac{D+1}{2},\tanh^2 u\right)\,.
\end{equation}
which is smooth as $u$ goes to zero. The full solution follows:

\begin{equation}
    \phi_{ij}(x)=\sum_l a_l \frac{f_l(u)}{f_l(u_0)}Y_{ij}^{(l)}(x)\,,\quad a_l=\int d^{D-1}x\sqrt{\hat{H}}\phi^{ij}(x)Y_{ij}^{(l)}(x)\,.
\end{equation}
We solved for the most general smooth solution to the tensor sector of the linearized Einstein's equations around AdS$_D$. The solution is given in terms of the decomposition of the boundary condition $\phi_{ij}(x)$ into tensor spherical harmonics, each modes propagating differently in the bulk.

\bibliographystyle{jhep}
\bibliography{HHinAdS}

@article{Takayanagi11,
      author         = "Takayanagi, Tadashi",
      title          = "{Holographic Dual of BCFT}",
      journal        = "Phys. Rev. Lett.",
      volume         = "107",
      year           = "2011",
      pages          = "101602",
      doi            = "10.1103/PhysRevLett.107.101602",
      eprint         = "1105.5165",
      archivePrefix  = "arXiv",
      primaryClass   = "hep-th",
      reportNumber   = "IPMU11-0091",
      SLACcitation   = "%%CITATION = ARXIV:1105.5165;%%"
}

@article{FTT11,
      author         = "Fujita, Mitsutoshi and Takayanagi, Tadashi and Tonni,
                        Erik",
      title          = "{Aspects of AdS/BCFT}",
      journal        = "JHEP",
      volume         = "11",
      year           = "2011",
      pages          = "043",
      doi            = "10.1007/JHEP11(2011)043",
      eprint         = "1108.5152",
      archivePrefix  = "arXiv",
      primaryClass   = "hep-th",
      reportNumber   = "IPMU-11-0136, MIT-CTP-4289",
      SLACcitation   = "%%CITATION = ARXIV:1108.5152;%%"
}

@article{GKP98,
      author         = "Gubser, S. S. and Klebanov, Igor R. and Polyakov,
                        Alexander M.",
      title          = "{Gauge theory correlators from noncritical string
                        theory}",
      journal        = "Phys. Lett.",
      volume         = "B428",
      year           = "1998",
      pages          = "105-114",
      doi            = "10.1016/S0370-2693(98)00377-3",
      eprint         = "hep-th/9802109",
      archivePrefix  = "arXiv",
      primaryClass   = "hep-th",
      reportNumber   = "PUPT-1767",
      SLACcitation   = "%%CITATION = HEP-TH/9802109;%%"
}

@article{Witten98,
      author         = "Witten, Edward",
      title          = "{Anti-de Sitter space and holography}",
      journal        = "Adv. Theor. Math. Phys.",
      volume         = "2",
      year           = "1998",
      pages          = "253-291",
      doi            = "10.4310/ATMP.1998.v2.n2.a2",
      eprint         = "hep-th/9802150",
      archivePrefix  = "arXiv",
      primaryClass   = "hep-th",
      reportNumber   = "IASSNS-HEP-98-15",
      SLACcitation   = "%%CITATION = HEP-TH/9802150;%%"
}

@article{Maldacena97,
      author         = "Maldacena, Juan Martin",
      title          = "{The Large N limit of superconformal field theories and
                        supergravity}",
      journal        = "Int. J. Theor. Phys.",
      volume         = "38",
      year           = "1999",
      pages          = "1113-1133",
      doi            = "10.1023/A:1026654312961, 10.4310/ATMP.1998.v2.n2.a1",
      note           = "[Adv. Theor. Math. Phys.2,231(1998)]",
      eprint         = "hep-th/9711200",
      archivePrefix  = "arXiv",
      primaryClass   = "hep-th",
      reportNumber   = "HUTP-97-A097, HUTP-98-A097",
      SLACcitation   = "%%CITATION = HEP-TH/9711200;%%"
}

@article{GH77,
    author = "Gibbons, G. W. and Hawking, S. W.",
    title = "{Cosmological Event Horizons, Thermodynamics, and Particle Creation}",
    doi = "10.1103/PhysRevD.15.2738",
    journal = "Phys. Rev. D",
    volume = "15",
    pages = "2738--2751",
    year = "1977"
}

@article{RS99-1,
    author = "Randall, Lisa and Sundrum, Raman",
    title = "{A Large mass hierarchy from a small extra dimension}",
    eprint = "hep-ph/9905221",
    archivePrefix = "arXiv",
    reportNumber = "MIT-CTP-2860, PUPT-1860, BUHEP-99-9",
    doi = "10.1103/PhysRevLett.83.3370",
    journal = "Phys. Rev. Lett.",
    volume = "83",
    pages = "3370--3373",
    year = "1999"
}

@article{RS99-2,
    author = "Randall, Lisa and Sundrum, Raman",
    title = "{An Alternative to compactification}",
    eprint = "hep-th/9906064",
    archivePrefix = "arXiv",
    reportNumber = "MIT-CTP-2874, PUPT-1867, BUHEP-99-13",
    doi = "10.1103/PhysRevLett.83.4690",
    journal = "Phys. Rev. Lett.",
    volume = "83",
    pages = "4690--4693",
    year = "1999"
}

@article{AKRTW21,
    author = "Akal, Ibrahim and Kawamoto, Taishi and Ruan, Shan-Ming and Takayanagi, Tadashi and Wei, Zixia",
    title = "{Page curve under final state projection}",
    eprint = "2112.08433",
    archivePrefix = "arXiv",
    primaryClass = "hep-th",
    reportNumber = "YITP-21-158, IPMU21-0086, YITP-21-158; IPMU21-0086",
    doi = "10.1103/PhysRevD.105.126026",
    journal = "Phys. Rev. D",
    volume = "105",
    number = "12",
    pages = "126026",
    year = "2022"
}

@article{AKTW20,
    author = "Akal, Ibrahim and Kusuki, Yuya and Takayanagi, Tadashi and Wei, Zixia",
    title = "{Codimension two holography for wedges}",
    eprint = "2007.06800",
    archivePrefix = "arXiv",
    primaryClass = "hep-th",
    reportNumber = "YITP-20-91, IPMU20-0079",
    doi = "10.1103/PhysRevD.102.126007",
    journal = "Phys. Rev. D",
    volume = "102",
    number = "12",
    pages = "126007",
    year = "2020"
}

@article{Wei24,
    author = "Wei, Zixia",
    title = "{Holographic dual of crosscap conformal field theory}",
    eprint = "2405.03755",
    archivePrefix = "arXiv",
    primaryClass = "hep-th",
    doi = "10.1007/JHEP03(2025)086",
    journal = "JHEP",
    volume = "03",
    pages = "086",
    year = "2025"
}

@article{AKRTW22,
    author = "Akal, Ibrahim and Kawamoto, Taishi and Ruan, Shan-Ming and Takayanagi, Tadashi and Wei, Zixia",
    title = "{Zoo of holographic moving mirrors}",
    eprint = "2205.02663",
    archivePrefix = "arXiv",
    primaryClass = "hep-th",
    reportNumber = "YITP-22-42, IPMU22-0023, YITP-22-42; IPMU22-0023",
    doi = "10.1007/JHEP08(2022)296",
    journal = "JHEP",
    volume = "08",
    pages = "296",
    year = "2022"
}

@article{KR00,
    author = "Karch, Andreas and Randall, Lisa",
    editor = "Duff, Michael J. and Liu, J. T. and Lu, J.",
    title = "{Locally localized gravity}",
    eprint = "hep-th/0011156",
    archivePrefix = "arXiv",
    reportNumber = "MIT-CTP-3099",
    doi = "10.1088/1126-6708/2001/05/008",
    journal = "JHEP",
    volume = "05",
    pages = "008",
    year = "2001"
}

@article{WY24,
    author = "Wei, Zixia and Yoneta, Yasushi",
    title = "{Crosscap Quenches and Entanglement Evolution}",
    eprint = "2412.18610",
    archivePrefix = "arXiv",
    primaryClass = "hep-th",
    month = "12",
    year = "2024"
}

@article{HH83,
    author = "Hartle, J. B. and Hawking, S. W.",
    editor = "Fang, Li-Zhi and Ruffini, R.",
    title = "{Wave Function of the Universe}",
    reportNumber = "PRINT-83-0937 (CAMBRIDGE)",
    doi = "10.1103/PhysRevD.28.2960",
    journal = "Phys. Rev. D",
    volume = "28",
    pages = "2960--2975",
    year = "1983"
}

@article{ILM24,
    author = "Ivo, Victor and Li, Yue-Zhou and Maldacena, Juan",
    title = "{The no boundary density matrix}",
    eprint = "2409.14218",
    archivePrefix = "arXiv",
    primaryClass = "hep-th",
    doi = "10.1007/JHEP02(2025)124",
    journal = "JHEP",
    volume = "02",
    pages = "124",
    year = "2025"
}

@article{Wei25,
    author = "Wei, Zixia",
    title = "{Observers and Timekeepers: From the Page-Wootters Mechanism to the Gravitational Path Integral}",
    eprint = "2506.21489",
    archivePrefix = "arXiv",
    primaryClass = "hep-th",
    month = "6",
    year = "2025"
}

@article{NU25,
    author = "Nomura, Yasunori and Ugajin, Tomonori",
    title = "{Nonperturbative quantum gravity in a closed Lorentzian universe}",
    eprint = "2505.20390",
    archivePrefix = "arXiv",
    primaryClass = "hep-th",
    reportNumber = "RIKEN-iTHEMS-Report-25",
    doi = "10.1007/JHEP10(2025)166",
    journal = "JHEP",
    volume = "10",
    pages = "166",
    year = "2025"
}

@article{Maldacena02,
    author = "Maldacena, Juan Martin",
    title = "{Non-Gaussian features of primordial fluctuations in single field inflationary models}",
    eprint = "astro-ph/0210603",
    archivePrefix = "arXiv",
    doi = "10.1088/1126-6708/2003/05/013",
    journal = "JHEP",
    volume = "05",
    pages = "013",
    year = "2003"
}

@article{BCT20,
    author = "Boruch, Jan and Caputa, Pawel and Takayanagi, Tadashi",
    title = "{Path-Integral Optimization from Hartle-Hawking Wave Function}",
    eprint = "2011.08188",
    archivePrefix = "arXiv",
    primaryClass = "hep-th",
    reportNumber = "YITP-20-147, IPMU20-0119",
    doi = "10.1103/PhysRevD.103.046017",
    journal = "Phys. Rev. D",
    volume = "103",
    number = "4",
    pages = "046017",
    year = "2021"
}

@article{BCGT21,
    author = "Boruch, Jan and Caputa, Pawel and Ge, Dongsheng and Takayanagi, Tadashi",
    title = "{Holographic path-integral optimization}",
    eprint = "2104.00010",
    reportNumber = "YITP-21-25, IPMU21-0022",
    doi = "10.1007/JHEP07(2021)016",
    journal = "JHEP",
    volume = "07",
    pages = "016",
    year = "2021",
    note = "[Erratum: JHEP 09, 111 (2022)]"
}

@article{Maldacena24,
    author = "Maldacena, Juan",
    title = "{Real observers solving imaginary problems}",
    eprint = "2412.14014",
    archivePrefix = "arXiv",
    primaryClass = "hep-th",
    month = "12",
    year = "2024"
}

@article{IMS25,
    author = "Ivo, Victor and Maldacena, Juan and Sun, Zimo",
    title = "{Physical instabilities and the phase of the Euclidean path integral}",
    eprint = "2504.00920",
    archivePrefix = "arXiv",
    primaryClass = "hep-th",
    doi = "10.1007/JHEP04(2026)118",
    journal = "JHEP",
    volume = "04",
    pages = "118",
    year = "2026"
}

@article{ST25,
    author = "Shi, Xiaoyi and Turiaci, Gustavo J.",
    title = "{The phase of the gravitational path integral}",
    eprint = "2504.00900",
    archivePrefix = "arXiv",
    primaryClass = "hep-th",
    doi = "10.1007/JHEP07(2025)047",
    journal = "JHEP",
    volume = "07",
    pages = "047",
    year = "2025"
}

@article{CJ25,
    author = "Cotler, Jordan and Jensen, Kristan",
    title = "{Norm of the no-boundary state}",
    eprint = "2506.20547",
    archivePrefix = "arXiv",
    primaryClass = "hep-th",
    doi = "10.1007/JHEP03(2026)180",
    journal = "JHEP",
    volume = "03",
    pages = "180",
    year = "2026"
}

@article{CSTY25,
    author = "Chen, Yiming and Stanford, Douglas and Tang, Haifeng and Yang, Zhenbin",
    title = "{On the phase of the de Sitter density of states}",
    eprint = "2511.01400",
    archivePrefix = "arXiv",
    primaryClass = "hep-th",
    doi = "10.1007/JHEP05(2026)068",
    journal = "JHEP",
    volume = "05",
    pages = "068",
    year = "2026"
}

@article{Polchinski88,
    author = "Polchinski, Joseph",
    title = "{The phase of the sum over spheres}",
    reportNumber = "UTTG-31-88",
    doi = "10.1016/0370-2693(89)90387-0",
    journal = "Phys. Lett. B",
    volume = "219",
    pages = "251--257",
    year = "1989"
}

@article{HJ18,
    author = "Harlow, Daniel and Jafferis, Daniel",
    title = "{The Factorization Problem in Jackiw-Teitelboim Gravity}",
    eprint = "1804.01081",
    archivePrefix = "arXiv",
    primaryClass = "hep-th",
    doi = "10.1007/JHEP02(2020)177",
    journal = "JHEP",
    volume = "02",
    pages = "177",
    year = "2020"
}

@article{Barvinsky05,
    author = "Barvinsky, A. O. and Nesterov, D. V.",
    title = "{Quantum effective action in spacetimes with branes and boundaries}",
    eprint = "hep-th/0512291",
    archivePrefix = "arXiv",
    doi = "10.1103/PhysRevD.73.066012",
    journal = "Phys. Rev. D",
    volume = "73",
    pages = "066012",
    year = "2006"
}

@article{Barvinsky06,
    author = "Barvinsky, A. O.",
    title = "{Quantum Effective Action in Spacetimes with Branes and Boundaries: Diffeomorphism Invariance}",
    eprint = "hep-th/0608004",
    archivePrefix = "arXiv",
    doi = "10.1103/PhysRevD.74.084033",
    journal = "Phys. Rev. D",
    volume = "74",
    pages = "084033",
    year = "2006"
}

@article{CJ23,
    author = "Chua, Wan Zhen and Jiang, Yikun",
    title = "{Hartle-Hawking state and its factorization in 3d gravity}",
    eprint = "2309.05126",
    archivePrefix = "arXiv",
    primaryClass = "hep-th",
    doi = "10.1007/JHEP03(2024)135",
    journal = "JHEP",
    volume = "03",
    pages = "135",
    year = "2024"
}

@article{MTY19,
    author = "Maldacena, Juan and Turiaci, Gustavo J. and Yang, Zhenbin",
    title = "{Two dimensional Nearly de Sitter gravity}",
    eprint = "1904.01911",
    archivePrefix = "arXiv",
    primaryClass = "hep-th",
    doi = "10.1007/JHEP01(2021)139",
    journal = "JHEP",
    volume = "01",
    pages = "139",
    year = "2021"
}

@article{CJM19,
    author = "Cotler, Jordan and Jensen, Kristan and Maloney, Alexander",
    title = "{Low-dimensional de Sitter quantum gravity}",
    eprint = "1905.03780",
    archivePrefix = "arXiv",
    primaryClass = "hep-th",
    doi = "10.1007/JHEP06(2020)048",
    journal = "JHEP",
    volume = "06",
    pages = "048",
    year = "2020"
}

@article{Maldacena24-1,
    author = "Maldacena, Juan",
    title = "{Comments on the no boundary wavefunction and slow roll inflation}",
    eprint = "2403.10510",
    archivePrefix = "arXiv",
    primaryClass = "hep-th",
    month = "3",
    year = "2024"
}

@article{Lehners23,
    author = "Lehners, Jean-Luc",
    title = "{Review of the no-boundary wave function}",
    eprint = "2303.08802",
    archivePrefix = "arXiv",
    primaryClass = "hep-th",
    doi = "10.1016/j.physrep.2023.06.002",
    journal = "Phys. Rept.",
    volume = "1022",
    pages = "1--82",
    year = "2023"
}

@article{CM08,
    author = "Compere, Geoffrey and Marolf, Donald",
    title = "{Setting the boundary free in AdS/CFT}",
    eprint = "0805.1902",
    archivePrefix = "arXiv",
    primaryClass = "hep-th",
    doi = "10.1088/0264-9381/25/19/195014",
    journal = "Class. Quant. Grav.",
    volume = "25",
    pages = "195014",
    year = "2008"
}

@article{KR16,
    author = "Krishnan, Chethan and Raju, Avinash",
    title = "{A Neumann Boundary Term for Gravity}",
    eprint = "1605.01603",
    archivePrefix = "arXiv",
    primaryClass = "hep-th",
    doi = "10.1142/S0217732317500778",
    journal = "Mod. Phys. Lett. A",
    volume = "32",
    number = "14",
    pages = "1750077",
    year = "2017"
}

@article{Witten18,
    author = "Witten, Edward",
    title = "{A note on boundary conditions in Euclidean gravity}",
    eprint = "1805.11559",
    archivePrefix = "arXiv",
    primaryClass = "hep-th",
    doi = "10.1142/S0129055X21400043",
    journal = "Rev. Math. Phys.",
    volume = "33",
    number = "10",
    pages = "2140004",
    year = "2021"
}

@article{HHR01,
    author = "Hawking, S. W. and Hertog, T. and Reall, H. S.",
    title = "{Trace anomaly driven inflation}",
    eprint = "hep-th/0010232",
    archivePrefix = "arXiv",
    reportNumber = "DAMTP-2000-92, QMW-PH-00-10",
    doi = "10.1103/PhysRevD.63.083504",
    journal = "Phys. Rev. D",
    volume = "63",
    pages = "083504",
    year = "2001"
}

@article{HHR00,
    author = "Hawking, S. W. and Hertog, T. and Reall, H. S.",
    title = "{Brane new world}",
    eprint = "hep-th/0003052",
    archivePrefix = "arXiv",
    reportNumber = "DAMTP-2000-25",
    doi = "10.1103/PhysRevD.62.043501",
    journal = "Phys. Rev. D",
    volume = "62",
    pages = "043501",
    year = "2000"
}

@article{Freidel08,
    author = "Freidel, Laurent",
    title = "{Reconstructing AdS/CFT}",
    eprint = "0804.0632",
    archivePrefix = "arXiv",
    primaryClass = "hep-th",
    month = "4",
    year = "2008"
}

@article{IKTV20,
    author = "Iliesiu, Luca V. and Kruthoff, Jorrit and Turiaci, Gustavo J. and Verlinde, Herman",
    title = "{JT gravity at finite cutoff}",
    eprint = "2004.07242",
    archivePrefix = "arXiv",
    primaryClass = "hep-th",
    doi = "10.21468/SciPostPhys.9.2.023",
    journal = "SciPost Phys.",
    volume = "9",
    pages = "023",
    year = "2020"
}

@article{ABM24,
    author = {Anninos, Dionysios and Baracco, Chiara and M{\"u}hlmann, Beatrix},
    title = "{Remarks on 2D quantum cosmology}",
    eprint = "2406.15271",
    archivePrefix = "arXiv",
    primaryClass = "hep-th",
    doi = "10.1088/1475-7516/2024/10/031",
    journal = "JCAP",
    volume = "10",
    pages = "031",
    year = "2024"
}

@article{AHK25,
    author = "Anninos, Dionysios and Hertog, Thomas and Karlsson, Joel",
    title = "{Quantum Liouville Cosmology}",
    eprint = "2512.15969",
    archivePrefix = "arXiv",
    primaryClass = "hep-th",
    month = "12",
    year = "2025"
}

@article{TW25,
    author = "Turiaci, Gustavo J. and Wu, Chih-Hung",
    title = "{The wavefunction of a quantum S$^{1}$ {\texttimes} S$^{2}$ universe}",
    eprint = "2503.14639",
    archivePrefix = "arXiv",
    primaryClass = "hep-th",
    doi = "10.1007/JHEP07(2025)158",
    journal = "JHEP",
    volume = "07",
    pages = "158",
    year = "2025"
}

@article{Maldacena2024-2,
    author = "Maldacena, Juan",
    title = "{Real observers solving imaginary problems}",
    eprint = "2412.14014",
    archivePrefix = "arXiv",
    primaryClass = "hep-th",
    month = "12",
    year = "2024"
}

@article{ElvangHad16,
    author = "Elvang, Henriette and Hadjiantonis, Marios",
    title = "{A Practical Approach to the Hamilton-Jacobi Formulation of Holographic Renormalization}",
    eprint = "1603.04485",
    archivePrefix = "arXiv",
    primaryClass = "hep-th",
    doi = "10.1007/JHEP06(2016)046",
    journal = "JHEP",
    volume = "06",
    pages = "046",
    year = "2016"
}

@article{AKST04,
    author = "Alishahiha, Mohsen and Karch, Andreas and Silverstein, Eva and Tong, David",
    editor = "Allen, R. E. and Nanopoulos, Dimitri V. and Pope, C. N.",
    title = "{The dS/dS correspondence}",
    eprint = "hep-th/0407125",
    archivePrefix = "arXiv",
    reportNumber = "SLAC-PUB-10540, SU-ITP-04-29, IPM-P-2004-31, MIT-CTP-3512, UW-PT-04-07",
    doi = "10.1063/1.1848341",
    journal = "AIP Conf. Proc.",
    volume = "743",
    number = "1",
    pages = "393--409",
    year = "2004"
}

@article{DHST10,
    author = "Dong, Xi and Horn, Bart and Silverstein, Eva and Torroba, Gonzalo",
    title = "{Micromanaging de Sitter holography}",
    eprint = "1005.5403",
    archivePrefix = "arXiv",
    primaryClass = "hep-th",
    reportNumber = "NSF-KITP-10-068, SU-ITP-10-19, SLAC-PUB-14146",
    doi = "10.1088/0264-9381/27/24/245020",
    journal = "Class. Quant. Grav.",
    volume = "27",
    pages = "245020",
    year = "2010"
}

@article{Cotler26,
    author = "Cotler, Jordan",
    title = "{Higher-loop norm of the no-boundary state}",
    eprint = "2601.20993",
    archivePrefix = "arXiv",
    primaryClass = "hep-th",
    month = "1",
    year = "2026"
}

@article{RO84,
    author = {Rubin, Mark A. and Ordóñez, Carlos R.},
    title = {Eigenvalues and degeneracies for n‐dimensional tensor spherical harmonics},
    journal = {Journal of Mathematical Physics},
    volume = {25},
    number = {10},
    pages = {2888-2894},
    year = {1984},
    month = {10},
    issn = {0022-2488},
    doi = {10.1063/1.526034},
    url = {https://doi.org/10.1063/1.526034}
}

@article{AKW22,
    author = "Araujo-Regado, Goncalo and Khan, Rifath and Wall, Aron C.",
    title = "{Cauchy slice holography: a new AdS/CFT dictionary}",
    eprint = "2204.00591",
    archivePrefix = "arXiv",
    primaryClass = "hep-th",
    doi = "10.1007/JHEP03(2023)026",
    journal = "JHEP",
    volume = "03",
    pages = "026",
    year = "2023"
}

@article{CKP20,
    author = "Caputa, Pawel and Kruthoff, Jorrit and Parrikar, Onkar",
    title = "{Building Tensor Networks for Holographic States}",
    eprint = "2012.05247",
    archivePrefix = "arXiv",
    primaryClass = "hep-th",
    doi = "10.1007/JHEP05(2021)009",
    journal = "JHEP",
    volume = "05",
    pages = "009",
    year = "2021",
    note = "[Erratum: JHEP 09, 112 (2022)]"
}

@article{Page86,
    author = "Page, Don N.",
    title = "{Density Matrix of the Universe}",
    reportNumber = "Print-86-0934 (PENN STATE)",
    doi = "10.1103/PhysRevD.34.2267",
    journal = "Phys. Rev. D",
    volume = "34",
    pages = "2267",
    year = "1986"
}

@article{DW25,
    author = {Dulac, Rapha{\"e}l and Wei, Zixia},
    title = "{No boundary density matrix in elliptic de Sitter dS/$\mathbb{Z}_2$}",
    eprint = "2512.00704",
    archivePrefix = "arXiv",
    primaryClass = "hep-th",
    doi = "10.1007/JHEP05(2026)022",
    journal = "JHEP",
    volume = "05",
    pages = "022",
    year = "2026"
}

@article{DQSY20,
    author = "Dong, Xi and Qi, Xiao-Liang and Shangnan, Zhou and Yang, Zhenbin",
    title = "{Effective entropy of quantum fields coupled with gravity}",
    eprint = "2007.02987",
    archivePrefix = "arXiv",
    primaryClass = "hep-th",
    doi = "10.1007/JHEP10(2020)052",
    journal = "JHEP",
    volume = "10",
    pages = "052",
    year = "2020"
}

@article{AKM20,
    author = "Anous, Tarek and Kruthoff, Jorrit and Mahajan, Raghu",
    title = "{Density matrices in quantum gravity}",
    eprint = "2006.17000",
    archivePrefix = "arXiv",
    primaryClass = "hep-th",
    doi = "10.21468/SciPostPhys.9.4.045",
    journal = "SciPost Phys.",
    volume = "9",
    number = "4",
    pages = "045",
    year = "2020"
}

@article{Wei26,
    author = {Wei, Zixia},
    title = "to appear"
}

@article{GHP78,
title = {Path integrals and the indefiniteness of the gravitational action},
journal = {Nuclear Physics B},
volume = {138},
number = {1},
pages = {141-150},
year = {1978},
issn = {0550-3213},
doi = {https://doi.org/10.1016/0550-3213(78)90161-X},
url = {https://www.sciencedirect.com/science/article/pii/055032137890161X},
author = {G.W. Gibbons and S.W. Hawking and M.J. Perry}
}

@article{CLM11,
    author = "Castro, Alejandra and Lashkari, Nima and Maloney, Alexander",
    title = "{A de Sitter Farey Tail}",
    eprint = "1103.4620",
    archivePrefix = "arXiv",
    primaryClass = "hep-th",
    doi = "10.1103/PhysRevD.83.124027",
    journal = "Phys. Rev. D",
    volume = "83",
    pages = "124027",
    year = "2011"
}

@article{ADLS20,
    author = "Anninos, Dionysios and Denef, Frederik and Law, Y. T. Albert and Sun, Zimo",
    title = "{Quantum de Sitter horizon entropy from quasicanonical bulk, edge, sphere and topological string partition functions}",
    eprint = "2009.12464",
    archivePrefix = "arXiv",
    primaryClass = "hep-th",
    doi = "10.1007/JHEP01(2022)088",
    journal = "JHEP",
    volume = "01",
    pages = "088",
    year = "2022"
}

@article{IT26,
    author = "Ivo, Victor and Tang, Haifeng",
    title = "{One-loop aspects of de Sitter axion wormholes}",
    eprint = "2603.02335",
    archivePrefix = "arXiv",
    primaryClass = "hep-th",
    month = "3",
    year = "2026"
}

@article{GS26,
    author = "Giombi, Simone and Sun, Zimo",
    title = "{The phase of de Sitter higher spin gravity}",
    eprint = "2601.15257",
    archivePrefix = "arXiv",
    primaryClass = "hep-th",
    month = "1",
    year = "2026"
}

@article{GIKY20,
    author = "Goel, Akash and Iliesiu, Luca V. and Kruthoff, Jorrit and Yang, Zhenbin",
    title = "{Classifying boundary conditions in JT gravity: from energy-branes to $\alpha$-branes}",
    eprint = "2010.12592",
    archivePrefix = "arXiv",
    primaryClass = "hep-th",
    doi = "10.1007/JHEP04(2021)069",
    journal = "JHEP",
    volume = "04",
    pages = "069",
    year = "2021"
}

@article{GMY08,
    author = "Giombi, Simone and Maloney, Alexander and Yin, Xi",
    title = "{One-loop Partition Functions of 3D Gravity}",
    eprint = "0804.1773",
    archivePrefix = "arXiv",
    primaryClass = "hep-th",
    doi = "10.1088/1126-6708/2008/08/007",
    journal = "JHEP",
    volume = "08",
    pages = "007",
    year = "2008"
}

@article{Suzuki2021,
    author = "Suzuki, Yu-ki",
    title = "{One-loop correction to the AdS/BCFT partition function in three-dimensional pure gravity}",
    eprint = "2106.00206",
    archivePrefix = "arXiv",
    primaryClass = "hep-th",
    reportNumber = "YITP-21-50",
    doi = "10.1103/PhysRevD.105.026023",
    journal = "Phys. Rev. D",
    volume = "105",
    number = "2",
    pages = "026023",
    year = "2022"
}

@article{Sun2020,
    author = "Sun, Zimo",
    title = "{AdS one-loop partition functions from bulk and edge characters}",
    eprint = "2010.15826",
    archivePrefix = "arXiv",
    primaryClass = "hep-th",
    doi = "10.1007/JHEP12(2021)064",
    journal = "JHEP",
    volume = "12",
    pages = "064",
    year = "2021"
}

@article{GKT14,
    author = "Giombi, Simone and Klebanov, Igor R. and Tseytlin, Arkady A.",
    title = "{Partition Functions and Casimir Energies in Higher Spin AdS$_{d+1}$/CFT$_d$}",
    eprint = "1402.5396",
    archivePrefix = "arXiv",
    primaryClass = "hep-th",
    reportNumber = "PUTP-2460, IMPERIAL-TP-AT-2014-01",
    doi = "10.1103/PhysRevD.90.024048",
    journal = "Phys. Rev. D",
    volume = "90",
    number = "2",
    pages = "024048",
    year = "2014"
}

@article{Witten21-2,
    author = "Witten, Edward",
    title = "{A Note On Complex Spacetime Metrics}",
    eprint = "2111.06514",
    archivePrefix = "arXiv",
    primaryClass = "hep-th",
    month = "11",
    year = "2021"
}

@article{Gubser99,
    author = "Gubser, Steven S.",
    title = "{AdS / CFT and gravity}",
    eprint = "hep-th/9912001",
    archivePrefix = "arXiv",
    reportNumber = "HUTP-99-A065",
    doi = "10.1103/PhysRevD.63.084017",
    journal = "Phys. Rev. D",
    volume = "63",
    pages = "084017",
    year = "2001"
}

@article{GT99,
    author = "Garriga, Jaume and Tanaka, Takahiro",
    title = "{Gravity in the brane world}",
    eprint = "hep-th/9911055",
    archivePrefix = "arXiv",
    reportNumber = "UAB-FT-476, OU-TAP-106",
    doi = "10.1103/PhysRevLett.84.2778",
    journal = "Phys. Rev. Lett.",
    volume = "84",
    pages = "2778--2781",
    year = "2000"
}

@article{SMS99,
    author = "Shiromizu, Tetsuya and Maeda, Kei-ichi and Sasaki, Misao",
    title = "{The Einstein equation on the 3-brane world}",
    eprint = "gr-qc/9910076",
    archivePrefix = "arXiv",
    reportNumber = "DAMTP-1999-150, OUTAP-103, UTAP-349, RESCEU-40-99",
    doi = "10.1103/PhysRevD.62.024012",
    journal = "Phys. Rev. D",
    volume = "62",
    pages = "024012",
    year = "2000"
}

@article{FKKT25,
    author = "Fujiki, Kosei and Kanda, Hiroki and Kohara, Michitaka and Takayanagi, Tadashi",
    title = "{Brane cosmology from AdS/BCFT}",
    eprint = "2501.05036",
    archivePrefix = "arXiv",
    primaryClass = "hep-th",
    reportNumber = "YITP-24-180",
    doi = "10.1007/JHEP03(2025)135",
    journal = "JHEP",
    volume = "03",
    pages = "135",
    year = "2025"
}

@article{ST22BCFT,
    author = "Suzuki, Kenta and Takayanagi, Tadashi",
    title = "{BCFT and Islands in two dimensions}",
    eprint = "2202.08462",
    archivePrefix = "arXiv",
    primaryClass = "hep-th",
    reportNumber = "YITP-22-14, IPMU22-0002",
    doi = "10.1007/JHEP06(2022)095",
    journal = "JHEP",
    volume = "06",
    pages = "095",
    year = "2022"
}

@article{NSS24,
    author = "Neuenfeld, Dominik and Svesko, Andrew and Sybesma, Watse",
    title = "{Liouville gravity at the end of the world:deformed defects in AdS/BCFT}",
    eprint = "2404.07260",
    archivePrefix = "arXiv",
    primaryClass = "hep-th",
    doi = "10.1007/JHEP07(2024)215",
    journal = "JHEP",
    volume = "07",
    pages = "215",
    year = "2024"
}

@article{LXH25,
    author = "Li, Yun-Ze and Xie, Yunfei and He, Song",
    title = "{Holographic correlators of boundary/crosscap CFTs in two dimensions}",
    eprint = "2501.18386",
    archivePrefix = "arXiv",
    primaryClass = "hep-th",
    doi = "10.1007/JHEP07(2025)010",
    journal = "JHEP",
    volume = "07",
    pages = "010",
    year = "2025"
}

@article{BBLST26,
    author = "Banihashemi, Batoul and Batra, Gauri and Law, Albert Y. T. and Silverstein, Eva and Torroba, Gonzalo",
    title = "{The yes boundaries wavefunctions of the universe}",
    eprint = "2604.10267",
    archivePrefix = "arXiv",
    primaryClass = "hep-th",
    month = "4",
    year = "2026"
}

@article{STW26,
    author = "Shi, Xiaoyi and Turiaci, Gustavo J. and Wu, Chih-Hung",
    title = "{The Fate of Nucleated Black Holes in de Sitter Quantum Gravity}",
    eprint = "2605.03015",
    archivePrefix = "arXiv",
    primaryClass = "hep-th",
    month = "5",
    year = "2026"
}

@article{NU26,
    author = "Nomura, Yasunori and Ugajin, Tomonori",
    title = "{Physical Predictions in Closed Quantum Gravity}",
    eprint = "2602.13387",
    archivePrefix = "arXiv",
    primaryClass = "hep-th",
    reportNumber = "RIKEN-iTHEMS-Report-26",
    month = "2",
    year = "2026"
}

@article{BC26,
    author = "Blommaert, Andreas and Chen, Chang-Han",
    title = "{Time in gravitational subregions and in closed universes}",
    eprint = "2602.22153",
    archivePrefix = "arXiv",
    primaryClass = "hep-th",
    month = "2",
    year = "2026"
}

@article{Godet2026,
    author = "Godet, Victor",
    title = "{Inflation and topology from the no-boundary state}",
    eprint = "2605.05317",
    archivePrefix = "arXiv",
    primaryClass = "hep-th",
    month = "5",
    year = "2026"
}

@article{AABILS26,
    author = "Abdalla, Ahmed I. and Antonini, Stefano and Bousso, Raphael and Iliesiu, Luca V. and Levine, Adam and Shahbazi-Moghaddam, Arvin",
    title = "{Consistent Evaluation of the No-Boundary Proposal}",
    eprint = "2602.02682",
    archivePrefix = "arXiv",
    primaryClass = "hep-th",
    month = "2",
    year = "2026"
}

@article{Harlow26,
    author = "Harlow, Daniel",
    title = "{Observers, $\alpha$-parameters, and the Hartle-Hawking state}",
    eprint = "2602.03835",
    archivePrefix = "arXiv",
    primaryClass = "hep-th",
    reportNumber = "MIT-CTP/5900",
    month = "2",
    year = "2026"
}

@article{Zhao26,
    author = "Zhao, Ying",
    title = "{''It from Bit'': The Hartle-Hawking state and quantum mechanics for de Sitter observers}",
    eprint = "2602.05939",
    archivePrefix = "arXiv",
    primaryClass = "hep-th",
    reportNumber = "MIT-CTP/6000",
    month = "2",
    year = "2026"
}

\end{document}